\documentclass[11pt,a4paper]{article}
\pdfoutput=1

\usepackage{jcappub}

\usepackage{mathrsfs}
\usepackage{bbm}
\usepackage{bm}
\usepackage{dsfont}
\usepackage{tensor}
\usepackage{xspace}
\usepackage{lmodern}
\usepackage{wasysym}
\usepackage{microtype}
\usepackage{empheq}
\usepackage{ifthen}
\usepackage{amsmath}
\usepackage{wrapfig}
\usepackage{cleveref}
\crefname{section}{sec.}{secs.}
\Crefname{section}{Section}{Sections}
\crefname{subsection}{subsec.}{subsecs.}
\Crefname{subsection}{Subsection}{Subsections}
\crefname{subsubsection}{subsubsec.}{subsubsecs.}
\Crefname{subsubsection}{Subsubsection}{Subsubsections}
\usepackage{xcolor}
\usepackage{import}
\usepackage{transparent}
\newcommand\ph{\ensuremath{\varphi}}
\newcommand\eps{\ensuremath{\varepsilon}}

\newcommand{\cst}{\mathrm{cst}}

\newcommand\define{\equiv}

\newcommand\vect[1]{\boldsymbol{#1}}
\newcommand\mat[1]{\boldsymbol{#1}}
\newcommand\cplx[1]{\underline{#1}}
\newcommand\ex[1]{\mathrm{e}^{#1}}
\newcommand\ii{\mathrm{i}}

\renewcommand\Re{\mathrm{Re}}
\renewcommand\Im{\mathrm{Im}}

\newcommand{\order}{\mathcal{O}}

\newcommand\e[1]{_{\mathrm{#1}}}
\newcommand\h[1]{^{\mathrm{#1}}}
\newcommand\U[1]{\:\mathrm{#1}}

\newcommand{\dd}{\mathrm{d}}

\newcommand{\pd}[3][]{\frac{\partial^{#1} #2}{\partial {#3}^{#1}}}
\newcommand{\ddf}[3][]{\frac{\dd^{#1} #2}{\dd {#3}^{#1}}}

\newcommand{\delimiters}[4][]{
\ifthenelse{ \equal{#1}{1} }{  #2 #3 #4  }
					{ \ifthenelse{\equal{#1}{2}}{ \big#2 #3 \big#4 }
						{ \ifthenelse{\equal{#1}{3}}{ \Big#2 #3 \Big#4 }
							{ \ifthenelse{\equal{#1}{4}}{ \bigg#2 #3 \bigg#4 }
								{ \ifthenelse{\equal{#1}{5}}{ \Bigg#2 #3 \Bigg#4 }
									{ \left#2 #3 \right#4 }
								}
							}
						}
					}
													}
													
\newcommand{\pa}[2][]{\delimiters[#1]{(}{#2}{)}}
\newcommand{\pac}[2][]{\delimiters[#1]{[}{#2}{]}}
\newcommand{\paac}[2][]{\delimiters[#1]{\{}{#2}{\}}}
\newcommand{\abs}[2][]{\delimiters[#1]{|}{#2}{|}}
\newcommand{\ev}[2][]{\delimiters[#1]{\langle}{#2}{\rangle}}

\definecolor{blue4}{RGB}{0,0,143}
\definecolor{red4}{RGB}{143,0,0}
\definecolor{orange}{RGB}{255,128,0}
\definecolor{darkcyan}{RGB}{0,128,128}
\definecolor{olive}{RGB}{0,128,0}
\definecolor{purple}{RGB}{128,0,128}
\definecolor{cyan2}{RGB}{0,255,255}
\definecolor{fushia}{RGB}{255,0,255}
\definecolor{mygray}{gray}{0.5}
\definecolor{lightgray}{gray}{0.85}

\newcommand{\amplification}{\mathcal{A}}
\newcommand{\unitcircle}{\mathscr{C}_1}
\newcommand{\F}{\mathcal{F}}
\newcommand{\G}{\mathcal{G}}

\title{Line-of-sight effects in strong gravitational lensing}

\author[a]{Pierre Fleury,}
\author[b]{Julien Larena,}
\author[c,d]{Jean-Philippe Uzan.}
\affiliation[a]{Instituto de Física Teórica UAM-CSIC,
Universidad Autónoma de Madrid,\\
Cantoblanco, 28049 Madrid, Spain}
\affiliation[b]{Department of Mathematics and Applied Mathematics, University of Cape Town,\\
Rondebosch 7701, South Africa}
\affiliation[c]{Institut d'Astrophysique de Paris, CNRS UMR 7095, Sorbonne Universités,\\
98 bis Boulevard Arago, 75014 Paris, France}
\affiliation[d]{Sorbonne Universit\'es, Institut Lagrange de Paris,\\
98 bis, Boulevard Arago, 75014 Paris, France}

\emailAdd{pierre.fleury@uam.es}
\emailAdd{julien.larena@uct.ac.za}
\emailAdd{uzan@iap.fr}

\abstract{
While most strong-gravitational-lensing systems may be roughly modelled by a single massive object between the source and the observer, in the details all the structures near the light path contribute to the observed images. These additional contributions, known as line-of-sight effects, are non-negligible in practice.
This article proposes a new theoretical framework to model the line-of-sight effects, together with very promising applications at the interface of weak and strong lensing.
Our approach relies on the dominant-lens approximation, where one deflector is treated as the main lens while the others are treated as perturbations. The resulting framework is technically simpler to handle than the multi-plane lensing formalism, while allowing one to consistently model any sub-critical perturbation. In particular, it is not limited to the usual external-convergence and external-shear parameterisation.
As a first application, we identify a specific notion of line-of-sight shear that is not degenerate with the ellipticity of the main lens, and which could thus be extracted from strong-lensing images. This result supports and improves the recent proposal that Einstein rings might be powerful probes of cosmic shear.
As a second application, we investigate the distortions of strong-lensing critical curves under line-of-sight effects, and more particularly their correlations across the sky. We find that such correlations may be used to probe, not only the large-scale structure of the Universe, but also the dark-matter halo profiles of strong lenses. This last possibility would be a key asset to improve the accuracy of the measurement of the Hubble-Lemaître constant via time-delay cosmography.
}


\preprint{IFT-UAM/CSIC-21-39}

\date{\today}

\begin{document}

\maketitle
\flushbottom

\section{Introduction}
\label{sec:introduction}

A remarkable feature of science is certainly its ability to make tools out of natural phenomena, tools which in turn enable us to further delve into the laws of nature. The history of physics is an inexhaustible source of examples of this process, and gravitational lensing is one of its most beautiful illustrations. Over the last century, our perspective on the gravitational deflection of light has evolved from an early test of general relativity in 1920~\cite{1920RSPTA.220..291D}, a supposedly non-observable curiosity in 1936~\cite{1936Sci....84..506E}, to an entire field at the interface of astrophysics, exo-planetary science and cosmology nowadays (e.g. ref.~\cite{2006glsw.conf.....M}).

We traditionally distinguish between the strong and weak regimes of gravitational lensing. \emph{Strong lensing}, on the one hand, was first observed in 1979 at the Jodrell Bank Observatory~\cite{1979Natur.279..381W}, and its current applications range from the structural analysis of galaxies~\cite{2010ARA&A..48...87T} and clusters of galaxies (e.g. ref.~\cite{Meneghetti:2020yif}), to the measurement of the cosmic expansion rate via time-delay cosmography~\cite{1964MNRAS.128..307R, Wong:2019kwg, 2020A&A...643A.165B}. \emph{Weak lensing}, on the other hand, was first detected in 2000 by the Canada-France-Hawaii Telescope~\cite{2000A&A...358...30V} and it is today one of the most prominent probes of the large-scale distribution of matter in the Universe~\cite{Asgari:2020wuj, Gatti:2019clj}.

The focus of this article is the possible synergies between the weak and strong regimes of gravitational lensing, via the so-called \emph{line-of-sight effects} in strong lensing. Specifically, we are interested in how strong-lensing systems may be perturbed by matter inhomogeneities near the line of sight, and how such an interaction may be used to learn about its protagonists.

The simplest description of a strong lens consists in distributing some matter on a plane and placing that plane in an otherwise empty (Minkowski) Universe~\cite{1992grle.book.....S}. This simple approach is adapted to the description of nearby lensing systems, typically in our own galaxy. At cosmological distances, however, the impact of the rest of the Universe cannot be neglected. This issue is traditionally addressed by embedding the observer, the source and the lens in a homogeneous and isotropic Friedmann-Lemaître-Robertson-Walker (FLRW) background rather than in a Minkowski background. This operation automatically allows for (i) aberration effects due to cosmic recession, and (ii) light focusing due to the space-time curvature produced by the cosmic matter density. Both effects are encoded in the expressions of the angular-diameter distances involved in the lens equation.

However, given the current quality of the strong-lensing data, the FLRW background does not suffice anymore; namely, the inhomogeneities of the Universe turn out to have a significant impact on the observed strong-lensing images~\cite{Sengul:2020yya}. The current theoretical state of the art proposes two options to deal with that problem. The first option consists in supplementing the original lens plane with a discrete set of additional planes placed along the line of sight. This multi-plane lensing framework~\cite{1986ApJ...310..568B} is suitable when more than one lens is required to explain some observation; it is, however, much harder to handle than the single-plane case, because it turns the lens equation into a recursion problem. The second option consists in treating the inhomogeneities along the line of sight as mere tidal perturbations~\cite{1987ApJ...316...52K, 1996ApJ...468...17B, Schneider:1997bq, Birrer:2016xku}, which supplement the main lens with external convergence and shear. Albeit less accurate than the multi-plane framework, the tidal approach has the advantage of simplicity. Both approaches may also be combined within hybrid frameworks, as was done in refs.~\cite{McCully:2013fga, Schneider:2014vka, Fleury:2020cal}.

Because of its practical simplicity, the tidal approach (external convergence/shear) has been widely used in the modelling of strong-lensing systems. In particular, the external convergence is known to be a key source of uncertainty in the measurement of the Hubble-Lemaître constant~$H_0$ from time-delay cosmography~\cite{2010ApJ...711..201S, McCully:2016yfe, 2020MNRAS.498.1406T, Li:2020fpq}, while the external shear is essential to explain the observed abundance of quadruply imaged quasars~\cite{Keeton:1996tq, 2021arXiv210208470L}. In those examples, line-of-sight perturbations are viewed as additional ingredients which improve strong-lensing models. But recently Birrer et al.~\cite{Birrer:2016xku, Birrer:2017sge, Kuhn:2020wpy} proposed to reverse that logic---what if strong-lensing images could be used to \emph{measure} the external shear, thereby constituting novel probes of weak lensing? Could the strong serve the weak?

This intriguing possibility is the main motivation of the work reported here. In particular, we wondered how to model line-of-sight perturbations \emph{beyond convergence and shear}, which indeed constitute a limited description of general weak-lensing effects~\cite{Fleury:2018odh}. From a more practical point of view, we were also interested in determining what such line-of-sight effects could tell us about both strong lenses and their perturbers if they were measured. This set of questions led us to develop a new theoretical framework allowing one to model the effect of general line-of-sight perturbations in strong lensing. This \emph{dominant-lens approximation} is comprehensively exposed in \cref{sec:dominant_lens}; it combines the generality of the multi-plane approach with the technical simplicity of the external convergence/shear approach, thereby constituting an ideal theoretical compromise.

Equipped with this brand new framework, we first revisit the standard tidal treatment of line-of-sight perturbations in \cref{sec:parametric} and refine the earlier results of refs.~\cite{Birrer:2016xku, Birrer:2017sge}. In particular, we identify a \emph{certain notion of external shear} that can be measured independently of the properties of the main lens, notably its intrinsic ellipticity. We also explore the possibility of including the line-of-sight flexion~\cite{2006MNRAS.365..414B}. As a second application of the dominant-lens formalism, we propose in \cref{sec:critical_curves} that the distortions of strong-lensing \emph{critical curves} due to line-of-sight effects could be a relevant cosmic probe. Perhaps even more interestingly, we find that the correlations of the shapes of critical curves across the sky could be a key probe of the mass profile of strong lenses. Our main results, together with the new perspectives that they offer, are summarised in \cref{sec:conclusion}.

\textbf{Conventions and notation.} Throughout the article, we adopt units such that the speed of light is unity, $c=1$. Bold symbols indicate two-dimensional Euclidean vectors ($\vect{\beta}, \vect{\theta}, \vect{x}, \ldots$), or $2\times 2$ matrices ($\mat{\Gamma}, \mat{\Sigma}, \mat{\amplification}$). Matrix products are indicated by a dot where there are ambiguities with the argument of a matrix, e.g., $\mat{\Gamma}\cdot(\mat{x}+\mat{y})$. Underlined symbols like $\cplx{x}$ indicate the complex counterpart of two-dimensional vectors, with $\vect{x}=(x_1, x_2) \mapsto \cplx{x}=x_1+\ii x_2$. A summary of the notation introduced in this article may be found in \cref{tab:notation}.

\begin{table}[p!]
\centering
\begin{tabular}{r|l|l}
Symbol & Description & Definition
\\ \hline
$\vect{\beta}$
& angular position of a source
& \cref{fig:multi-plane}
\\
$\vect{\theta}$
& angular position of an image
& \cref{fig:multi-plane}
\\
$\vect{\alpha}$
& lensing displacement angle
& $\vect{\alpha}\define \vect{\theta}-\vect{\beta}$
\\
$\vect{x}_l$
& transverse position of a light ray in $l\h{th}$ plane
& \cref{fig:multi-plane}
\\
$\Sigma_l$
& projected density in the $l\h{th}$ plane
& \cref{subsubsec:lenses}
\\
$\hat{\psi}_l$
& twice the projected potential in $l\h{th}$ plane
& \cref{eq:projected_potential}
\\
$\hat{\vect{\alpha}}_l$
& deflection angle in $l\h{th}$ plane
& \cref{eq:deflection_angle}
\\
$\mat{\Sigma}_l$
& density-quadrupole matrix in $l\h{th}$ plane
& \cref{eq:Sigma_l_and_Sigma_crit}
\\
$Q_l$
& complex density quadrupole in the $l\h{th}$ plane
& \cref{eq:density-quadrupole_matrix}
\\
$o, d, s$
& indices for the observer, dominant lens, and source
& \cref{fig:multi-plane_dominant}
\\
$\eps$
& book-keeping parameter in DL approximation
& \cref{subsubsec:dominant_lens_def}
\\
$\chi$
& radial comoving distance
& -
\\
$D_{ij}$
& angular-diameter distance to $j$ as seen from $i$
& \cref{eq:D_ij}
\\
$\tau_{ij}$
& time-delay scale for a lens at $i$ and a source at $j$
& \cref{eq:time-delay_scale}
\\
$\vect{\beta}_{ij}$
& angle under which $\vect{x}_j$ is seen from $i$
& \cref{eq:beta_ij}
\\
$\vect{\alpha}_{ilj}$
& lensing displacement for observer $i$, lens $l$, source $j$
& \cref{eq:partial_displacement_def}
\\
$\Sigma\h{crit}_{ilj}$
& critical lensing density for observer $i$, lens $l$, source $j$
& \cref{eq:Sigma_l_and_Sigma_crit}
\\
$\vect{\Gamma}_{ilj}$
& partial shear matrix for observer $i$, lens $l$, source $j$
& \cref{eq:partial_amplification}
\\
$\kappa_{ilj}$
& partial convergence for observer $i$, lens $l$, source $j$
& \cref{eq:convergence_shear_matrix}
\\
$\gamma_{ilj}$
& partial shear for observer $i$, lens $l$, source $j$
& \cref{eq:convergence_shear_matrix}
\\
$\vect{\alpha}_{ij}$
& cumulative non-dominant displacement from $i$ to $j$
& \cref{eq:alpha_od,eq:alpha_os}
\\
$\vect{\Gamma}_{ij}$
& cumulative non-dominant shear matrix from $i$ to $j$
& \cref{subsubsec:shear_matrix}
\\
$\kappa_{ij}, \gamma_{ij}$
& cumulative convergence and shear from $i$ to $j$
& \cref{subsubsec:shear_matrix}
\\
$\gamma\e{LOS}$
& measurable combination of external shears
& \cref{eq:gamma_LOS}
\\
$\hat{\F}_l, \hat{\G}_l$
& potential dipole and hexapole in the $l\h{th}$ plane
& \cref{eq:potential_dipole_hexapole}
\\
$\F_{ilj}$
& partial type-$\F$ flexion for observer $i$, lens $l$, source $j$
& \cref{eq:partial_F_flexion}
\\
$\G_{ilj}$
& partial type-$\G$ flexion for observer $i$, lens $l$, source $j$
& \cref{eq:partial_G_flexion}
\\
${}^{(1,2)}\F_{ij}$
& cumulative non-dominant $\F$-type flexions from $i$ to $j$
& \cref{eq:F_od,eq:F_os,eq:F_ds_1,eq:F_ds_2}
\\
${}^{(1,2)}\G_{ij}$
& cumulative non-dominant $\G$-type flexions from $i$ to $j$
& same as ${}^{(1,2)}\F_{ij}$
\\
$\vect{\theta}\e{cc}$
& critical curve
& \cref{subsubsec:critical_curve_general}
\\
$c_n$
& critical modes, i.e., Fourier modes of critical curves
& \cref{eq:critical_modes}
\\
$\mathcal{D}_n(\cplx{w})$
& complex integral for direct crit. curve perturbations
& \cref{eq:complex_integral_direct}
\\
$\mathcal{C}_n(\cplx{w})$
& complex integral for lens-lens coupling perturbations
& \cref{eq:complex_integral_coupling}
\\
$W_{\mathcal{D}}(\chi)$
& line-of-sight weight for $\mathcal{D}_n$
& \cref{eq:W_D}
\\
$W_{\mathcal{C}}(\chi)$
& line-of-sight weight for $\mathcal{C}_n$
& \cref{eq:W_C}
\\
$r(\chi)$
& comoving size of the critical beam at $\chi$
& \cref{eq:comoving_radius_critical_beam}
\\
$\bar{c}_n(\vect{\vartheta})$
& effective critical modes in direction $\vect{\vartheta}$
& \cref{eq:critical_modes_effective}
\\
$\vect{\Pi}$
& parameters marginalised over in $\bar{c}_n$
& $\vect{\Pi}\define (\theta\e{E}, \kappa\e{E}, \chi_d, \chi_s)$
\\
$\theta\e{E}$
& Einstein radius of the dominant lens
& $\mu_{ods}^{-1}(\theta\e{E})=0$
\\
$\kappa\e{E}$
& convergence of dominant lens at Einstein radius
& $\kappa\e{E}=\kappa_{ods}(\theta\e{E})$
\\
$\xi_{n_1 n_2}^\pm(\vartheta)$
& correlation functions of $\bar{c}_{n_1}$ and $\bar{c}_{n_2}$
& \cref{eq:correlation_functions_pm}
\\
$P_{n_1 n_2}(\ell)$
& power spectrum of $\bar{c}_{n_1}$ and $\bar{c}_{n_2}$
& \cref{eq:power_spectrum_result}
\\
$q_n(\chi, \ell;\vect{\Pi})$
& weight of inhomogeneities at $\chi$ in $P_{n_1 n_2}(\ell)$ for $\vect{\Pi}$
& \cref{eq:q_n}
\\
$\bar{q}_n(\chi, \ell)$
& weight marginalised over $\vect{\Pi}$
& \cref{eq:q_n_bar}
\\
\hline
\end{tabular}
\caption{Summary of the notation used in this article.}
\label{tab:notation}
\end{table}

\section{Multi-plane lensing with a dominant lens}
\label{sec:dominant_lens}

In this section, we lay out the general framework to model lensing by multiple deflectors, when one of them overwhelms the effect of the others. This \emph{dominant-lens approximation} is based on the multi-plane lensing formalism~\cite{1986ApJ...310..568B}, whose main results are summarised in \cref{subsec:multi-plane_lensing}. We then derive the lens equation (\cref{subsec:dominant_lens_equation}), the distortion matrix for infinitesimal images (\cref{subsec:amplification_dominant_lens}) and time delays (\cref{subsec:time_delays}) in the dominant-lens regime.

\begin{figure}[t]
    \centering
    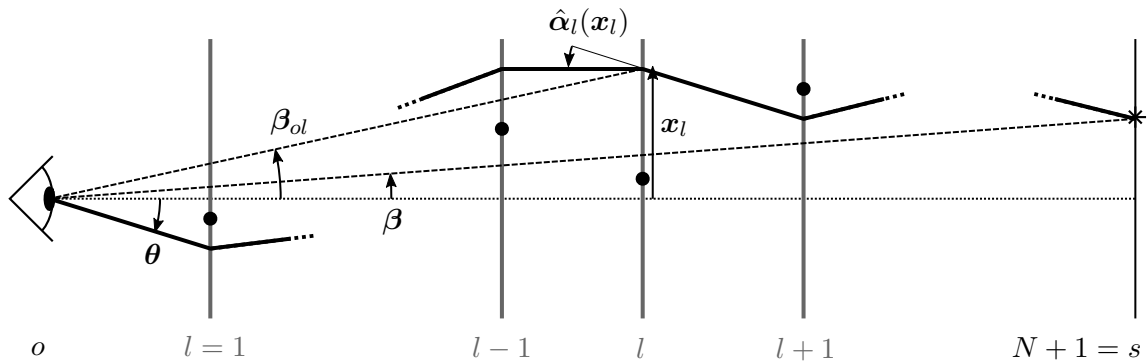
    \caption{Multi-plane lensing on an FLRW background. The optical axis (dotted line) is a straight line and defines the origin of the $N$ lens planes, labelled by $l$. The physical ray (thick solid line) intersects the $l\h{th}$ plane at a physical transverse position $\vect{x}_l$. This position has an angular position $\vect{\beta}_{ol}=\vect{x}_l/D_{ol}$ from the observer's point of view ($o$), where $D_{ol}$ is the angular-diameter distance to $l$ as seen by $o$ in the absence of any lens, i.e., in the FLRW background. In the $l\h{th}$ plane, the physical ray is deflected by an angle $\hat{\vect{\alpha}}_l$, as measured in the rest frame of the $l\h{th}$ lens. The angle $\vect{\theta}=\vect{\beta}_{o1}$ is the observed image position, and $\vect{\beta}\define\vect{\beta}_{os}$ is the unlensed position of the source, i.e. the direction in which it would be seen by the observer in the absence of any lens.}
    \label{fig:multi-plane}
\end{figure}

\subsection{Multi-plane lensing formalism}
\label{subsec:multi-plane_lensing}

Consider a light source ($s$) and an observer ($o$). Throughout this article, we shall assume that the space-time through which light propagates from $s$ to $o$ is well modelled by an FLRW cosmology supplemented with a number of thin matter planes, called \emph{lens planes}, which are orthogonal to an arbitrary optical axis. When light crosses a lens plane, its path is suddenly deflected (see \cref{fig:multi-plane}). This framework thus assumes that all the relevant inhomogeneities of the Universe can be treated as thin lenses; see refs.~\cite{Schneider:2014vka, Fleury:2020cal} for a more general approach.

\subsubsection{Lenses}
\label{subsubsec:lenses}

Each lens plane $l$ is characterised by a surface density $\Sigma_l(\vect{x})$, from which we define
\begin{equation}
\label{eq:projected_potential}
\hat{\psi}_l(\vect{x}) \define \int \dd^2\vect{y} \; 4G\Sigma_l(\vect{y}) \, \ln|\vect{x}-\vect{y}| \ ,
\end{equation}
where $G$ is Newton's constant. Physically speaking, $\hat{\psi}_l$ is twice the projected gravitational potential generated by $\Sigma_l$; it satisfies the projected Poisson equation $\Delta\hat{\psi}_l=8\pi G \Sigma_l$, where $\Delta$ is the two-dimensional Laplacian. Note that, since the background FLRW space-time represents a spatially averaged Universe, \emph{our lens planes may have positive or negative densities}; see appendix A of ref.~\cite{Fleury:2018cro} for a discussion on that specific point.

As a light ray crosses the plane $l$ at $\vect{x}_l$, it is deflected by an angle
\begin{equation}
\label{eq:deflection_angle}
\hat{\vect{\alpha}}_l(\vect{x}_l)
\define \ddf{\hat{\psi_l}}{\vect{x}_l}
= \int \dd^2\vect{x} \; 4G\Sigma_l(\vect{x}) \,
                        \frac{\vect{x}_l-\vect{x}}{|\vect{x}_l-\vect{x}|^2} \ .
\end{equation}
The deflection angle $\hat{\vect{\alpha}}_l$ is defined in the rest frame of the lens.

\subsubsection{Lens equation and recursion}

We denote with $\vect{\theta}$ the position of an image, i.e., the angular separation between the direction in which the source is actually observed and the optical axis. We call $\vect{\beta}$ the unlensed position of the source, i.e., the direction in which it would be seen without any lens. These two quantities are related by the multi-plane lens equation~\cite{1986ApJ...310..568B}
\begin{equation}
\label{eq:lens_equation}
\vect{\beta} = \vect{\theta} - \vect{\alpha} \ ,
\qquad
\vect{\alpha}
= \sum_{l=1}^N \frac{D_{ls}}{D_{os}} \,
    \hat{\vect{\alpha}}_l(\vect{x}_l) \ ,
\end{equation}
where $\vect{\alpha}$ is called the displacement angle. The notation $D_{ij}$ refers to the unlensed angular diameter distance to $j$ as seen from $i$; its expression reads\footnote{\Cref{eq:D_ij} actually holds only if $i$ is comoving with cosmic expansion. In order to account for the peculiar velocity of $i$, and more generally to any phenomenon that affects the frequency of light received by $i$, \cref{eq:D_ij} must be corrected by the ratio $\omega_i/\bar{\omega}_i$ between observed and background frequencies.}
\begin{equation}
\label{eq:D_ij}
D_{ij} = \frac{f_K(\chi_j-\chi_i)}{1+z_j} \ ,
\end{equation}
where $z_j$ is the cosmological redshift of $j$, $\chi_i, \chi_j$ are the comoving distances of $i,j$ from the observer, and $f_K(\chi)\define \sin(\sqrt{K}\chi)/\sqrt{K}$, $K$ being the FLRW spatial-curvature parameter. For the sake of readability, we used in \cref{eq:lens_equation} the index $o\define 0$ for the observer plane and $s\define N+1$ for the source plane.

In \cref{eq:lens_equation}, the argument $\vect{x}_l$ of $\hat{\vect{\alpha}}_l$ is the physical position where light pierces the $l\h{th}$ plane, with respect to the optical axis. The presence of the $N$ variables $\vect{x}_l$ makes the analysis of the lens equation~\eqref{eq:lens_equation} considerably harder than the standard single-lens case. Indeed, the positions~$\vect{x}_l$ must be determined iteratively from the recursion relation~\cite{1986ApJ...310..568B}
\begin{equation}
\label{eq:lens_recursion}
\forall l\in\{1,\ldots, N+1\}
\qquad
\vect{x}_l = D_{ol}\,\vect{\theta}
- \sum_{m=1}^{l-1} D_{ml} \, \hat{\vect{\alpha}}_m(\vect{x}_m) \ .
\end{equation}
\Cref{eq:lens_equation} corresponds to the case $l=N+1=s$, after dividing by $D_{os}$.

\subsubsection{Partial displacement, convergence and shear}
\label{subsubsec:partial_displacements_convergence_shear}

Let us now introduce a few quantities and notation which will prove convenient in the remainder of this article. First of all, we call
\begin{equation}
\label{eq:beta_ij}
\vect{\beta}_{ij} \define \frac{\vect{x}_j}{D_{ij}}
\end{equation}
the direction in which the point $\vect{x}_j$ would be seen from $i$ without lenses. This notation may be applied to any couple of planes $i<j$; in particular $\vect{\beta}_{os}=\vect{\beta}$. Second, we shall call \emph{partial displacement} and denote with $\vect{\alpha}_{ols}$ each term of the sum of \cref{eq:lens_equation}, with
\begin{equation}
\label{eq:partial_displacement_def}
\vect{\alpha}_{ilj}(\vect{x}_l)
\define
\frac{D_{lj}}{D_{ij}} \, \hat{\vect{\alpha}}_l(\vect{x}_l) \ .  
\end{equation}
That notation may be applied to any triplet of planes $i<l<j$.

The gradient of $\vect{\alpha}_{ilj}$ with respect to $\vect{\beta}_{il}$ defines a notion of \emph{shear matrix} such that
\begin{equation}
\label{eq:partial_amplification}
\mat{\Gamma}_{ilj}
\define \ddf{\vect{\alpha}_{ilj}}{\vect{\beta}_{il}}
= \frac{D_{il}D_{lj}}{D_{ij}} \,
    \frac{\dd^2\hat{\psi}_l}{\dd\vect{x}_l^2}
= \frac{\mat{\Sigma}_l}{\Sigma\h{crit}_{ilj}} \ ,
\end{equation}
with
\begin{align}
\label{eq:Sigma_l_and_Sigma_crit}
\mat{\Sigma}_l
\define \frac{\dd^2}{\dd\vect{x}_l^2}
	\int \frac{\dd^2\vect{y}}{\pi} \; \Sigma_l(\vect{y}) \ln|\vect{x}_l-\vect{y}| \ ,
\qquad
\Sigma\h{crit}_{ilj}
\define \pa{\frac{4\pi G D_{il}D_{lj}}{D_{ij}}}^{-1} .
\end{align}
Geometrically speaking, the matrix $\mat{\amplification}_{ilj}=\mat{1}-\mat{\Gamma}_{ilj}$ characterises the distortions of an infinitesimal source that would be placed at $j$, due to the lens $l$ alone, and as observed at $i$; in other words, it is a partial distortion matrix. We note from \cref{eq:partial_amplification} that $\mat{\Gamma}_{ilj}$ is a symmetric matrix; it may thus be decomposed as
\begin{equation}
\label{eq:convergence_shear_matrix}
\mat{\Gamma}_{ilj}
\define
\kappa_{ilj} \, \mat{1}
+
\begin{bmatrix}
\Re(\gamma_{ilj}) & \Im(\gamma_{ilj}) \\
\Im(\gamma_{ilj}) & -\Re(\gamma_{ilj})
\end{bmatrix} ,
\end{equation}
thereby defining the partial convergence $\kappa_{ilj}$ and partial complex shear $\gamma_{ilj}$.

As for the other two quantities defined in \cref{eq:Sigma_l_and_Sigma_crit}, the matrix $\mat{\Sigma}_l$ characterises the distribution of matter in plane $l$. It enjoys a similar decomposition as $\mat{\Gamma}_{ilj}$, namely
\begin{equation}
\label{eq:density-quadrupole_matrix}
\mat{\Sigma}_l
\define
\Sigma_l \, \mat{1}
+
\begin{bmatrix}
\Re(Q_l) & \Im(Q_l) \\
\Im(Q_l) & -\Re(Q_l)
\end{bmatrix} .
\end{equation}
The trace of $\mat{\Sigma}_l$ is indeed $2\Sigma_l$ by virtue of the projected Poisson equation satisfied by $\hat{\psi}_l$, while the complex number $Q_l$ may be understood as the quadrupole of the projected density. Finally, the proportionality factor $\Sigma\h{crit}_{ilj}$ fully encodes the dependence of $\Gamma_{ilj}$ in the ``observer''~$i$ and ``source''~$j$; it represents the density scale over which the set-up $(ilj)$ is critical, i.e., could lead to strong-lensing effects. Since $\mat{\Gamma}_{ilj}=\mat{\Sigma}_l/\Sigma\h{crit}_{ilj}$, we also have
\begin{equation}
\kappa_{ilj} = \frac{\Sigma_l}{\Sigma\h{crit}_{ilj}} \ ,
\qquad
\gamma_{ilj} = \frac{Q_l}{\Sigma\h{crit}_{ilj}} \ .
\end{equation}

\subsection{Lens equation in the dominant-lens regime}
\label{subsec:dominant_lens_equation}

The multi-plane lensing formalism is difficult to handle in practice and in full generality, because it is a recursion problem with as many variables as there are lenses. In this sub-section, we shall demonstrate that the problem drastically simplifies if one of the lenses dominates, while the others are treated as perturbations. This approach, which is depicted in \cref{fig:multi-plane_dominant}, will be referred to as the \emph{dominant-lens (DL) approximation}.

\subsubsection{Definition of the dominant-lens approximation}
\label{subsubsec:dominant_lens_def}

We assume that there exists a lens plane $l=d$ that dominates the lens equation, while all the others can be treated as perturbations. This implies, in particular, that only the plane $d$ is allowed to be super-critical ($\kappa_{ods}>1$); all the other planes must be sub-critical. In the DL approximation, we even assume that the secondary planes are amply sub-critical,
\begin{equation}
\forall l\neq d \quad \forall i<l<j \qquad
\kappa_{ilj} \ll 1 \ ,
\end{equation}
over the relevant area of the sky. This implies that all the lensing quantities associated with the secondary lenses $l\neq d$ are small ($|\gamma_{ilj}|, |\vect{\alpha}_{ilj}|\ll 1$) and can be treated perturbatively. Since we opt for a perturbative treatment, it is convenient to introduce the book-keeping parameter $\eps$, such that for any $l\not=d$, $\kappa_{ilj}=\order(\eps^2)$. For localised mass distributions, $\eps$ may be thought of as the typical Einstein radius of the non-dominant lenses, whence the square.

In practice, the DL approximation will consist in expanding the lens equation (and recursion relations for $\vect{x}_l$) at first order in $\eps^2$. Since the displacement angles $\vect{\alpha}_{ilj}$ are controlled by $\kappa_{ilj}$, this implies that we will work at linear order in $\vect{\alpha}_{ilj}$ if $l\neq d$, and non-perturbatively in $\vect{\alpha}_{idj}$ throughout. This programme will lead us to adapt the Born approximation in order to consistently account for the perturbative displacements as $l\neq d$ and the non-perturbative displacement at $d$.

\subsubsection{Derivation of the lens equation}
\label{subsubsec:derivation_DL_equation}

\begin{figure}
    \centering
    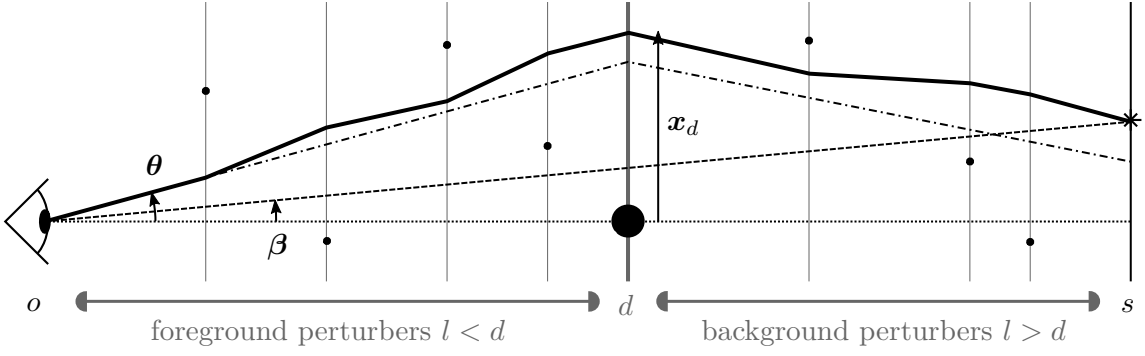
    \caption{Multi-plane lensing with a dominant lens (DL). The total displacement angle~$\vect{\theta}-\vect{\beta}$ is dominated by the effect of one lens ($d$), while the other lenses ($l\not=d$) are treated as perturbations. The optical axis (dotted line) is conventionally aligned with the centre of the dominant lens. The thick solid line represents the physical ray; the dashed line is the unlensed ray (without any lens); the dot-dashed line indicates the dominant-lens-only ray, i.e., the ray that would be followed in the presence of the dominant lens, but in the absence of perturbers. In the DL approximation, the effect of the perturbers is evaluated along the dominant-lens-only ray, while the effect of the dominant lens is evaluated on the real position $\vect{x}_d$ of the ray in the $d$ plane.}
    \label{fig:multi-plane_dominant}
\end{figure}

In the expression of the displacement angle~\eqref{eq:lens_equation} one may identify the contributions of the dominant lens, $\vect{\alpha}_{ods}$, and those of the foreground lenses ($l<d$) and background lenses ($l>d$),
\begin{equation}
\label{eq:foreground_dominant_background}
\vect{\theta}-\vect{\beta}
=
\vect{\alpha}
= \underbrace{
    \sum_{l<d} \vect{\alpha}_{ols}(\vect{\beta}_{ol}) 
    }_{\text{foreground}}
    +
    \underbrace{
    \vect{\alpha}_{ods}(\vect{\beta}_{od})
    }_{\text{dominant}}
    +
    \underbrace{
    \sum_{l>d} \vect{\alpha}_{ols}(\vect{\beta}_{ol})
    }_{\text{background}}
    \ ,
\end{equation}
where we chose to express the partial displacements~$\vect{\alpha}_{ols}$ as functions of the angles $\vect{\beta}_{ol}=\vect{x}_l/D_{ol}$ instead of the positions $\vect{x}_l$. From \cref{eq:lens_recursion}, the various $\vect{\beta}_{ol}$ are related to each other as
\begin{equation}
\label{eq:lens_recursion_beta}
\vect{\beta}_{ol}
= \vect{\theta} - \sum_{m<l} \vect{\alpha}_{oml}(\vect{\beta}_{om}) \ .
\end{equation}
The goal of the forthcoming calculation is to use the DL approximation to expand the terms of \cref{eq:foreground_dominant_background}, so as to reduce it to an equation for $\vect{\theta}$ only.

\begin{description}
\item[Foreground displacements] For $l<d$, the difference between $\vect{\beta}_{ol}$ and $\vect{\theta}$ does not involve the main lens; thus $\vect{\beta}_{ol}=\vect{\theta}+\order(\eps^2)$, and hence each partial displacement reads
\begin{equation}
\vect{\alpha}_{ols}(\vect{\beta}_{ol})
= \vect{\alpha}_{ols}(\vect{\theta}) + \order(\eps^4) \ .
\end{equation}
In other words, Born's approximation applies to the foreground lenses.

\item[Main displacement] Unlike the other terms of \cref{eq:foreground_dominant_background}, the main-lens displacement~$\vect{\alpha}_{ods}$ is not $\order(\eps^2)$, and hence we must go beyond the Born approximation for this term. Applying \cref{eq:lens_recursion_beta} to $l=d$ and Taylor-expanding $\vect{\alpha}_{ods}$ at first order yields
\begin{equation}
\label{eq:main_displacement_Gamma}
\vect{\alpha}_{ods}(\vect{\beta}_{od})
= \vect{\alpha}_{ods}(\vect{\theta}) 
    - \mat{\Gamma}_{ods}(\vect{\theta})
        \sum_{m<d} \vect{\alpha}_{omd}(\vect{\theta})
    + \order(\eps^4) \ .
\end{equation}
This expansion is licit as long as $\vect{\theta} \mapsto \vect{\alpha}_{ods}(\vect{\theta})$ is differentiable, i.e., as long as its partial derivatives encoded in $\mat{\Gamma}_{ods}(\vect{\theta})$ remain finite. This restrictive assumption is generically satisfied, except near the centre of pathological lens models.\footnote{For example, the point-lens model~$\vect{\alpha}\e{PL}(\vect{\theta})\propto \vect{\theta}/|\vect{\theta}|^2$ and the singular isothermal sphere~$\vect{\alpha}\e{SIS}(\vect{\theta})\propto \vect{\theta}/|\vect{\theta}|$ are non-differentiable at $\vect{\theta}=0$.} Fortunately, such problematic locations in the image plane are practically irrelevant, because they have vanishing surface brightness---they come from sources far from the line of sight.

For some of the forthcoming applications, and also to keep expressions shorter, we may prefer to keep the perturbation inside the argument of $\vect{\alpha}_{ods}$ as
\begin{equation}
\label{eq:main_displacement}
\vect{\alpha}_{ods}(\vect{\beta}_{od})
= \vect{\alpha}_{ods}
    \pac{
        \vect{\theta}
        - \vect{\alpha}_{od}(\vect{\theta})
        }
    + \order(\eps^4) \ ,
\qquad
\vect{\alpha}_{od}(\vect{\theta})
\define \sum_{l<d} \vect{\alpha}_{old}(\vect{\theta}) \ .
\end{equation}
Note the difference between $\vect{\alpha}_{old}$ with a three-letter subscript, which represents a partial displacement, and $\vect{\alpha}_{od}$ with a two-letter subscript, which is the sum of such partial displacements for $o<l<d$. \Cref{eq:main_displacement} is simpler than its exact counterpart because $\vect{\alpha}_{od}$ is evaluated along the unperturbed path. Under the assumption of differentiability for $\vect{\theta}\mapsto\vect{\alpha}_{ods}(\vect{\theta})$, \cref{eq:main_displacement_Gamma,eq:main_displacement} are strictly equivalent up to $\order(\eps^4)$ terms, so they are interchangeable in the DL approximation; otherwise none of them is valid.

\item[Background displacements] Contrary to the foreground case, for $l>d$ the angle $\vect{\beta}_{ol}$ does receive a contribution from the main lens, namely $\vect{\beta}_{ol}=\vect{\theta}-\vect{\alpha}_{odl}(\vect{\theta})+\order(\eps^2)$. Thus,
\begin{equation}
\vect{\alpha}_{ols}(\vect{\beta}_{ol})
= \vect{\alpha}_{ols}\pac{\vect{\theta}-\vect{\alpha}_{odl}(\vect{\theta})} + \order(\eps^4) \ .
\end{equation}
We cannot simplify that expression further because the correction to the argument of $\vect{\alpha}_{ols}$ is a dominant-lens term, which is non-perturbative.
\end{description}

Putting everything together yields a lens equation for $\vect{\theta}$ only,
\begin{empheq}[box=\fbox]{align}
\label{eq:lens_equation_dominant}
\vect{\beta} &= \vect{\theta} - \vect{\alpha}(\vect{\theta}) \ ,
\\
\label{eq:displacement_dominant}
\vect{\alpha}(\vect{\theta})
&= \underbrace{
    \vect{\alpha}_{ods}(\vect{\theta})
    }_{\text{dominant}}
    - 
    \underbrace{
    \mat{\Gamma}_{ods}(\vect{\theta})
        \sum_{l<d} \vect{\alpha}_{old}(\vect{\theta})
    }_{\text{post Born for dominant}}
    +
    \underbrace{
    \sum_{l<d} \vect{\alpha}_{ols}(\vect{\theta})
    }_{\text{foreground}}
    +
    \underbrace{
    \sum_{l>d} \vect{\alpha}_{ols}
    \pac{\vect{\theta}-\vect{\alpha}_{odl}(\vect{\theta})}
    }_{\text{background}}
    + \order(\eps^4).
\end{empheq}
The first term on the right-hand side of \cref{eq:displacement_dominant} would be the contribution of the dominant lens in the absence of perturbations; the next ones encode, respectively, departure from the Born approximation in the main-lens plane, and the direct corrections due to foreground and background lenses. These corrections are evaluated along the path that would be followed by light in the absence of perturbers (dot-dashed line in \cref{fig:multi-plane_dominant}).

\subsubsection{Comparison with the critical-sheet Born approximation}
\label{subsubsec:CSB}

In ref.~\cite{Birrer:2016xku}, Birrer et al. formulated the \emph{critical-sheet Born (CSB) approximation} in order to simplify the lens equation near its critical curve. Although ref.~\cite{Birrer:2016xku} only treated tidal perturbations to a main deflector (see \cref{subsec:tidal}), the idea of the CSB approximation is applicable to general perturbations as considered here.

One difficulty of \cref{eq:displacement_dominant} is that the background ($l>d$) terms~$\vect{\alpha}_{ols} \pac{\vect{\theta}-\vect{\alpha}_{odl}(\vect{\theta})}$ depend on the properties of the main deflector through $\vect{\alpha}_{odl}(\vect{\theta})$. Yet, we notice that at first order in $\eps^2$, this main-lens dependence may be removed by substituting the lens equation,
\begin{equation}
\vect{\alpha}_{odl}(\vect{\theta})
= \frac{D_{dl}}{D_{ol}} \, \hat{\vect{\alpha}}_d(D_{od}\vect{\theta})
= \frac{D_{dl} D_{os}}{D_{ol} D_{ds}} \, \vect{\alpha}_{ods}(\vect{\theta})
= \frac{D_{dl} D_{os}}{D_{ol} D_{ds}} \, (\vect{\theta}-\vect{\beta}) + \order(\eps^2) \ ,
\end{equation}
so that for $l>d$,
\begin{equation}
\vect{\alpha}_{ols}[\vect{\theta}-\vect{\alpha}_{odl}(\vect{\theta})]
= \vect{\alpha}_{ols}\pac{\vect{\theta}-\frac{D_{dl} D_{os}}{D_{ol} D_{ds}} \, (\vect{\theta}-\vect{\beta})} + \order(\eps^4) \ .
\end{equation}

In general, the above substitution is not particularly useful, because it trades the main-lens dependence in $\vect{\alpha}_{ols}$ for a $\vect{\beta}$ dependence, which dramatically affects the structure of the lens equation---$\vect{\beta}$ now appears on both sides of it. However, near the critical curve of an axially symmetric main deflector, one can consider $\vect{\beta}\approx \vect{0}$, so that $\vect{\alpha}_{ods}$ only depends on $\vect{\theta}$ and on the distance ratios. This is the CSB approximation. Its name comes from the fact that substituting $\vect{\alpha}_{odl}(\vect{\theta})$ with $\frac{D_{dl} D_{os}}{D_{ol} D_{ds}}\vect{\theta}$ is equivalent to substituting the main lens with a critical mass sheet ($\kappa_{ods}=1$), for which $\vect{\alpha}_{ods}(\vect{\theta})=\vect{\theta}$ by definition. The deflections due to background perturbers ($l>d$) are then evaluated along this idealised path, whence ``Born''.

The initial motivation of the CSB approximation was to decouple the perturbers from the main lens. However, as seen in \cref{subsubsec:derivation_DL_equation}, such a coupling is anyway unavoidable due to the post-Born corrections in the main-displacement term $\vect{\alpha}_{ods}(\vect{\theta}-\vect{\alpha}_{od})$. Therefore, the CSB approximation ends up being less accurate than the DL approximation proposed here, while bringing no qualitative simplification of the problem. For those reasons, we recommend the use of DL over CSB in practice.

Let us finally note that our DL approximation may correspond, in the terminology of ref.~\cite{Birrer:2016xku}, to the ``strong-lens deflected Born'' (SLB) approximation. However, it is not entirely clear whether the SLB approach as envisaged in ref.~\cite{Birrer:2016xku} would account for the post-Born corrections at the main deflector, i.e. the second term in \cref{eq:displacement_dominant}.

\subsection{Distortions of infinitesimal images}
\label{subsec:amplification_dominant_lens}

In \cref{subsubsec:partial_displacements_convergence_shear}, we have introduced the partial shear matrices~$\mat{\Gamma}_{ilj}=\dd\vect{\alpha}_{ilj}/\dd\vect{\beta}_{il}$ which characterise the distortions caused by the sole lens $l$ to an infinitesimal source placed at $j$ and as observed from $i$. Let us now see how all those partial distortions act together in the framework of the DL approximation.

The total shear matrix~$\mat{\Gamma}(\vect{\theta})$ derives from the total displacement angle~$\vect{\alpha}(\vect{\theta})$ as
\begin{equation}
\mat{\Gamma}(\vect{\theta}) \define \ddf{\vect{\alpha}}{\vect{\theta}} \ .
\end{equation}
This matrix characterises the distortions of infinitesimal images in the $s$ plane as seen from $o$, because it is related to the Jacobian matrix of the lensing map $\vect{\theta}\mapsto\vect{\beta}(\vect{\theta})$, also called distortion matrix~$\mat{\amplification}(\vect{\theta})\define\dd\vect{\beta}/\dd\vect{\theta}=\mat{1}-\mat{\Gamma}(\vect{\theta})$. This subsection is dedicated to the explicit calculation of $\mat{\Gamma}(\vect{\theta})$ in the DL approximation, and of its geometric features---convergence, shear, rotation, and magnification.

\subsubsection{Shear matrix}
\label{subsubsec:shear_matrix}

Taking the derivative of $\vect{\alpha}(\vect{\theta})$ as given in \cref{eq:displacement_dominant} yields,
\begin{equation}
\label{eq:shear_matrix_dominant_initial}
\mat{\Gamma}
=
\textcolor{red4}{
\mat{\Gamma}_{od s}
- \pa[5]{ \sum_{l<d}\vect{\alpha}_{old} \cdot \ddf{}{\vect{\theta}}}
    \mat{\Gamma}_{ods}
}
-
\textcolor{purple}{
\mat{\Gamma}_{od s} \sum_{l<d} \mat{\Gamma}_{old}
}
+
\textcolor{blue4}{
\sum_{l<d} \mat{\Gamma}_{ols}
}
+ \sum_{l>d} \pa{
                \textcolor{blue4}{
                \mat{\Gamma}_{ols}
                }
                -
                \textcolor{purple}{
                \mat{\Gamma}_{ols}\mat{\Gamma}_{od l}
                }
                } ,
\end{equation}
at linear order in $\eps^2$, where all the quantities are evaluated at $\vect{\theta}$, except $\mat{\Gamma}_{ols}$ for $l>d$, which must be evaluated at $\vect{\theta}-\vect{\alpha}_{odl}(\vect{\theta})$ to allow for the non-perturbative main displacement.

Let us now interpret the various terms composing the right-hand side of \cref{eq:shear_matrix_dominant_initial}. The \textcolor{red4}{first two terms} encode the direct contribution of the main deflector, evaluated at a position that is slightly displaced due to foreground lenses. This displacement may be written
\begin{equation}
\label{eq:alpha_od}
\vect{\alpha}_{od}(\vect{\theta})
\define \sum_{l<d}\vect{\alpha}_{old}(\vect{\theta}) \ ,
\end{equation}
and represents the displacement of a source that would be located in the main deflector's plane. Because of that interpretation, we may gather those two terms into \textcolor{red4}{$\mat{\Gamma}_{ods}[\vect{\theta}-\vect{\alpha}_{od}(\vect{\theta})]$}.

The \textcolor{purple}{third term} of \cref{eq:shear_matrix_dominant_initial} encodes the non-linear coupling between the foreground and main lenses, although its physical origin is the departure from the Born approximation just like the second term. In this term, the sum over $l$ may be re-written as
\begin{equation}
\mat{\Gamma}_{od}(\vect{\theta})
\define \sum_{l<d} \mat{\Gamma}_{old}(\vect{\theta}) \ ;
\end{equation}
this shear matrix would describe the distortions of an infinitesimal source located in the main deflector's plane due to the foreground lenses.

The \textcolor{blue4}{fourth and fifth terms} of \cref{eq:shear_matrix_dominant_initial} are the direct contributions of the non-dominant lenses, they may be gathered into
\begin{equation}
\mat{\Gamma}_{os}(\vect{\theta})
\define
\sum_{l<d} \mat{\Gamma}_{ols}(\vect{\theta})
+ \sum_{l>d} \mat{\Gamma}_{ols}[\vect{\theta}-\vect{\alpha}_{od l}(\vect{\theta})]
\ .
\end{equation}
This matrix would describe the distortions of an infinitesimal source lying in the source plane, due to all the lenses but the dominant one. However, since the $\mat{\Gamma}_{ols}$ for $l>d$ are evaluated along the main-deflected path, $\mat{\Gamma}_{os}(\vect{\theta})$ is actually not independent of the main lens.

Finally the \textcolor{purple}{last term} of \cref{eq:shear_matrix_dominant_initial} is analogous to the third term, in that it encodes non-linear couplings between the dominant and background lenses. This is more easily seen if we rewrite it as
\begin{equation}
\label{eq:redistribution_distance_factors}
\mat{\Gamma}_{ols} \mat{\Gamma}_{od l}
= \frac{\mat{\Sigma}_l}{\Sigma\h{crit}_{ol s}} \,
    \frac{\mat{\Sigma}_d}{\Sigma\h{crit}_{od l}}
= \frac{\mat{\Sigma}_l}{\Sigma\h{crit}_{d l s}} \,
    \frac{\mat{\Sigma}_d}{\Sigma\h{crit}_{od s}}
= \mat{\Gamma}_{d ls} \mat{\Gamma}_{od s}
\end{equation}
which may be checked from the definition of $\Sigma_{ilj}\h{crit}$ in \cref{eq:Sigma_l_and_Sigma_crit}. When the dominant-lens term $\mat{\Gamma}_{ods}$ is factored out of the sum, the remainder may be re-written as
\begin{equation}
\mat{\Gamma}_{ds}(\vect{\theta})
\define
\sum_{l>d} \mat{\Gamma}_{dls}[\vect{\theta}-\vect{\alpha}_{odl}(\vect{\theta})] \ ,
\end{equation}
which would describe the distortions of an infinitesimal source in the source plane, but as observed from the main-lens plane.

The final result reads, up to $\order(\eps^4)$ terms,
\begin{empheq}[box=\fbox]{align}
\mat{\Gamma}(\vect{\theta})
&=
\mat{\Gamma}_{ods}[\vect{\theta}-\vect{\alpha}_{od}(\vect{\theta})]
- \mat{\Gamma}_{ods}(\vect{\theta}) \mat{\Gamma}_{od}(\vect{\theta})
- \mat{\Gamma}_{ds}(\vect{\theta}) \mat{\Gamma}_{ods}(\vect{\theta})
+ \mat{\Gamma}_{os}(\vect{\theta})
\\
&=
\pac{\mat{1}-\mat{\Gamma}_{ds}(\vect{\theta})} \,
\mat{\Gamma}_{ods}[\vect{\theta}-\vect{\alpha}_{od}(\vect{\theta})] \,
\pac{\mat{1}-\mat{\Gamma}_{od}(\vect{\theta})}
+ \mat{\Gamma}_{os}(\vect{\theta}) \ .
\label{eq:shear_matrix_dominant_final}
\end{empheq}
The last expression is formally equivalent to the shear matrix of a strong lens in the presence of purely tidal line-of-sight perturbations; see, e.g., eq.~(25) of ref.~\cite{Fleury:2020cal}. We stress, nevertheless, that \cref{eq:shear_matrix_dominant_final} is more general because the perturbations $\vect{\alpha}_{od}, \mat{\Gamma}_{od}, \mat{\Gamma}_{os}, \mat{\Gamma}_{ds}$ are allowed to vary across the image plane.

\subsubsection{Convergence, shear, and rotation}

Contrary to the partial shear matrices~\eqref{eq:convergence_shear_matrix}, the complete shear matrix~$\mat{\Gamma}$ is generally not symmetric. This may be seen directly from \cref{eq:shear_matrix_dominant_final}, since the product of two symmetric matrices is generally not symmetric. Thus, we shall decompose $\mat{\Gamma}$ according to\footnote{Other parameterisations of $\mat{\Gamma}$ are possible, see e.g. \S~2.2.2 of ref.~\cite{Fleury:2015hgz}. However, the convergence-shear-rotation decomposition is particularly adapted to the present perturbative set-up.}
\begin{equation}
\label{eq:convergence-shear-rotation_matrix}
\mat{\Gamma}
=
\begin{bmatrix}
\kappa+\Re(\gamma) & \Im(\gamma)-\omega \\
\Im(\gamma)+\omega & \kappa-\Re(\gamma)
\end{bmatrix} ,
\end{equation}
where $\omega$ essentially encodes the rotation of infinitesimal images with respect to their sources. The explicit computation of $\kappa, \gamma, \omega$ may be performed directly from \cref{eq:shear_matrix_dominant_final}, but a more elegant method relies on the complex notation.

\paragraph{Complex notation} Because we are working with two-dimensional quantities on a ``flat sky'', we may express all the relevant quantities involved hitherto using complex numbers. Specifically, if $(\vect{e}_1, \vect{e}_2)$ denotes an orthonormal basis for the flat sky, then we shall canonically associate complex numbers to vectors as
\begin{equation}
\vect{\theta}=\theta_1\vect{e}_1+\theta_2\vect{e}_2
\mapsto
\cplx{\theta}=\theta_1 + \ii\theta_2 \ .
\end{equation}
Besides, if $\mat{\Gamma}$ is parameterised as in \cref{eq:convergence-shear-rotation_matrix}, then its action on a vector $\vect{u}$ has a quite simple complex counterpart
\begin{equation}
\vect{v} = \mat{\Gamma} \vect{u}
\mapsto
\cplx{v} = (\kappa+\ii\omega)\cplx{u} + \gamma\cplx{u}^* \ ,
\end{equation}
where a star denotes complex conjugation.

\paragraph{Complex derivatives} The complex function $\cplx{\alpha}$ associated with a vector field $\vect{\alpha}(\vect{\theta})$ must be considered a function of two independent variables, namely $\cplx{\theta}, \cplx{\theta}^*$. Partial derivatives with respect to these are
\begin{equation}
\pd{}{\cplx{\theta}}
\define
\frac{1}{2} \pa{ \pd{}{\theta_1}-\ii\pd{}{\theta_2} } \ ,
\qquad
\pd{}{\cplx{\theta}^*}
\define
\frac{1}{2} \pa{ \pd{}{\theta_1}+\ii\pd{}{\theta_2} } \ ,
\end{equation}
from which it follows that $[\partial\cplx{\alpha}/\partial\cplx{\theta}]^*=\partial\cplx{\alpha}^*/\partial\cplx{\theta}^*$. 
The first-order Taylor expansion of a complex function must generally account for both variables $\cplx{\theta}, \cplx{\theta}^*$ in the sense that
\begin{equation}
\cplx{\alpha}(\cplx{\theta}+\delta\cplx{\theta})
=
\cplx{\alpha}(\cplx{\theta})
+ \pd{\cplx{\alpha}}{\cplx{\theta}} \, \delta\cplx{\theta}
+ \pd{\cplx{\alpha}}{\cplx{\theta}^*} \, \delta\cplx{\theta}^*
+ \order(\delta\theta^2) \ .
\end{equation}
This double dependence must also be accounted for in the chain rule, for instance
\begin{equation}
\pd{}{\cplx{\theta}}
\paac{
\cplx{\alpha}[\cplx{\beta}(\cplx{\theta})]
}
=
\pd{\cplx{\alpha}}{\cplx{\beta}} \pd{\cplx{\beta}}{\cplx{\theta}}
+ \pd{\cplx{\alpha}}{\cplx{\beta}^*} \pd{\cplx{\beta}^*}{\cplx{\theta}} \ .
\end{equation}
Finally, the complex counterpart of the gradient of a scalar function $f(\vect{\theta})$ is $2\partial f/\partial\cplx{\theta}^*$.

\paragraph{Computing $\kappa,\gamma,\omega$ from the complex lens equation} With the above definitions, and recalling that $\mat{\Gamma}=\dd\vect{\alpha}/\dd\vect{\theta}$, it is straightforward to check that
\begin{equation}
\label{eq:kappa_gamma_complex_notation}
\kappa+\ii\omega
=\pd{\cplx{\alpha}}{\cplx{\theta}} \ ,
\qquad
\gamma
= \pd{\cplx{\alpha}}{\cplx{\theta}^*} \ .
\end{equation}
This also applies to the partial convergences~$\kappa_{ilj}=\partial\cplx{\alpha}_{ilj}/\partial\cplx{\beta}_{il}$ and shears~$\gamma_{ilj}=\partial\cplx{\alpha}_{ilj}/\partial\cplx{\beta}_{il}^*$.

Starting from the complex counterpart of the displacement vector~\eqref{eq:displacement_dominant},
\begin{equation}
\cplx{\alpha}(\cplx{\theta})
=
\cplx{\alpha}_{od s} \pac{\cplx{\theta}-\cplx{\alpha}_{od}(\cplx{\theta})}
+ \sum_{l<d} \cplx{\alpha}_{ols}(\cplx{\theta})
+ \sum_{l>d} \cplx{\alpha}_{ols}[\cplx{\theta}-\cplx{\alpha}_{odl}(\cplx{\theta})] \ ,
\end{equation}
we obtain
\begin{empheq}[box=\fbox]{align}
\label{eq:convergence-rotation_dominant_approximation}
\kappa(\vect{\theta})+\ii\omega(\vect{\theta})
&= \pa{1-\kappa_{od}-\kappa_{ds}}
    \kappa_{ods}(\vect{\theta}-\vect{\alpha}_{od})
    - \pa{\gamma_{od}^*\gamma_{ods}
        + \gamma_{ds} \gamma_{ods}^*}
    + \kappa_{os} \ ,
\\
\label{eq:shear_dominant_approximation}
\gamma(\vect{\theta})
&= \pa{1-\kappa_{od}-\kappa_{ds}}
    \gamma_{ods}(\vect{\theta}-\vect{\alpha}_{od})
    - (\gamma_{od}+\gamma_{ds}) \kappa_{ods} + \gamma_{os} \ ,
\end{empheq}
where the various convergences and shears are naturally associated with the corresponding shear matrices involved in \cref{eq:shear_matrix_dominant_final}. For example,
\begin{equation}
\gamma_{os}(\vect{\theta})
\define
\sum_{l<d} \gamma_{ol s}(\vect{\theta})
+ \sum_{l>d} \gamma_{ol s}\pac{\vect{\theta}-\vect{\alpha}_{odl}(\vect{\theta})} \ .
\end{equation}
We stress again that the convergences and shears involved in \cref{eq:convergence-rotation_dominant_approximation,eq:shear_dominant_approximation} are generally not constant across the image plane; the quantities for which no argument is specified are implicitly evaluated at $\vect{\theta}$ to alleviate notation.

\paragraph{Remarks} A given perturber $l\neq d$ contributes differently to the convergences $\kappa_{od}, \kappa_{ds}, \kappa_{os}$ and shears $\gamma_{od}, \gamma_{ds}, \gamma_{os}$, because of the distance ratios controlling its efficiency. Specifically, the $(od)$-efficiency of $l$ peaks roughly half-way between $o$ and $d$; its $(ds)$-efficiency half-way between $d$ and $s$; and its $(os)$-efficiency half-way between $o$ and $s$.

The expressions~\eqref{eq:convergence-rotation_dominant_approximation} and \eqref{eq:shear_dominant_approximation} of $\kappa+\ii\omega$ and $\gamma$ exhibit two different classes of terms. On the one hand, $\kappa_{os}, \gamma_{os}$ are \emph{direct contributions} of the perturbers; they simply translate the fact that light is further deflected in their presence. All the other contributions are \emph{lens-lens coupling} terms of the form $(\text{dominant})\times(\text{perturber})$; they are all due to departures from the Born approximation. As expected, the rotation $\omega=\Im[(\gamma_{od}-\gamma_{ds})\gamma_{ods}^*]$ entirely consists of lens-lens coupling terms.

\subsubsection{Magnification}

The image-plane magnification~$\mu(\vect{\theta})$ is the relative increase in the angular size of an infinitesimal image at $\vect{\theta}$ due to lensing, $\mu(\vect{\theta})\define \dd^2\vect{\theta}/\dd^2\vect{\beta}=1/\det\mat{\amplification}$. Its inverse may thus be expressed in terms of the convergence, shear, and rotation as
\begin{equation}
\mu^{-1}(\vect{\theta})
= |1-\kappa-\ii\omega|^2-|\gamma|^2
\ .
\end{equation}
From the expressions \eqref{eq:convergence-rotation_dominant_approximation} and \eqref{eq:shear_dominant_approximation} of $\kappa+\ii\omega$ and $\gamma$, we find
\begin{empheq}[box=\fbox]{multline}
\label{eq:magnification_dominant_lens}
\mu^{-1}(\vect{\theta})
=
\pac{
    1-2\pa{\kappa_{od}+\kappa_{ds}}
    }
\mu^{-1}_{ods}(\vect{\theta}-\vect{\alpha}_{od})
\\
+ 2(1-\kappa_{ods}) (\kappa_{od}+\kappa_{ds}-\kappa_{os})
+ 2\Re\pac{\gamma_{ods}^*(\gamma_{od}+\gamma_{ds}-\gamma_{os})} ,
\end{empheq}
where $\mu^{-1}_{ods}\define(1-\kappa_{ods})^2-|\gamma_{ods}|^2$ would be the inverse magnification of the dominant lens alone. The quantities whose argument is not specified are implicitly evaluated at $\vect{\theta}$.

\subsection{Time delays and Fermat potential}
\label{subsec:time_delays}

Although it is not the prime focus of the present article, let us finally say a word about time delays for the sake of completeness. In the multi-plane formalism, the delay between two images A and B of the same source $\vect{\beta}$, characterised by the paths $\{\vect{x}_l\h{A}\}$ and $\{\vect{x}_l\h{B}\}$ through the $N$ planes, reads $\Delta t_{AB}= T(\{\vect{x}_l\h{A}\})-T(\{\vect{x}_l\h{B}\})$, with~\cite{1992grle.book.....S,Fleury:2020cal}
\begin{align}
\label{eq:time_delay_multi-plane}
T(\{\vect{x}_l\})
&=
\sum_{l=1}^N
\pac{
    \frac{1}{2} \, \tau_{l(l+1)} |\vect{\beta}_{o(l+1)}-\vect{\beta}_{ol}|^2
    - (1+z_l) \hat{\psi}_l(\vect{x}_l)
    } ,
\\
\label{eq:time-delay_scale}
\tau_{ij}
&\define
(1+z_i) \, \frac{D_{oi}D_{oj}}{D_{ij}} \ ,
\end{align}
and where we may recall that $\vect{\beta}_{ol}=\vect{x}_l/D_{ol}$.

In the dominant-lens regime, the arguments of $T$ reduce to two variables $(\vect{\theta}, \vect{\beta})$ and its expression becomes
\begin{empheq}[box=\fbox]{multline}
\label{eq:time_delay_dominant}
T(\vect{\theta},\vect{\beta})
=
\frac{1}{2} \tau_{ds}\abs{\vect{\theta}-\vect{\alpha}_{od}-\vect{\beta}}^2
- (1+z_{d}) \hat{\psi}_{d}\pac{D_{od}(\vect{\theta}-\vect{\alpha}_{od})}
\\
- \sum_{l<d} (1+z_l)\hat{\psi}_l(D_{ol}\vect{\theta})
- \sum_{l>d} (1+z_l)\hat{\psi}_l\pac{D_{ol}(\vect{\theta}-\vect{\alpha}_{odl})}
+ \order(\eps^4) \ .
\end{empheq}
The standard single-lens case without perturbations is recovered by setting $\vect{\alpha}_{od}, \hat{\psi}_{l\neq d}$ to zero. The calculation leading from \cref{eq:time_delay_multi-plane} to \cref{eq:time_delay_dominant} is tedious and not relevant to main point of this article; we nevertheless refer the interested reader to \cref{app:time_delays} for details.

Fermat's principle states that, for physical rays, the time delay must be stationary with respect to infinitesimal changes of the light path~\cite{1992grle.book.....S}. In the DL regime, since the light path is entirely controlled by $\vect{\theta}$, this means that imposing $\partial T/\partial\vect{\theta}=\vect{0}$ should yield the lens equation~\eqref{eq:lens_equation_dominant}. However, this does not immediately work here, because the expression of $T$ as given in \cref{eq:time_delay_dominant} has already been evaluated ``on shell''; in other words, the lens equation has already been substituted in \cref{eq:time_delay_dominant} so as to simplify its expression. The Fermat potential~$\phi$ that would yield the lens equation via $\partial\phi/\partial\vect{\theta}=\vect{0}$ actually reads
\begin{equation}
\phi(\vect{\theta}, \vect{\beta})
\define
T(\vect{\theta}, \vect{\beta})
+ \pac{\vect{\theta}-\vect{\beta}-\vect{\alpha}(\vect{\theta})} \cdot
    \sum_{l>d} \frac{\tau_{ds}\tau_{ls}}{\tau_{dl}}
            \vect{\alpha}_{ols}(\vect{\theta}-\vect{\alpha}_{odl}) \ ,
\end{equation}
where the second term on the right-hand side vanishes for physical rays, whereas its derivative with respect to $\vect{\theta}$ does not. Again, we refer the interested reader to \cref{app:time_delays} for details about those statements and derivations.

\section{Parameterising the line of sight: convergence, shear, and beyond}
\label{sec:parametric}

The lens equation~\eqref{eq:lens_equation_dominant} and time delay~\eqref{eq:time_delay_dominant} derived in \cref{sec:dominant_lens} provide an accurate description of line-of-sight corrections in strong-lensing systems. However, their application requires a good knowledge of the properties of the secondary deflectors---such as their positions and mass profiles---which is practically difficult.

In this section we show how, under additional assumptions, line-of-sight effects may be reduced to a finite number of parameters. This approach is thus well suited to parameterised lens modelling. In \cref{subsec:tidal} we present the well-known case where secondary deflectors can be reduced to tidal fields. We use this opportunity to discuss the degeneracy between line-of-sight corrections and the properties of the main lens in \cref{subsec:azimuthal_degeneracies}. In \cref{subsec:flexion}, we show how the tidal approximation may be supplemented with flexion parameters.

\subsection{Tidal approximation: external convergence and shear}
\label{subsec:tidal}

The most common approach to line-of-sight corrections in strong lensing consists in adding exterior convergence and shear parameters to the main lens~\cite{1987ApJ...316...52K, 1994A&A...287..349S, Schneider:1997bq, 1996ApJ...468...17B, Keeton:1996tq, Birrer:2016xku, Fleury:2020cal}. This approach is relevant if the secondary deflectors can be treated in the tidal regime. Concretely, the tidal regime applies if the secondary deflectors are either some smoothly distributed matter component on the line of sight (external convergence), or matter lumps lying far from the line of sight (external shear). In the terminology of refs.~\cite{Schneider:2014vka, Fleury:2020cal}, the secondary deflectors are assumed to produce a smooth gravitational field at the beam's scale.

\subsubsection{Defining the tidal regime}

A lens~$l\neq d$ is in the tidal regime if the separation between a physical ray and the optical axis is very small (infinitesimal) compared to the typical scale over which the lens's gravitational field varies appreciably. In other words, the space-time curvature produced by the lens may be considered constant across the region of the sky under consideration. In practice, this means that $\mat{\Sigma}_l$ can be treated as if it were homogeneous across the $l\h{th}$ plane. In such conditions, the deflection angle~$\hat{\vect{\alpha}}_l(\vect{x}_l)$ is linear in $\vect{x}_l$; for any couple of position~$\vect{x}_l,\vect{x}'_l$,
\begin{equation}
\hat{\vect{\alpha}}_l(\vect{x}_l)
= \hat{\vect{\alpha}}_l(\vect{x}'_l) + 4\pi G \mat{\Sigma}_l \cdot (\vect{x}_l-\vect{x}'_l) \ ,
\end{equation}
and hence any associated partial displacement angle
$
\vect{\alpha}_{ilj}(\vect{\beta}_{il})
= (D_{lj}/D_{ij})
    \hat{\vect{\alpha}}_l(\vect{x}_l/D_{il})
$
reads
\begin{equation}
\label{eq:partial_displacement_tidal}
\vect{\alpha}_{ilj}(\vect{\beta}_{il})
= \vect{\alpha}_{ilj}(\vect{\beta}_{il}')
 + \mat{\Gamma}_{ilj}\cdot
    \pa{\vect{\beta}_{il}-\vect{\beta}_{il}'} \ ,
\end{equation}
where $\mat{\Gamma}_{ilj}=\mat{\Sigma}_l/\Sigma\h{crit}_{ilj}$ is homogeneous. Similarly, the projected potential~$\hat{\psi}_l(\vect{x}_l)$ is quadratic,
\begin{equation}
\label{eq:projected_potential_tidal}
\hat{\psi}_l(\vect{x}_l)
= \hat{\psi}_l(\vect{x}'_l)
    + \hat{\vect{\alpha}}_l(\vect{x}'_l)
        \cdot(\vect{x}_l-\vect{x}'_l)
    + \frac{1}{2} (\vect{x}_l-\vect{x}'_l)
        \cdot 4\pi G\mat{\Sigma}_l \cdot
        (\vect{x}_l-\vect{x}'_l) \ .
\end{equation}

\subsubsection{Lens equation and shear matrix}

Applying \cref{eq:partial_displacement_tidal} to the various terms of \cref{eq:displacement_dominant}, we find that the lens equation in the DL approximation and tidal regime reads $\vect{\beta}=\vect{\theta}-\vect{\alpha}(\vect{\theta})$, with
\begin{align}
\label{eq:displacement_tidal_1}
\vect{\alpha}(\vect{\theta})
&=
\vect{\alpha}_{ods}
    \pac{\vect{\theta}
        -\vect{\alpha}_{od}(\vect{0})
        - \mat{\Gamma}_{od}\vect{\theta}}
+ \sum_{l<d}
    \paac{
    \vect{\alpha}_{ols}(\vect{0})
    + \mat{\Gamma}_{ols}\vect{\theta}
    }
\nonumber\\ &\hspace{5cm}
+ \sum_{l>d}
    \paac{
    \vect{\alpha}_{ols}(\vect{0})
    + \mat{\Gamma}_{ols}\cdot
        \pac{\vect{\theta}
            -\vect{\alpha}_{odl}(\vect{\theta})}
    }
\\
\label{eq:displacement_tidal_2}
&= \pa{\mat{1}-\mat{\Gamma}_{ds}}\cdot
    \vect{\alpha}_{ods}
        \pac{
            \pa{\mat{1}-\mat{\Gamma}_{od}}\vect{\theta}
            -\vect{\alpha}_{od}(\vect{0})}
    + \vect{\alpha}_{os}(\vect{0})
    + \mat{\Gamma}_{os}\vect{\theta} \ ,
\end{align}
where the ``integral'' tidal matrices $\mat{\Gamma}_{od}, \mat{\Gamma}_{os}, \mat{\Gamma}_{ds}$ were defined in \cref{subsubsec:shear_matrix}, and $\vect{\alpha}_{os}$ is defined as the displacement of an image due to all lenses but the dominant one,
\begin{equation}
\label{eq:alpha_os}
\vect{\alpha}_{os}(\vect{0}) \define \sum_{l\not= d} \vect{\alpha}_{ols}(\vect{0}) \ .
\end{equation}
When going from \cref{eq:displacement_tidal_1} to \cref{eq:displacement_tidal_2}, we used that $\mat{\Gamma}_{ols}\vect{\alpha}_{odl}=\mat{\Gamma}_{dls}\vect{\alpha}_{ods}$, which comes from a redistribution of the distance factors similarly to \cref{eq:redistribution_distance_factors}.

In the absence of line-of-sight perturbations, \cref{eq:displacement_tidal_2} would reduce to $\vect{\alpha}(\vect{\theta})=\vect{\alpha}_{ods}(\vect{\theta})$. The presence of the secondary deflectors thus adds 9 real parameters to the lens model: 3 convergences $(\kappa_{od}, \kappa_{os}, \kappa_{ds})$ plus 3 complex shears $(\gamma_{od}, \gamma_{os}, \gamma_{ds})$, which are the constituents of the three shear matrices $\mat{\Gamma}_{od}, \mat{\Gamma}_{os}, \mat{\Gamma}_{ds}$. The constant displacements $\vect{\alpha}_{od}(\vect{0})$ and $\vect{\alpha}_{os}(\vect{0})$ can be absorbed as a re-definition of the origin of the source and main lens plane, and hence do not count as extra parameters. We discuss their geometric meaning in \cref{subsubsec:comparison_literature}.

The shear matrix~$\mat{\Gamma}(\vect{\theta})$ associated with $\vect{\alpha}(\vect{\theta})$, which describes the distortions of an infinitesimal image by the dominant lens and tidal line-of-sight perturbations, is identical to \cref{eq:shear_dominant_approximation}, except that the argument of $\mat{\Gamma}_{ods}$ is replaced with $\vect{\theta}-\vect{\alpha}_{od}(\vect{\theta})
\rightarrow\vect{\theta}-\vect{\alpha}_{od}(\vect{0})-\mat{\Gamma}_{od}\vect{\theta}$. This may be checked by taking the derivative of \cref{eq:displacement_tidal_2} with respect to $\vect{\theta}$.

\subsubsection{Comparison with the literature}
\label{subsubsec:comparison_literature}

\paragraph{Eliminating the homogeneous displacements} The problem of tidal line-of-sight perturbations to strong lenses has been well studied for more than 30 years. The associated lens equation can be found in, e.g., refs.~\cite{1987ApJ...316...52K, 1996ApJ...468...17B}. The careful reader may have noticed the absence of the constant displacements $\vect{\alpha}_{od}(\vect{0}), \vect{\alpha}_{os}(\vect{0})$ in those earlier results---eq.~(6.7) in ref.~\cite{1987ApJ...316...52K} and eq.~(14) in ref.~\cite{1996ApJ...468...17B}. Yet, such terms have no reason to be quantitatively negligible compared to the linear, tidal, terms involving shear matrices. 

The reason for the usual omission of the constant terms is that they have no observational consequences, and hence can be absorbed in suitable re-definitions of the origins of the source plane and main-lens plane. In \cref{eq:displacement_tidal_2}, $\vect{\alpha}_{os}(\vect{0})$ causes a global displacement of images with respect to their unlensed counterpart, without changing their apparent shape or luminosity. Since the unlensed position of sources is unobservable, this global shift may be eliminated via the substitution $\vect{\beta}\rightarrow\vect{\beta}-\vect{\alpha}_{os}(\vect{0})$.

As for $\vect{\alpha}_{od}(\vect{0})$ which appears in the argument of $\vect{\alpha}_{ods}$, it encodes the apparent displacement of the dominant plane due to foreground lenses. As such, it can be absorbed by shifting the origin of that plane as $\vect{x}_d\mapsto\vect{x}_d'\define \vect{x}_d-D_{od}\vect{\alpha}_{od}(\vect{0})$. This homogeneous displacement is not measurable, because it also applies to the apparent position of the main lens. For example, if the image were an Einstein ring, then its centre would be shifted by $\vect{\alpha}_{od}(\vect{0})$, and so would the apparent position of the lens; thus, the latter would appear at the centre of the ring, regardless of $\vect{\alpha}_{od}(\vect{0})$. Thus, we may simply omit $\vect{\alpha}_{od}(\vect{0})$ in the following.

\begin{figure}[t]
    \centering
    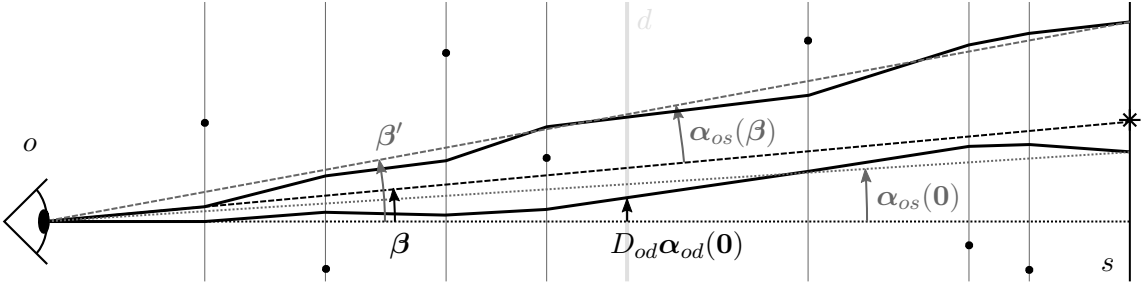
    \caption{Same as \cref{fig:multi-plane_dominant} but without the action of the dominant lens. Thick solid lines indicate light rays that are only affected by the secondary deflectors. In that sense the top thick line is the perturbed counterpart of the unlensed ray, while the bottom thick line is the perturbed counterpart of the fiducial ray. The angle $\vect{\beta}'=\vect{\beta}+\vect{\alpha}_{os}(\vect{\beta})$ [resp. $\vect{\alpha}_{os}(\vect{0})$] represents the position of a fictitious source that would be observed in the direction $\vect{\beta}$ (resp. $\vect{0}$) in the absence of the dominant lens. The position $D_{od}\vect{\alpha}_{od}(\vect{0})$ is the intersection of the perturbed fidiucial ray with the dominant lens plane, and can be taken as an alternative origin of that plane. Note that the secondary deflectors depicted in this figure cannot be treated in the tidal regime.}
    \label{fig:definition_unlensed}
\end{figure}

The above may be re-formulated as a redefinition of the fiducial ray. Indeed, the origin of the lens planes and source plane are traditionally set by their intersection with an unlensed ray playing the role of the optical axis. As shown in \cref{fig:definition_unlensed}, the shifts $\vect{x}_d\mapsto\vect{x}_d'\define \vect{x}_d-D_{od}\vect{\alpha}_{od}(\vect{0})$ and $\vect{x}_s=D_{os}\vect{\beta}\mapsto\vect{x}_s'\define \vect{x}_s-D_{os}\vect{\alpha}_{os}(\vect{0})$ correspond to defining the origin of the $d,s$ planes as their intersections with the weakly lensed counterpart of the fiducial ray, i.e., the ray that would be observed at $\vect{\theta}=\vect{0}$ in the presence of the secondary deflectors only.

\paragraph{What does ``unlensed'' mean?} The previous discussion raises the more general question of how to define unlensed rays. So far we have adopted the standard convention for which the unlensed ray propagates in the FLRW background, i.e., in the absence of all lenses.

Another possibility consists in including the secondary deflectors in the background, and calling ``lensing'' the sole effect of the main deflector. This alternative approach is necessary to describe lensing within arbitrary space-times~\cite{Fleury:2020cal}. In that case, the ``unlensed position'' of the source is the direction~$\vect{\beta}'=\vect{\beta}+\vect{\alpha}_{os}(\vect{\beta})=\vect{\beta}+\vect{\alpha}_{os}(\vect{\beta}')+\order(\eps^4)$ in which it would be seen without the main lens (see \cref{fig:definition_unlensed}). Besides, the origin of the main lens plane is now at $D_{od}\vect{\alpha}_{od}(\vect{0})$. With such conventions, the lens equation is found to read
\begin{equation}
\label{eq:lens_equation_dominant_tidal_prime}
\vect{\beta}'=\vect{\theta} - \vect{\alpha}'(\vect{\theta}) \ ,
\qquad
\vect{\alpha}'(\vect{\theta})
= (\mat{1}+\mat{\Gamma}_{os}-\mat{\Gamma}_{ds})
     \vect{\alpha}_{ods}[(\mat{1}-\mat{\Gamma}_{od})\vect{\theta}] \ ,
\end{equation}
in the tidal regime, in agreement with eq.~(7) of ref.~\cite{Fleury:2020cal}. Note that, contrary to \cref{eq:displacement_tidal_2}, $\vect{\alpha}'(\vect{\theta})$ does not involve the homogeneous displacement $\vect{\alpha}_{os}(\vect{0})$ by construction.

\paragraph{Comparison with the CSB approach} It is instructive to compare our \cref{eq:displacement_tidal_2} with its counterpart when the CSB (rather than DL) approximation is applied. As discussed in \cref{subsubsec:CSB}, the CSB approximation consists in making the substitution $\vect{\theta}-\vect{\alpha}_{odl}(\vect{\theta})\rightarrow[1-(D_{dl}D_{os})/(D_{ds}D_{ol})]\vect{\theta}$ in the argument of the background displacement angles~$\vect{\alpha}_{ols}$. In the tidal regime, this eventually corresponds to the substitution $\mat{\Gamma}_{ds}\vect{\alpha}_{ods}\rightarrow \mat{\Gamma}_{ds}\vect{\theta}$. Thus, the displacement angles corresponding to the DL and CSB approximations compare as
\begin{align}
\vect{\alpha}\e{DL}(\vect{\theta})
&= \pa{\mat{1}-\textcolor{red4}{\mat{\Gamma}_{ds}}}\vect{\alpha}_{ods}
        \pac{\pa{\mat{1}-\mat{\Gamma}_{od}}\vect{\theta}}
    + \mat{\Gamma}_{os}\vect{\theta}
    + \vect{\alpha}_{os}(\vect{0})\ ,
\label{eq:displacement_DL_simple}
\\
\vect{\alpha}\e{CSB}(\vect{\theta})
&= \vect{\alpha}_{ods}
        \pac{\pa{\mat{1}-\mat{\Gamma}_{od}}\vect{\theta}}
    + \pa{\mat{\Gamma}_{os}-\textcolor{red4}{\mat{\Gamma}_{ds}}}\vect{\theta}
    + \vect{\alpha}_{os}(\vect{0})\ .
\label{eq:displacement_CSB}
\end{align}
\Cref{eq:displacement_CSB} corresponds to eq.~(3.7) of ref.~\cite{Birrer:2016xku} with our notation.

\Cref{eq:displacement_DL_simple,eq:displacement_CSB} are structurally similar and they involve the same line-of-sight parameters. Nevertheless, while \cref{eq:displacement_CSB} only applies in the immediate vicinity of the critical curve, \cref{eq:displacement_DL_simple} holds across a more extended region of the image plane. The DL approximation is thus expected to be more accurate than the CSB approximation for the description of images departing from the critical curve, which includes thick Einstein rings.

\subsubsection{Time delays}

In the tidal regime, the time delay~\eqref{eq:time_delay_dominant} between a physical signal observed at $\vect{\theta}$ and its unlensed counterpart at $\vect{\beta}$ is found to read $T(\vect{\theta},\vect{\beta})=T_d(\vect{\theta},\vect{\beta}')+\delta T(\vect{\beta}')$, with
\begin{align}
T_d(\vect{\theta},\vect{\beta}')
&= \frac{1}{2} \tau_{ds}
    (\vect{\theta}-\vect{\beta}')\cdot
    \pa{\mat{1}-\mat{\Gamma}_{os}-\mat{\Gamma}_{od}+\mat{\Gamma}_{ds}}
    (\vect{\theta}-\vect{\beta}')
    - (1+z_d) \hat{\psi}_d\pac{D_{od}(\mat{1}-\mat{\Gamma}_{od})\vect{\theta}} ,
\label{eq:time_delay_tidal_main}
\\
\delta T(\vect{\beta}')
&= - \sum_{l\not=d} (1+z_l) \hat{\psi}_l(D_{ol}\vect{\beta}') \ ,
\label{eq:time_delay_tidal_correction}
\end{align}
where $\vect{\beta}'=\vect{\beta}+\vect{\alpha}_{os}(\vect{\beta}')$ is the partially unlensed source position discussed in \cref{subsubsec:comparison_literature}.

Physically speaking, $T_d(\vect{\theta},\vect{\beta}')$ represents the time delay between the physical ray and a ray affected by the secondary lenses only. Its geometrical component involves the usual time-delay scale $\tau_{ds}=(1+z_d)D_{od}D_{os}/D_{ds}$. \Cref{eq:time_delay_tidal_main} agrees with refs.~\cite{1992grle.book.....S, Schneider:1997bq, 1996ApJ...468...17B, Fleury:2020cal}. The second term $\delta T(\vect{\beta}')$ represents the time delay between the partially unlensed signal ($\vect{\beta}'$) and the fully unlensed one ($\vect{\beta}$). It involves only potential terms because geometrical terms would be of order $|\vect{\beta}'-\vect{\beta}|^2 = |\vect{\alpha}_{os}|^2 = \order(\eps^4)$. The derivation of \cref{eq:time_delay_tidal_main,eq:time_delay_tidal_correction} being quite tedious, we refer the interested reader to \cref{subsec:derivation_time_delay_tidal} for details.

For practical purposes, $\delta T(\vect{\beta}')$ is irrelevant, because the only measurable quantity is the time delay between different images of the same source, $\Delta t(\vect{\theta}\e{A},\vect{\theta}\e{B})=T(\vect{\theta}\e{A},\vect{\beta})-T(\vect{\theta}\e{B},\vect{\beta})$, where $\delta T$ cancels. However, it is important to insert the correct value of $\vect{\beta}'$ in the main term $T_d(\vect{\theta},\vect{\beta}')$. In particular, one should not omit the constant displacement $\vect{\alpha}_{os}(\vect{0})$ here. In practice, the easiest solution consists in substituting the lens equation~\eqref{eq:lens_equation_dominant_tidal_prime} so as to express everything in terms of a unique shear matrix~$\mat{\Gamma}\e{LOS}$ and an effective main-lens model~$\psi\e{eff}$,
\begin{equation}
\label{eq:time_delay_tidal_model}
T(\vect{\theta})
= \tau_{ds}
    \pac{
        \frac{1}{2}\,
        \vect{\alpha}\e{eff}(\vect{\theta})
        \cdot (\mat{1}+\mat{\Gamma}\e{LOS})
        \vect{\alpha}\e{eff}(\vect{\theta})
        - \psi\e{eff}(\vect{\theta})
        } ,
\end{equation}
with
\begin{align}
\label{eq:Gamma_LOS_def}
\mat{\Gamma}\e{LOS} &\define \mat{\Gamma}_{od} + \mat{\Gamma}_{os} - \mat{\Gamma}_{ds} \ ,
\\
\label{eq:psi_eff_def}
\psi\e{eff}(\vect{\theta})
&\define \psi_{ods}\pac{(1-\mat{\Gamma}_{od})\vect{\theta}} ,
\qquad
\psi_{ods}(\vect{\beta}_{od})
\define \frac{D_{ds}}{D_{od}D_{os}} \, \hat{\psi}_d(D_{od}\vect{\beta}_{od}) \ ,
\\
\vect{\alpha}\e{eff}(\vect{\theta})
&= \ddf{\psi\e{eff}}{\vect{\theta}}
= (\mat{1}-\mat{\Gamma}_{od}) \, \vect{\alpha}_{ods}\pac{(1-\mat{\Gamma}_{od})\vect{\theta}}
= (\mat{1}-\mat{\Gamma}\e{LOS})\vect{\alpha}'(\vect{\theta})\ ,
\end{align}
and we omitted the irrelevant $\delta T$. This suggests that when analysing strong-lensing time delays, there is (i) a degeneracy between $\mat{\Gamma}_{od}$ and the intrinsic properties of the main lens within $\psi\e{eff}$; and (ii) a degeneracy between the external convergences and shears within $\mat{\Gamma}\e{LOS}$. We shall see in \cref{subsec:azimuthal_degeneracies} that this is also true for the analysis of strong-lensing images.

\subsection{Is the line-of-sight shear measurable?}
\label{subsec:azimuthal_degeneracies}

We have seen that when line-of-sight effects are modelled by tidal pertubations, their effect in the lensing displacement~\eqref{eq:displacement_tidal_2} is encoded in three external convergences~$\kappa_{os}, \kappa_{od}, \kappa_{ds}$, and three external complex shears~$\gamma_{os}, \gamma_{od}, \gamma_{ds}$. An important question is then whether those parameters can be observationally distinguished from the properties of the main lens.

The answer for the convergence is notoriously ``no''. This issue is known as the (external) mass-sheet degeneracy~\cite{1985ApJ...289L...1F,2020A&A...643A.165B}. It is indeed clear from \cref{eq:displacement_tidal_2} that the external convergences may be absorbed by a redefinition of the main-lens model. This degeneracy is a key source of uncertainty when determining the Hubble constant~$H_0$ from time-delay cosmography~\cite{2010ApJ...711..201S,2020MNRAS.498.1406T}.

The case of the external shear is a priori more promising, because it is harder to mimic a homogeneous shear than a homogeneous convergence with a realistic localised mass distribution---even a non-spherical one. This intuition was put in practice in ref.~\cite{Birrer:2016xku}, which empirically concluded that the sole observation of an Einstein ring allowed one to measure $\gamma_{od}$ and $\gamma_{os}-\gamma_{ds}$ with a relative precision of a few percent\footnote{In ref.~\cite{Birrer:2016xku}, the difference $\gamma_{os}-\gamma_{ds}$ is denoted with $\gamma_{s}$.}. In the following, we argue that their conclusions were probably too optimistic due to inherent degeneracies between parameters. We show, however, that the results of ref.~\cite{Birrer:2016xku} should remain valid for the combination $\gamma\e{LOS}\define\gamma_{od}+\gamma_{os}-\gamma_{ds}$ rather than for $\gamma_{od}$ and $\gamma_{os}-\gamma_{ds}$ taken separately.

\subsubsection{Minimal lens model and azimuthal degeneracies}

Let us take the lens equation in the DL approximation and tidal regime, in which we omit the constant terms~$\vect{\alpha}_{od}(\vect{0}), \vect{\alpha}_{os}(\vect{0})$ as discussed in \cref{subsubsec:comparison_literature},
\begin{equation}
\label{eq:lens_equation_DL_tidal}
\vect{\beta}
= \pa{\mat{1}-\mat{\Gamma}_{os}}\vect{\theta}
    - \pa{\mat{1}-\mat{\Gamma}_{ds}}\cdot
    \vect{\alpha}_{ods}
        \pac{
            \pa{\mat{1}-\mat{\Gamma}_{od}}\vect{\theta}} .
\end{equation}
Since the source position is unknown, and hence a free parameter of the model, we may apply any transformation to $\vect{\beta}$ without affecting the relevance of the resulting lens equation. This procedure is known as a source-position transformation (SPT)~\cite{2014A&A...564A.103S, 2018A&A...620A..86W}. In particular, as we multiply \cref{eq:lens_equation_DL_tidal} with the matrix $\mat{1}-\mat{\Gamma}_{od}+\mat{\Gamma}_{ds}$, we get the \emph{minimal lens model}
\begin{empheq}[box=\fbox]{equation}
\label{eq:minimal_lens_model}
\tilde{\vect{\beta}}
= \pa{1-\mat{\Gamma}\e{LOS}}\vect{\theta} - \ddf{\psi\e{eff}}{\vect{\theta}}
\ ,
\end{empheq}
where $\tilde{\vect{\beta}}=(\mat{1}-\mat{\Gamma}_{od}+\mat{\Gamma}_{ds})\,\vect{\beta}$, while $\mat{\Gamma}\e{LOS}$ and $\psi\e{eff}(\vect{\theta})$ were defined in \cref{eq:Gamma_LOS_def,eq:psi_eff_def}. \Cref{eq:minimal_lens_model} is formally identical to the lens equation of a dominant lens with Fermat potential~$\psi\e{eff}$, perturbed by external tides~$\mat{\Gamma}\e{LOS}$ within the same plane.

This model is minimal in the sense that it involves the minimal number of free parameters, thereby fully encoding the degeneracies of the problem.\footnote{Interestingly, the lens equations obtained from the DL and CSB approximations are related by an SPT; this can be easily checked by comparing \cref{eq:displacement_CSB,eq:displacement_DL_simple}. Thus, they are formally associated with the same minimal lens model. The only difference is the definition of $\tilde{\vect{\beta}}$ in each case, which is irrelevant in practice.} Let us focus on azimuthal degeneracies, i.e., degeneracies between the three external shears and the azimuthal structure of the main lens. We may identify two classes of degeneracies:

\paragraph{Internal azimuthal degeneracy} Since $\psi\e{eff}(\vect{\theta})= \psi_{ods}[(1-\mat{\Gamma}_{od})\vect{\theta}]$, the properties of the dominant lens are entangled with the foreground perturbations. In particular, it is practically impossible to distinguish between the ellipticity of the main lens and a foreground shear from a strong-lensing image only (see example in \cref{subsubsec:example_SIE}).

\paragraph{External azimuthal degeneracy} Since the three external shears~$\gamma_{od}, \gamma_{os}, \gamma_{ds}$ do not strike individually but together in the combination~$\gamma\e{LOS}=\gamma_{od}+\gamma_{os}-\gamma_{ds}$, it is impossible to measure any of them separately. However, it may be possible to measure the whole $\gamma\e{LOS}$ independently of the properties of the dominant lens.

\subsubsection{Example: singular isothermal ellipsoid with external shear}
\label{subsubsec:example_SIE}

\paragraph{Singular isothermal ellipsoid (SIE)} The SIE~\cite{1994A&A...284..285K} is a flattened version of the singular isothermal sphere (SIS)---its projected density has elliptical contours and decreases proportionally to the inverse of the distance to the lens. If the semi-minor axis of these contours makes an angle $\ph_0$ with the horizontal direction, then the projected density reads
\begin{equation}
\kappa_{ods}(\vect{\theta})
= \kappa\e{SIE}(\vect{\theta})
\define \frac{\theta\e{E}}{2\theta}\,\sqrt{\frac{f}{\cos^2(\ph-\ph_0)+f^2\sin^2(\ph-\ph_0)}}
\end{equation}
where $\theta\e{E}$ is the Einstein radius of the SIE, $\vect{\theta}=\theta(\cos\ph,\sin\ph)$, and $f=b/a\leq 1$ is the ratio of the semi-minor axis~$b$ and semi-major axis~$a$ of the iso-density contours. When $f=1$ the SIS model $\kappa\e{SIS}(\vect{\theta})=\theta\e{E}/(2\theta)$ is recovered.

Let us consider the weak-ellipticity case, $f\approx 1$. For that purpose, it is convenient to introduce the complex parameter $e=|e|\ex{2\ii\ph_0}$, where $|e|=(a-b)/(a+b)=(1-f)/(1+f)$ is an ellipticity measure sometimes called the third flattening. This definition matches the one used in ref.~\cite{Birrer:2016xku}. For $|e|\ll 1$, we then have $f\approx 1-2|e|$ and the SIE projected density takes the simpler form
\begin{equation}
\kappa_{ods}(\vect{\theta})
= \frac{\theta\e{E}}{2\theta} \pac{ 1 - \Re\pa{e\,\ex{-2\ii\ph}}} + \order(|e|^2) \ .
\end{equation}
Using the complex formalism, it is then straightforward to derive the Fermat potential of the weakly elliptical SIE as
\begin{equation}
\label{eq:Fermat_potential_SIE}
\psi_{ods}(\vect{\theta})
= \theta\e{E}\theta\pac{1+\frac{1}{3}\,\Re\pa{e\,\ex{-2\ii\ph}}}
    + \order(|e|^2) \ .
\end{equation}

\begin{figure}[t]
    \centering
    \includegraphics[width=\columnwidth]{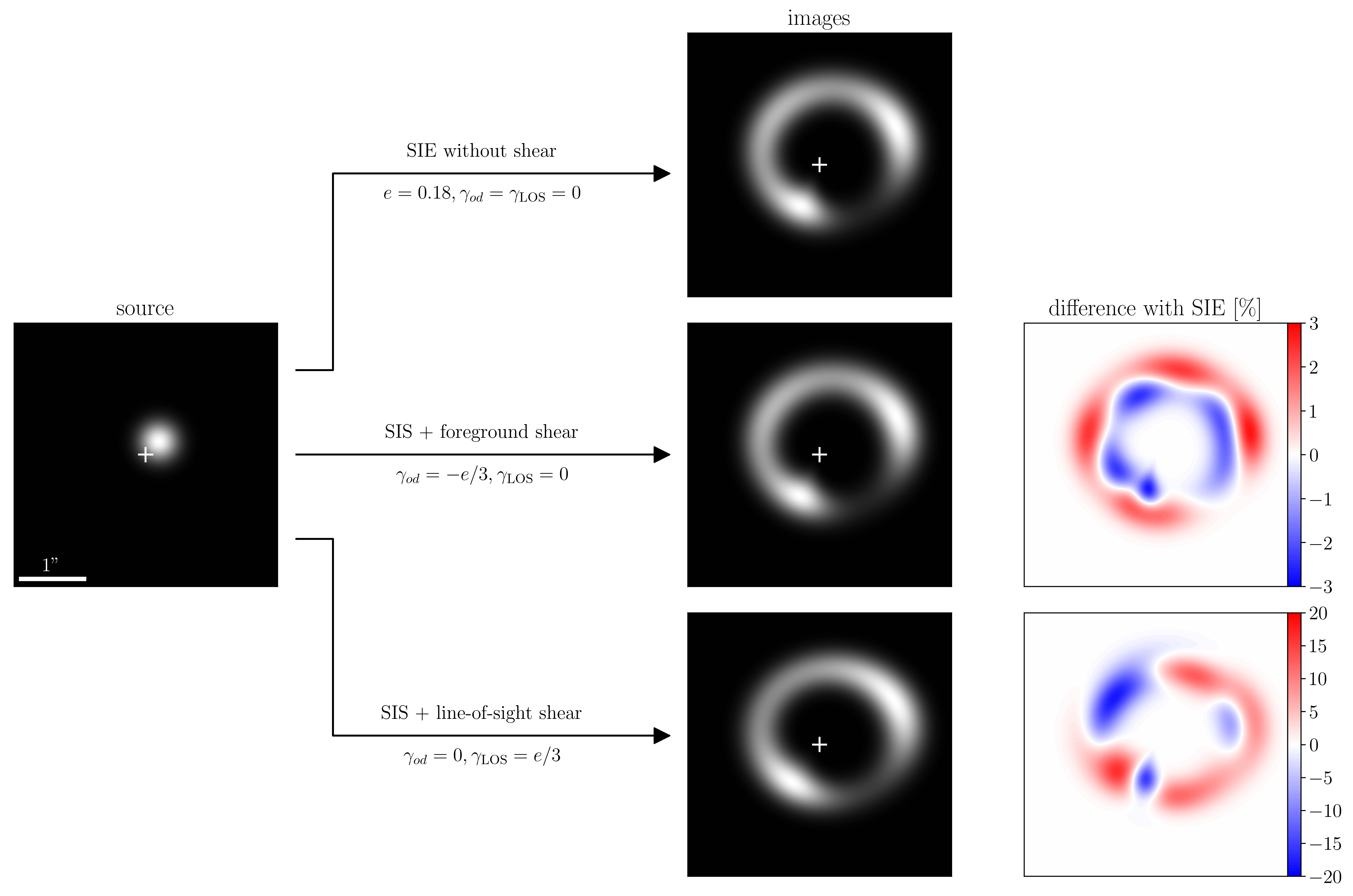}
    \caption{Illustrating the ellipticity-shear degeneracy. \textit{Left}: source with Gaussian surface brightness $I\e{s}(\vect{\beta})=I_0\exp\pac{-|\vect{\beta}-\vect{\beta}_0|^2/(2\sigma^2)}$, $\vect{\beta}_0=(0.2",0.2")$, $\sigma=0.2"$; a white cross indicates the centre of the main lens. \textit{Middle}: images produced by, from top to bottom, an SIE ($\theta\e{E}=1"$, $f=0.7\Leftrightarrow e\approx 0.18$) without external shear; a SIS ($\theta\e{E}=1"$) with the equivalent foreground shear ($\gamma_{od}=-e/3$, $\gamma\e{LOS}=0$); and a SIS ($\theta\e{E}=1"$) with no foreground shear but with the closest LOS attempt ($\gamma\e{LOS}=e/3$). \textit{Right}: normalised surface-brightness difference~$\Delta I/I_0$ between the SIS + shear models and the SIE.}
    \label{fig:ellipticity_shear}
\end{figure}

\paragraph{Degeneracy between ellipticity and foreground shear.} Consider the effective Fermat potential~$\psi\e{eff}(\vect{\theta})$ involved in the minimal lens model~\eqref{eq:minimal_lens_model}, in the simplified case of pure-shear foreground perturbations ($\kappa_{od}=0$, $\gamma_{od}\neq 0$). At linear order in $\gamma_{od}$ and in the main-lens ellipticity $e$, we find
\begin{align}
\psi\e{eff}(\vect{\theta})
&= \psi_{ods}[(\mat{1}-\mat{\Gamma}_{od})\vect{\theta}]
\\
&= \theta\e{E}\theta
    \paac{
        1 + \frac{1}{3}\,\Re\pac{(e-3\gamma_{od})\,\ex{-2\ii\ph}}
        }
    + \order(|e|^2, |e\,\gamma_{od}|, |\gamma_{od}|^2) \ ,
\end{align}
which is formally identical to \cref{eq:Fermat_potential_SIE} with the substitution~$e\rightarrow e-3\gamma_{od}$. At first order, the main-lens ellipticity is thus degenerate with the foreground shear. This degeneracy is illustrated in \cref{fig:ellipticity_shear}; the first two images (SIE without shear and SIS + foreground shear) differ by a maximum of $3\%$ in surface brightness, even though the ellipticity is $e\approx 0.18\not\ll 1$.

Albeit small, such higher-order differences between ellipticity and foreground shear may be picked up by numerical experiments of lens-parameter inference~\cite{Birrer:2016xku}---see also this more recent \href{https://github.com/sibirrer/lenstronomy_extensions/blob/494b0ececae8e48626ea3647549c1fdbed4fc0bf/lenstronomy_extensions/Notebooks/EinsteinRingShear_simulations.ipynb}{notebook}. Such a conclusion is, nevertheless, dangerously model-dependent. For instance, if the dominant-lens model is such that its iso-potential contours are elliptical---instead of its iso-density contours like the SIE---then the degeneracy between ellipticity and foreground shear becomes \emph{exact}. More generally, given the strong entanglement of $\psi_{ods}$ and $\mat{\Gamma}_{od}$ within $\psi\e{eff}$, we recommend not to attempt to distinguish between them, unless one has extra information on either of those quantities.

\paragraph{Non-degeneracy between ellipticity and line-of-sight shear.} Unlike the foreground shear~$\gamma_{od}$, the line-of-sight combination~$\gamma\e{LOS}$ is not degenerate with the ellipticity~$e$ of the main lens, even at linear order in those parameters. This statement may be checked explicitly by examining the complex form of the minimal lens model~\eqref{eq:minimal_lens_model},
\begin{equation}
\label{eq:minimal_lens_model_SIE}
\tilde{\cplx{\beta}}
=
\cplx{\theta}
- \theta\e{E}\ex{\ii\ph}
- \gamma\e{LOS} \theta\ex{-\ii\ph}
- \theta\e{E}
    \pac{
        \frac{1}{2} \, (e-3\gamma_{od}) \, \ex{-\ii\ph}
        - \frac{1}{6} \, (e-3\gamma_{od})^* \ex{3\ii\ph}
        } ,
\end{equation}
where we have set $\kappa_{od}=\kappa_{os}=\kappa_{ds}=0$ for simplicity. Clearly, $\gamma\e{LOS}$ has a distinct role from $e$ and $\gamma_{od}$ in \cref{eq:minimal_lens_model_SIE}. The non-degeneracy between $\gamma\e{LOS}$ and $e$ is illustrated in the bottom row of \cref{fig:ellipticity_shear}, where we can see that the best attempt to imitate $e$ with $\gamma\e{LOS}$, namely setting $\gamma\e{LOS}=e/3$,\footnote{In practice, we have set $\gamma_{os}=e/3$ and $\gamma_{od}=\gamma_{ds}=0$ here. This is not the only option that leads to $\gamma_{od}=0, \gamma\e{LOS}=e/3$, but it ensures that $\tilde{\vect{\beta}}=\vect{\beta}$ so that we do not need to distort the source.} leads to a $20\%$ difference between the corresponding images. This is much larger than the $3\%$ difference achieved with foreground shear. We conclude that
\begin{empheq}[box=\fbox]{equation}
\label{eq:gamma_LOS}
\gamma\e{LOS} \define \gamma_{od} + \gamma_{os} - \gamma_{ds}
\end{empheq}
is largely non-degenerate with $e$; it is thus \emph{measurable} in practice. The generality of this statement, suggested by \cref{eq:minimal_lens_model}, will be fully investigated in future work using a forward-modelling approach~\cite{2015ApJ...813..102B} and the \href{https://github.com/sibirrer/lenstronomy}{\textsc{Lenstronomy}} package~\cite{2018ascl.soft04012B, Birrer:2021wjl}.

\subsubsection{Cosmic shear with strong lenses}

Based on the results of ref.~\cite{Birrer:2016xku}, the same group proposed that Einstein rings could be used as \emph{standard shapes} to measure cosmic shear~\cite{Birrer:2017sge}. They showed that the high precision of such shear measurements could compensate their small number, and could become a competitive probe of weak lensing in future surveys. A first attempt to cross-correlate strong-lensing-based and galaxy-shape-based measurements of cosmic shear was proposed in ref.~\cite{Kuhn:2020wpy}.

\begin{figure}[h]
\centering
\begin{minipage}{0.55\textwidth}
\centering
\includegraphics[width=\columnwidth]{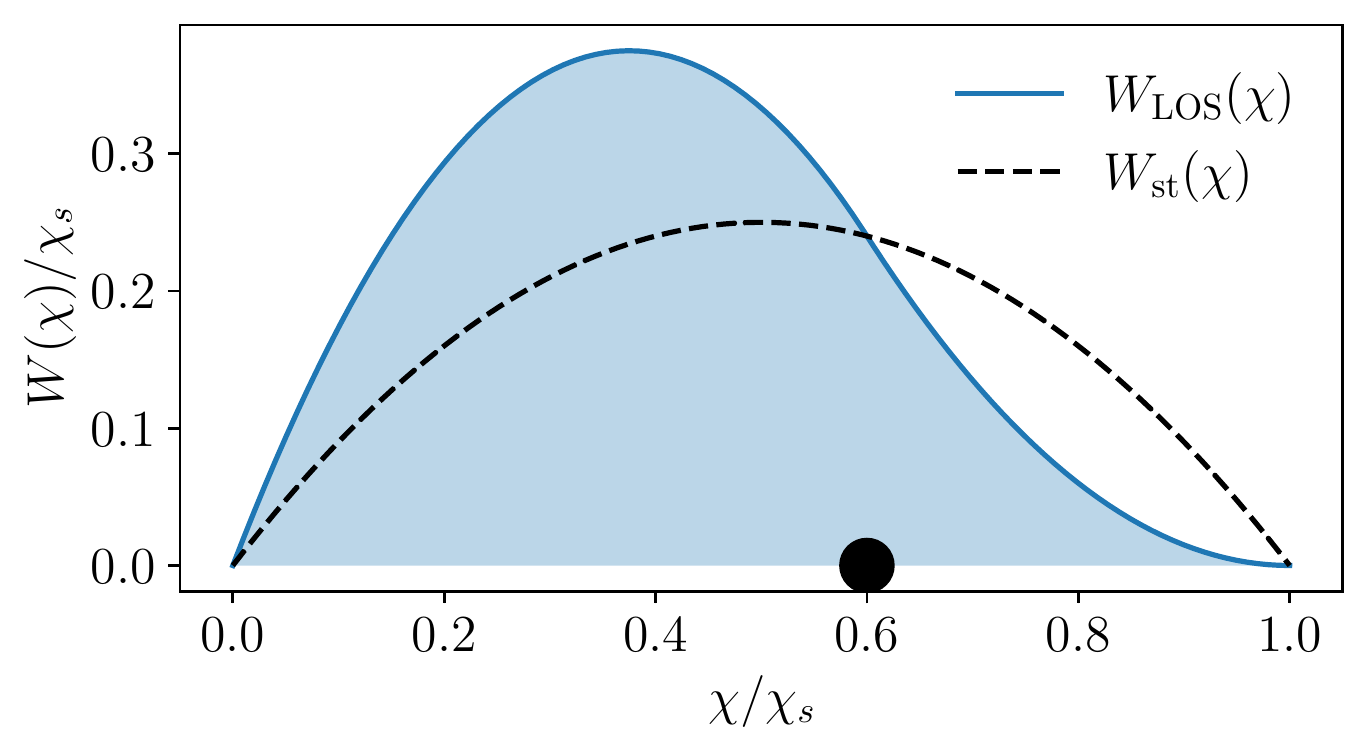}
\end{minipage}
\hfill
\begin{minipage}{0.4\textwidth}
\vspace*{-1.5cm}
\caption{Weight function~$W\e{LOS}(\chi)$ indicating the relative significance of inhomogeneities at $\chi$ along the line of sight in $\gamma\e{LOS}$. Here $K=0$ and the dominant lens was placed at $\chi_d=0.6\chi_s$. The standard cosmic-shear weight~$W\e{st}(\chi)$, which is only the first term of \cref{eq:W_LOS} is indicated as a dashed line for comparison.}
\label{fig:W_LOS}
\end{minipage}
\end{figure}

The results of ref.~\cite{Birrer:2016xku} were probably too optimistic regarding the individual inference of $\gamma_{od}$ and $\gamma_{os}-\gamma_{ds}$. Indeed, distinguishing between $\gamma_{od}$ and the azimuthal structure of the main lens is a strongly model-dependent operation, which exposes one to the risk of over-constraining the model parameters. However, we have found that the combination $\gamma\e{LOS}=\gamma_{od}+\gamma_{os}-\gamma_{ds}$ eludes this degeneracy with the main-lens structure. Therefore, the idea proposed in ref.~\cite{Birrer:2017sge} remains relevant modulo a slight change of its observable. In particular, the lensing kernel for $\gamma\e{LOS}$ is different from the standard weak-lensing $\gamma_{os}$ or even from the kernel of $\gamma_{os}-\gamma_{ds}$. The explicit expression of $\gamma\e{LOS}$ is
\begin{equation}
\label{eq:gamma_LOS_explicit}
\gamma\e{LOS}(\vect{\theta})
= -\frac{3}{2} \Omega\e{m0} H_0^2
	\int_{0}^{\chi_s} \dd\chi \; (1+z) \, W\e{LOS}(\chi)
	\int_{\mathbb{R}^2} \frac{\dd^2\vect{x}}{\pi x^2} \; \ex{2\ii\xi}\,
	    \delta[\eta_0-\chi, \chi, f_K(\chi)\vect{\theta}+\vect{x}] \ ,
\end{equation}
with
\begin{equation}
\label{eq:W_LOS}
W\e{LOS}(\chi) \define 
\begin{cases}
\displaystyle
\frac{f_K(\chi_s-\chi)f_K(\chi)}{f_K(\chi_s)} 
+ \frac{f_K(\chi_d-\chi)f_K(\chi)}{f_K(\chi_d)} & \chi\leq\chi_d \ , \\[3mm]
\displaystyle
\frac{f_K(\chi_s-\chi)f_K(\chi)}{f_K(\chi_s)}
- \frac{f_K(\chi_s-\chi)f_K(\chi-\chi_d)}{f_K(\chi_s-\chi_d)} & \chi\geq\chi_d \ .
\end{cases}
\end{equation}
In \cref{eq:gamma_LOS_explicit}, $\delta(\eta, \chi, \vect{x})$ is the matter density contrast at conformal time~$\eta$, comoving radius~$\chi$, and transverse position $\vect{x}=x(\cos\xi,\sin\xi)$ from the optical axis. The weighting due to ~$W\e{LOS}(\chi)$ of inhomogeneities at $\chi$ in $\gamma\e{LOS}$ differs from the one of standard cosmic shear, which would only feature the first terms of each line in \cref{eq:W_LOS}; see \cref{fig:W_LOS} for illustration. We see that foreground perturbers have more impact than the background perturbers.

While ref.~\cite{Birrer:2017sge} focused on Einstein rings as \emph{standard circles}, another option would consist in exploiting quadruply imaged quasars as \emph{standard crosses}. Indeed, for a single point source, \cref{eq:minimal_lens_model_SIE} has 4 free parameters ($\tilde{\cplx{\beta}},\theta\e{E}, e-3\gamma_{od}, \gamma\e{LOS}$) which can thus be fully determined from the 4 image positions with respect to the centre of the lens. This idea is supported by the recent findings of ref.~\cite{2021arXiv210208470L}.\footnote{Ref.~\cite{2021arXiv210208470L} showed that the positions of four images, combined with the position of the main lens, allows one to distinguish between the ellipticity of the latter and an external shear in the main lens plane. In our language, that set-up corresponds to $\gamma_{od}=\gamma_{ds}=0$ and $\gamma_{os}\not= 0$, so that $\gamma\e{LOS}=\gamma_{os}$.} The resulting measurement of $\gamma\e{LOS}$ would presumably be less precise than the one obtained from a full Einstein ring, but this precision loss may be counter-balanced by greater statistics.

Let us finally point out possible caveats. Our illustration of the non-degeneracy of $\gamma\e{LOS}$ and the azimuthal structure of the dominant lens was based on a simplistic model (SIE). It remains to be proved that its conclusions hold for more general models. For example, it is crucial to evaluate the influence of the main lens's satellites, and the offset between baryonic matter and dark matter within the lens~\cite{Gomer:2021gio}. Neglecting such details might lead one to bias, over-constrain, or mis-interpret the shear, similarly to how one may over-constrain $H_0$ by relying on simplistic lens models in time-delay cosmography~\cite{Kochanek:2019ruu,Kochanek:2020crs}. We shall address this issue in future work.

\subsection{Beyond external convergence and shear: flexion}
\label{subsec:flexion}

Flexion is a type of lensing distortions inducing skewed and arc-like images~\cite{2005ApJ...619..741G,2007ApJ...660..995O}; it is provoked by the inhomogeneity of the convergence and shear fields beyond the tidal regime. The weak correlated flexion induced by cosmological structures is expected to contain much information about non-linear cosmic scales~\cite{2006MNRAS.365..414B,Fleury:2018odh}. However, its signal is notoriously difficult to extract from galaxy surveys, due to their large shape noise. Since strong-lensing images, such as Einstein rings, may be good standard shapes to measure the line-of-sight shear, they may also allow one to measure the line-of-sight flexion.

\subsubsection{Defining flexion}

While the convergence and shear are due to tidal fields, i.e., second derivatives of the projected gravitational potential, flexion is associated with its third derivatives. Expanding the potential of the $l\h{th}$ lens in the vicinity of the fiducial ray yields
\begin{equation}
\hat{\psi}_l(\vect{x}_l)
= \hat{\psi}_l(\vect{0})
    + \underbrace{x_l^a \hat{\psi}_{l,a}}_{\text{deflection}}
    + \underbrace{\frac{1}{2!} \, x_l^a x_l^b \, \hat{\psi}_{l,ab}}_{\text{tidal distortions}}
    + \underbrace{\frac{1}{3!} \, x_l^a x_l^b x_l^c \, \hat{\psi}_{l,abc}}_{\text{flexion}}
    + \order(|\vect{x}_l|^4) \ .
\end{equation}
In other words, flexion is due to gradients of tidal fields. As noted in ref.~\cite{2006MNRAS.365..414B}, flexion is conveniently described in terms of complex numbers. Considered as a function of $\cplx{x}_l$ and $\cplx{x}_l^*$, $\hat{\psi}_l$ has only two independent third complex derivatives, which we denote with
\begin{align}
\label{eq:potential_dipole_hexapole}
\hat{\F}_l \define 2\,\pd{}{\cplx{x}_l} \pd{}{\cplx{x}_l^*} \pd{}{\cplx{x}_l^*} \, \hat{\psi}_l \ ,
\qquad
\hat{\G}_l \define 2\,\pd{}{\cplx{x}_l^*} \pd{}{\cplx{x}_l^*} \pd{}{\cplx{x}_l^*} \, \hat{\psi}_l \ .
\end{align}
We easily see that the other two combinations of $\partial/\partial \cplx{x}_l$ and $\partial/\partial\cplx{x}_l^*$ are merely complex conjugations of $\hat{\F}_l, \hat{\G}_l$. We may then define what is traditionally referred to as the type-$\F$ and type-$\G$ flexions in terms of derivatives of the convergence and shear,
\begin{align}
\label{eq:partial_F_flexion}
\F_{ilj}
&\define \pd{\gamma_{ilj}}{\cplx{\beta}_{il}}
= \pd{\kappa_{ilj}}{\cplx{\beta}_{il}^*}
= \frac{D_{il}^2 D_{lj}}{D_{ij}} \, \hat{\F}_l \ ,
\\
\label{eq:partial_G_flexion}
\G_{ilj}
&\define \pd{\gamma_{ilj}}{\cplx{\beta}_{il}^*}
= \frac{D_{il}^2 D_{lj}}{D_{ij}} \, \hat{\G}_l \ ,
\end{align}
where $i,l,j$ respectively play the role of observer, deflector, and source. \emph{Beware, our definitions differ from the standard ones~\cite{2006MNRAS.365..414B} by a factor $2$}, $\F\e{standard}=2\F\e{us}$ and $\G\e{standard}=2\G\e{us}$.

Note that, contrary to convergence and shear, flexion is not a scale-invariant quantity in the sense that $[\F]=[\G]=\text{angle}^{-1}$. For example, a point-like deflector with Einstein radius~$\eps$ is characterised by a displacement angle $\cplx{\alpha}(\cplx{\theta})=\eps^2/\cplx{\theta}^*$. Its flexion thus reads $\F=0$ and $\G=2\eps^2/(\cplx{\theta}^*)^3$. We see that flexion decreases very fast as one moves away from the deflector---at a distance of $10$ times the deflector's Einstein radius, one has $\G=2\times 10^{-3}\eps^{-1}$.

The tidal regime (\cref{subsec:tidal}) consisted in expanding all displacement angles at first order, thereby modelling convergence and shear as being homogeneous near the line of sight. The \emph{flexion regime} then consists in going one order further in the expansion, i.e., assuming that the flexion coefficients~$\F_{ilj}, \G_{ilj}$ are homogeneous:
\begin{equation}
\label{eq:partial_displacement_flexion}
\cplx{\alpha}_{ilj}(\cplx{\theta})
= \cplx{\alpha}_{ilj}(0)
    + \kappa_{ilj}(0) \, \cplx{\theta}
    + \gamma_{ilj}(0) \, \cplx{\theta}^*
    + \frac{1}{2} \pac{ \F_{ilj}^*\cplx{\theta}^2
                        + 2\F_{ilj}\cplx{\theta}\cplx{\theta}^* 
                        + \G_{ilj}(\cplx{\theta}^*)^2} \ .
\end{equation}
Following the discussion of \cref{subsubsec:comparison_literature}, we shall drop the homogeneous displacement $\cplx{\alpha}_{ilj}(0)$; besides, we shall omit the $(0)$ argument in $\kappa_{ilj}, \gamma_{ilj}$ for short. \Cref{eq:partial_displacement_flexion} then matches eq.~(7) of ref.~\cite{2008A&A...485..363S} modulo the conventional factor 2.

\begin{figure}[t]
    \centering
    \includegraphics[width=0.75\columnwidth]{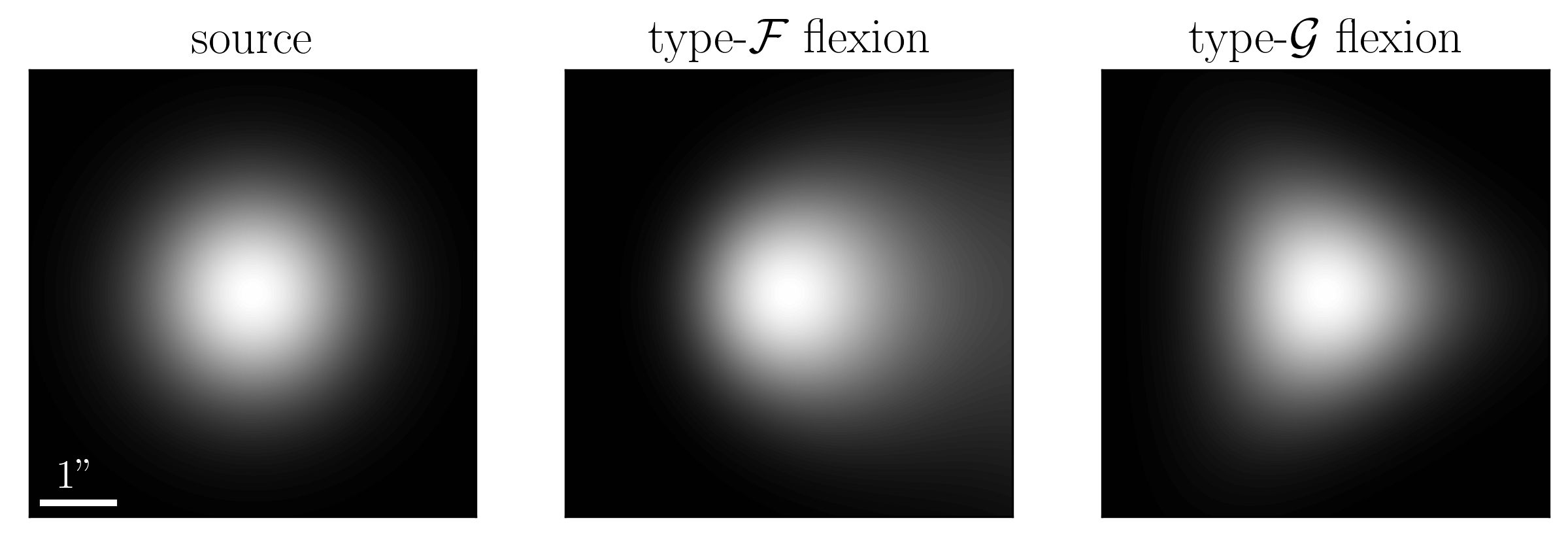}
    \caption{Illustrating the sole effects of a type-$\F$ and a type-$\G$ flexion in the absence of strong lensing. From left to right: Gaussian source with surface brightness $I\e{s}(\vect{\beta})=I_0\ex{-|\vect{\beta}|^2/(2\sigma^2)}$, $\sigma=1"$; image by a pure flexion plane with $\F=0.1\sigma^{-1}$; and image by a pure flexion plane with $\G=0.1\sigma^{-1}$.}
    \label{fig:pure_flexion}
\end{figure}

We illustrate the sole effect of the type-$\F$ and type-$\G$ flexions in \cref{fig:pure_flexion}. The figure shows the images of a Gaussian source by a pure-$\F$ and a pure-$\G$ flexion mode. Precisely, the images are obtained by a single ``flexion plane'' $f$ with $\kappa_{ofs}=\gamma_{ofs}=0$ and either $\F_{ofs}=0.1\U{arcsec}^{-1}$ or $\G_{ofs}=0.1\U{arcsec}^{-1}$. The type-$\F$ flexion is seen to induce a skewness in the brightness profile of the image, while the type-$\G$ flexion induces a triangular, or arc-like shape.

\subsubsection{Dominant lens equation with flexion}

We now turn to the effect of flexion as a line-of-sight correction to a dominant lens. Inserting \cref{eq:partial_displacement_flexion} in the general expression~\eqref{eq:displacement_dominant} of the lensing displacement, and reorganising the various terms leads to 
\begin{align}
\label{eq:displacement_flexion}
\cplx{\alpha}(\cplx{\theta})
&= \cplx{\alpha}_{ods}\paac{(1-\kappa_{od})\cplx{\theta}-\gamma_{od}\cplx{\theta}^*
                            - \frac{1}{2} \pac{ \F_{od}^*\cplx{\theta}^2
                                                + 2\F_{od}\cplx{\theta}\cplx{\theta}^* 
                                                + \G_{od}(\cplx{\theta}^*)^2}} \nonumber\\
&\quad + \kappa_{os}\cplx{\theta}
        + \gamma_{os}\cplx{\theta}^*
        + \frac{1}{2}
            \pac{
                \F_{os}^*\cplx{\theta}^2
                + 2\F_{os}\cplx{\theta}\cplx{\theta}^* 
                + \G_{os}(\cplx{\theta}^*)^2
                } \nonumber\\
&\quad - \pac{
                \kappa_{ds}
                + {}^{(2)}{\F}_{ds}^*\cplx{\theta}
                + {}^{(2)}{\F}_{ds}\cplx{\theta}^*
                } \cplx{\alpha}_{ods}(\cplx{\theta})
        - \pac{
                \gamma_{ds}
                + {}^{(2)}{\F}_{ds}\cplx{\theta}
                + {}^{(2)}{\G}_{ds}\cplx{\theta}^*
                } \cplx{\alpha}_{ods}^*(\cplx{\theta}) \nonumber\\
&\quad  + \frac{1}{2} \paac{
                {}^{(1)}\F_{ds}^*\cplx{\alpha}_{ods}^2(\cplx{\theta})
                + 2 \, {}^{(1)}\F_{ds}\cplx{\alpha}_{ods}^*(\cplx{\theta})\cplx{\alpha}_{ods}(\cplx{\theta})
                + {}^{(1)}\G_{ds}[\cplx{\alpha}_{ods}^*(\cplx{\theta})]^2
                } \ ,
\end{align}
where we omitted the homogeneous displacements $\cplx{\alpha}_{os}(0), \cplx{\alpha}_{od}(0)$ for simplicity. Compared to its counterpart in the tidal regime, the flexion regime features 8 new complex parameters: $\F_{od}, \F_{os}, {}^{(1)}\F_{ds}, {}^{(2)}\F_{ds}, \G_{od}, \G_{os}, {}^{(1)}\G_{ds}, {}^{(2)}\G_{ds}$. Their explicit expressions are
\begin{align}
\label{eq:F_od}
\F_{od} 
&\define \sum_{l<d} \F_{old}
= \sum_{l<d} \frac{D_{ol}^2 D_{ld}}{D_{od}} \, \hat{\F}_{l} \ ,
\\
\label{eq:F_os}
\F_{os} 
&\define \sum_{l\neq d} \F_{ols}
= \sum_{l\neq d} \frac{D_{ol}^2 D_{ls}}{D_{os}} \, \hat{\F}_{l} \ ,
\\
\label{eq:F_ds_1}
{}^{(1)}\F_{ds} 
&\define \sum_{l>d} \frac{D_{os}}{D_{ds}} \, \F_{dls}
= \sum_{l>d} \frac{D_{dl}^2 D_{ls} D_{os}}{D_{ds}^2} \, \hat{\F}_{l} \ ,
\\
\label{eq:F_ds_2}
{}^{(2)}\F_{ds} 
&\define \sum_{l>d} \frac{D_{ol}}{D_{dl}} \, \F_{dls}
= \sum_{l>d} \frac{D_{ol} D_{dl} D_{ls}}{D_{ds}} \, \hat{\F}_{l} \ ,
\end{align}
and similarly for the $\G$s. The first two lines of \cref{eq:displacement_flexion} are easily understood; the argument of $\cplx{\alpha}_{ods}$ is simply $\cplx{\theta}-\cplx{\alpha}_{od}(\cplx{\theta})$, while the second line is $\cplx{\alpha}_{os}(\cplx{\theta})$ in the presence of flexion. The interpretation of the third and fourth lines is less straightforward. The terms associated with ${}^{(1)}\F_{ds}, {}^{(1)}\G_{ds}$ may be seen as the flexion contribution to the physical separation, in the source plane, of two rays separated by the deflection angle~$\hat{\vect{\alpha}}_d$ in the main lens plane. The terms associated with ${}^{(2)}\F_{ds}, {}^{(2)}\G_{ds}$ are post-Born corrections to the background convergence and shear when evaluated in the direction $\vect{\theta}$ rather than $\vect{0}$.

\begin{figure}
    \centering
    \includegraphics[width=0.49\columnwidth]{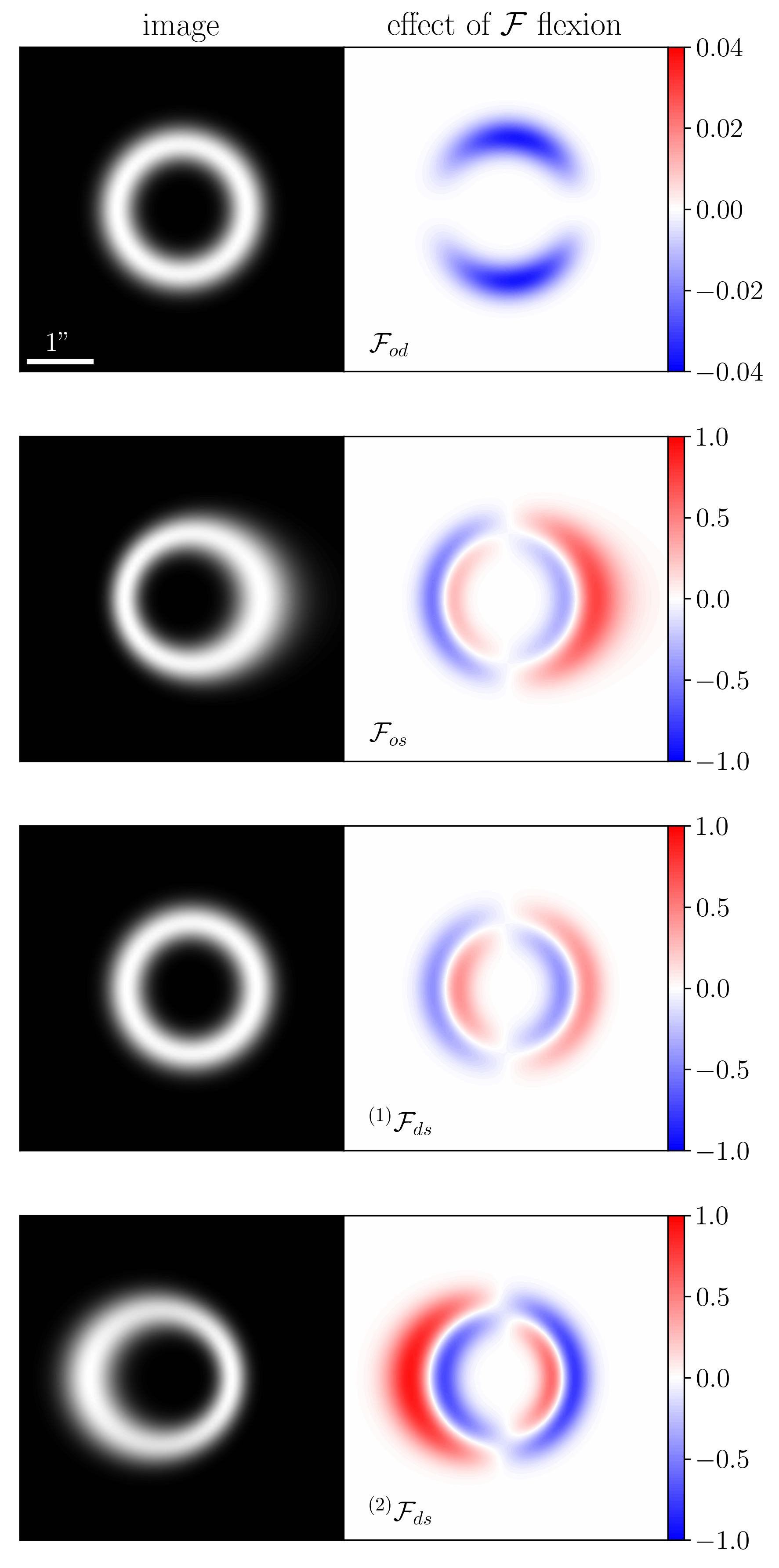}
    \hfill
    \includegraphics[width=0.49\columnwidth]{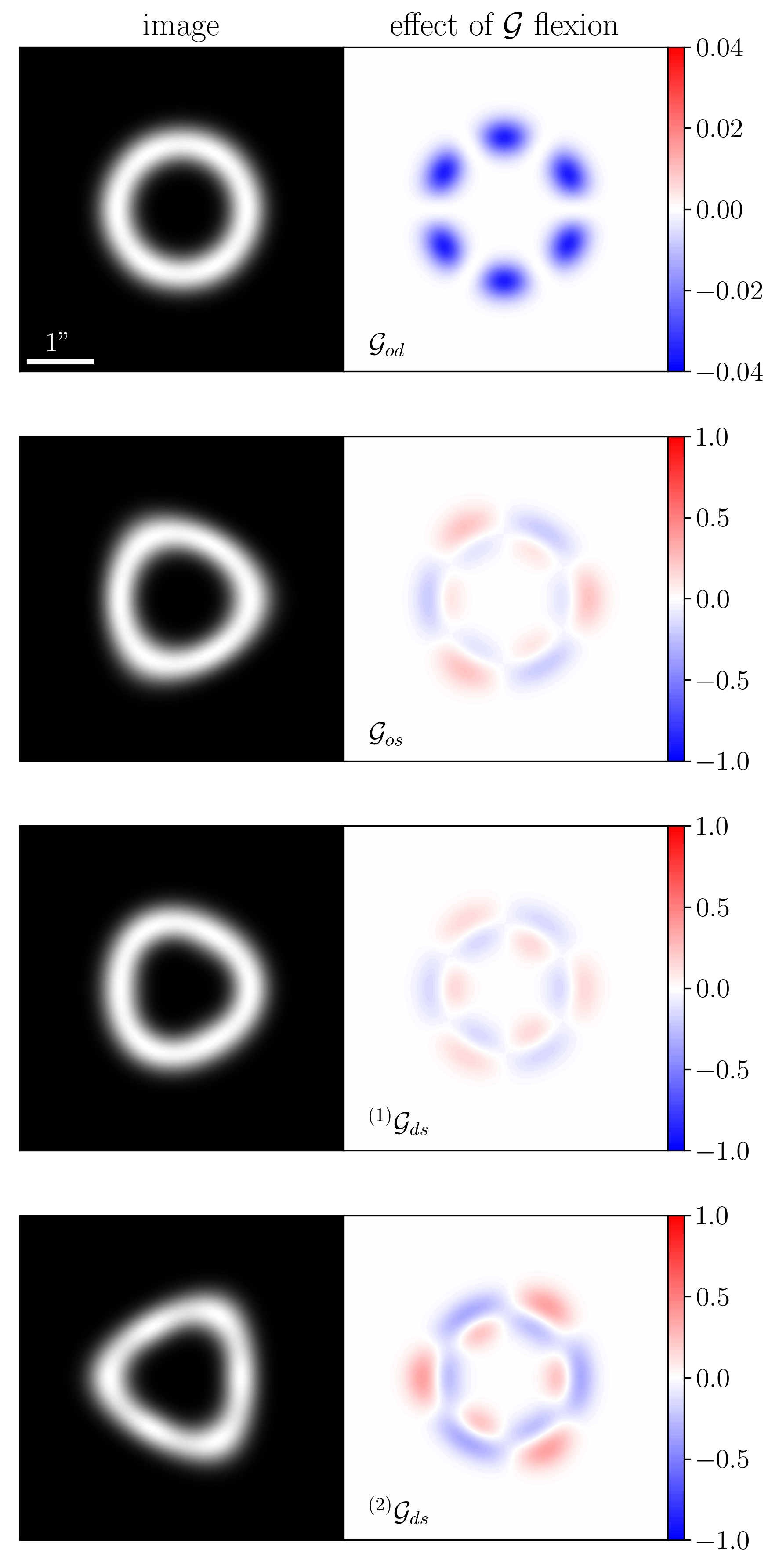}
    \caption{Individual effect of each flexion mode in \cref{eq:displacement_flexion} on an Einstein ring. The source is Gaussian, $I\e{s}(\vect{\beta})=I_0\ex{-|\vect{\beta}|^2/(2\sigma^2)}$ with $\sigma=0.2"$; it is aligned with the main lens taken to be a SIS with Einstein radius $\theta\e{E}=1"$. In each panel, the left part shows the image~$I(\vect{\theta})$ in the presence of a line-of-sight flexion, while the right part shows its difference $[I(\vect{\theta})-I\e{SIS}(\vect{\theta})]/I_0$ with the flexion-free case (SIS only). The amplitude of each individual flexion parameter is set to $0.1\theta\e{E}^{-1}$.}
    \label{fig:flexion_Einstein_ring}
\end{figure}

We illustrate the effect of each individual flexion parameter on a circular Einstein ring in \cref{fig:flexion_Einstein_ring}. We can see that the general interpretation of type-$\F$ and type-$\G$ flexion mostly holds---the former tends to skew the rings, while the latter gives them a triangular shape. An exception is the pure-foreground case, $\F_{od}, \G_{od}$, whose effect is much smaller than the others, and leads to spin-2 and spin-6 corrections instead of spin-1 and spin-3.

\subsubsection{Flexion degeneracies}

\paragraph{External degeneracy} Just like convergence and shear, flexion is subject to an external parameter degeneracy if one has no information about the intrinsic shape of the source. Let us perform the following transformation of the source position
\begin{equation}
\tilde{\cplx{\beta}} \define \cplx{\beta}
- \frac{1}{2} \pac{ \F\e{ext}^*\cplx{\beta}^2
                        + 2\F\e{ext}\cplx{\beta}\cplx{\beta}^* 
                        + \G\e{ext}(\cplx{\beta}^*)^2} ,
\end{equation}
where $\F\e{ext}, \G\e{ext}$ are arbitrary (small) complex parameters. We find that, if $\cplx{\theta}$ is a solution of the lens equation $\cplx{\beta}=\cplx{\theta}-\cplx{\alpha}(\cplx{\theta})$, then it is also a solution of $\tilde{\cplx{\beta}}=\cplx{\theta}-\tilde{\cplx{\alpha}}(\cplx{\theta})$, where $\tilde{\cplx{\alpha}}$ is formally identical to \cref{eq:displacement_flexion}, but with modified flexion parameters
\begin{equation}
\tilde{\F}_{os} = \F_{os} + \F\e{ext} \ , \qquad
{}^{(1)}\tilde{\F}_{ds} = {}^{(1)}\F_{ds} + \F\e{ext} \ , \qquad
{}^{(2)}\tilde{\F}_{ds} = {}^{(2)}\F_{ds} + \F\e{ext} \ ,
\end{equation}
and similarly for the type-$\G$ flexions. Like for convergence and shear, the pure-foreground terms~$\F_{od}, \G_{od}$ are left unchanged. Therefore, the image of a flexed source with no line-of-sight flexion is identical to a non-flexed source plus flexion on the line of sight with parameters $\F_{os}={}^{(1)}\F_{ds}={}^{(2)}\F_{ds}$ (and similarly for $\G$).

\paragraph{Internal degeneracy} Due to its peculiar geometry, we do not expect flexion to be degenerate with the internal parameters of simple elliptical lens models. However, they may exhibit degeneracies with more realistic models. For instance, we intuitively expect type-$\F$ perturbations to appear if, e.g., the centre of mass of the main lens's baryonic matter does not coincide with the centre of mass of its dark-matter halo.\footnote{Such a mismatch would be even more pronounced in alternative theories of gravity with screening mechanisms, which may effectively break the weak equivalence principle~\cite{Hui:2009kc}. In such scenarios, the stars of a galaxy (screened) and the gas or dark-matter halo (unscreened) would not experience the same external gravity fields, and hence would fall differently~\cite{Desmond:2018euk,Desmond:2018sdy}.} Type-$\G$ perturbations may be mimicked by massive satellites of the main lens. The precise investigation of such degeneracies, which conditions the feasibility of flexion measurements from Einstein rings, is beyond the scope of the present article, and is left for future work.

\section{Weak lensing with critical curves}
\label{sec:critical_curves}

In the previous section, we have shown how line-of-sight perturbations to strong lensing could be encapsulated in extra parameters (convergences, shears, flexions, \ldots) within lens models. The main advantage of that common approach is its conceptual simplicity, although, as seen with flexion, the number of extra parameters grows rapidly with the refinement of the line-of-sight model. Nevertheless, the parameterised approach is based on Taylor-expansions of the perturbers' displacement angles. By essence, this method is therefore not adapted to the description of small perturbers that are close to the line of sight. In the latter situation, the strong-lensing field must be treated as a \emph{finite beam}~\cite{Fleury:2017owg, Fleury:2018cro, Fleury:2018odh}.

In the present section, we propose another, complementary, approach solely based on strong-lensing \emph{critical curves}. For a given lens model~$\vect{\alpha}(\vect{\theta})$, the critical curve traces the loci~$\vect{\theta}$ where magnification is formally infinite, e.g. the theoretical Einstein ring for axially symmetric lenses. Albeit not directly observable, the critical curve may be reconstructed from a phenomenological lens model or inferred via machine learning~\cite{Hezaveh:2017sht, 2019ApJ...883...14M, 2019MNRAS.488..991P}. We shall demonstrate that, if measured, the distortions of such critical curves would constitute a powerful probe of weak lensing and galactic dark matter.

\subsection{Perturbed critical curves}

The DL formalism developed in \cref{sec:dominant_lens} is well suited to determining the impact of weak perturbers on strong-lensing critical curves to a high degree of generality. Let us first discuss the general set-up before proceeding with the special case of an axially symmetric main lens.

\subsubsection{General case}
\label{subsubsec:critical_curve_general}

A critical curve is defined as the set of points~$\vect{\theta}\e{cc}(\lambda)$ in the image plane where the magnification is formally infinite, $\mu^{-1}(\vect{\theta}\e{cc})=0$; here, $\lambda$ denotes an arbitrary parameter along the curve. The presence of weak perturbers affects the shape of critical curves in a non-trivial way. We may write~$\vect{\theta}\e{cc}=\bar{\vect{\theta}}\e{cc}+\delta\vect{\theta}\e{cc}$, where $\bar{\vect{\theta}}\e{cc}(\lambda)$ is the critical curve of the main lens alone, $\mu^{-1}_{ods}(\bar{\vect{\theta}}\e{cc})=0$.

Using the expression~\eqref{eq:magnification_dominant_lens} of $\mu^{-1}$ in the DL approximation, and expanding the equation $\mu^{-1}(\bar{\vect{\theta}}\e{cc}+\delta\vect{\theta}\e{cc})=0$ at first order in $\delta\vect{\theta}\e{cc}$ and $\eps^2$, we get 
\begin{equation}
\label{eq:cc_perturbation_general}
0 =
\pa{\delta\vect{\theta}\e{cc} - \vect{\alpha}_{od}}
\cdot \ddf{\mu_{ods}^{-1}}{\vect{\theta}}
+ 2(1-\kappa_{ods})(\kappa_{od}+\kappa_{ds}-\kappa_{os})
+ 2\Re\pac{\gamma_{ods}^*(\gamma_{od}+\gamma_{ds}-\gamma_{os})} .
\end{equation}
Recall that, contrary to \cref{sec:parametric}, the convergences and shears are not necessarily constant here (we are not in the tidal regime). In \cref{eq:cc_perturbation_general}, the main-lens and line-of-sight quantities are evaluated on the unperturbed ray~$\mathscr{R}(\lambda)$ associated with $\bar{\vect{\theta}}\e{cc}(\lambda)$, as depicted in \cref{fig:cc_perturbation_general}.

\newcommand{\bcc}{\bar{\vect{\theta}}\e{cc}(\lambda)}
\newcommand{\dcc}{\delta\vect{\theta}\e{cc}}
\begin{figure}[t]
    \centering
\begingroup%
  \makeatletter%
  \providecommand\color[2][]{%
    \errmessage{(Inkscape) Color is used for the text in Inkscape, but the package 'color.sty' is not loaded}%
    \renewcommand\color[2][]{}%
  }%
  \providecommand\transparent[1]{%
    \errmessage{(Inkscape) Transparency is used (non-zero) for the text in Inkscape, but the package 'transparent.sty' is not loaded}%
    \renewcommand\transparent[1]{}%
  }%
  \providecommand\rotatebox[2]{#2}%
  \newcommand*\fsize{\dimexpr\f@size pt\relax}%
  \newcommand*\lineheight[1]{\fontsize{\fsize}{#1\fsize}\selectfont}%
  \ifx\svgwidth\undefined%
    \setlength{\unitlength}{436.26958149bp}%
    \ifx\svgscale\undefined%
      \relax%
    \else%
      \setlength{\unitlength}{\unitlength * \real{\svgscale}}%
    \fi%
  \else%
    \setlength{\unitlength}{\svgwidth}%
  \fi%
  \global\let\svgwidth\undefined%
  \global\let\svgscale\undefined%
  \makeatother%
  \begin{picture}(1,0.33573721)%
    \lineheight{1}%
    \setlength\tabcolsep{0pt}%
    \put(0,0){\includegraphics[width=\unitlength,page=1]{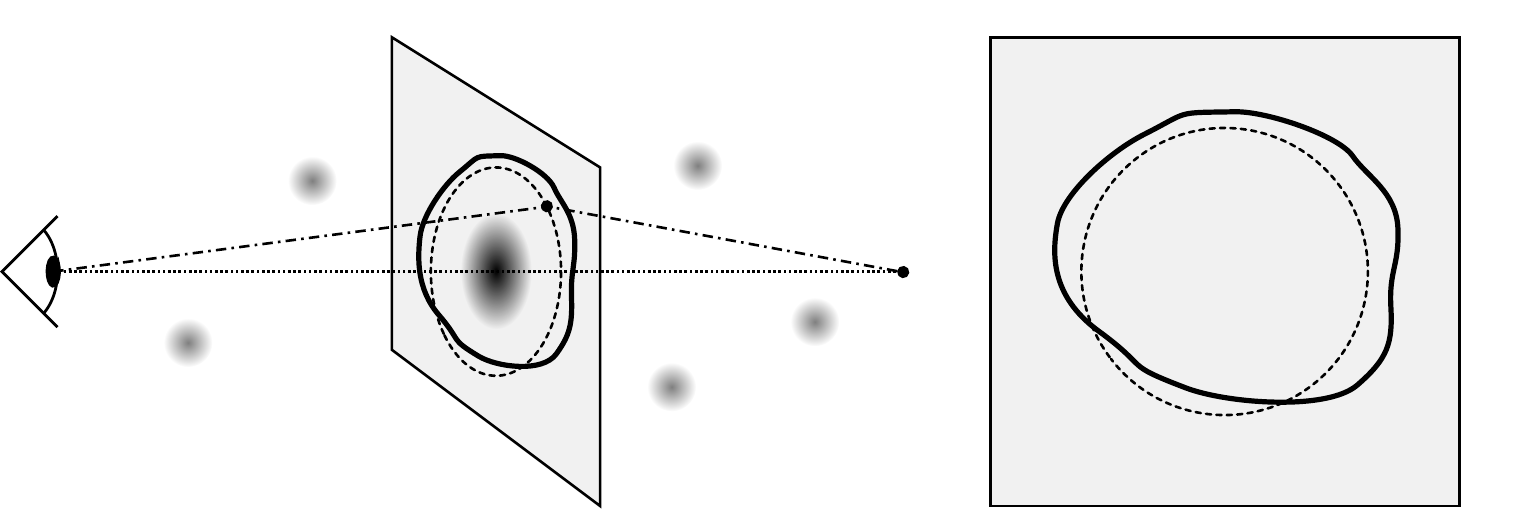}}%
    \put(0.10360712,0.17825258){\rotatebox{7.2456394}{\makebox(0,0)[lt]{\lineheight{1.25}\smash{\begin{tabular}[t]{l}$\mathscr{R}(\lambda)$\end{tabular}}}}}%
    \put(0.26550798,0.31877016){\rotatebox{-31.938726}{\makebox(0,0)[lt]{\lineheight{1.25}\smash{\begin{tabular}[t]{l}dominant lens\end{tabular}}}}}%
    \put(0,0){\includegraphics[width=\unitlength,page=2]{critical_curves.pdf}}%
    \put(0.80732812,0.17499309){\rotatebox{32.479241}{\makebox(0,0)[lt]{\lineheight{1.25}\smash{\begin{tabular}[t]{l}$\bcc$\end{tabular}}}}}%
    \put(0.90856435,0.22643883){\makebox(0,0)[lt]{\lineheight{1.25}\smash{\begin{tabular}[t]{l}$\dcc$\end{tabular}}}}%
    \put(0,0){\includegraphics[width=\unitlength,page=3]{critical_curves.pdf}}%
  \end{picture}%
\endgroup%

    \caption{Distortion of a strong-lensing critical curve by line-of-sight perturbers. In the DL approximation, the perturbation $\delta\vect{\theta}\e{cc}(\lambda)$ of the critical curve at $\lambda$ is evaluated along the unperturbed critical ray~$\mathscr{R}(\lambda)$.}
    \label{fig:cc_perturbation_general}
\end{figure}

We note that \cref{eq:cc_perturbation_general} is not an explicit expression for $\delta\vect{\theta}\e{cc}$. This is due to the invariance of the critical curve under re-parameterisations; in other words, one must specify how the unperturbed and perturbed critical curves are mapped for $\delta\vect{\theta}\e{cc}$ to be defined unambiguously. A natural choice consists in imposing that $\delta\vect{\theta}\e{cc}$ is orthogonal to the unperturbed critical curve, $\delta\vect{\theta}\e{cc}=\delta\theta\e{cc}\,\vect{n}$, where $\vect{n}\propto \dd\mu_{ods}^{-1}/\dd\vect{\theta}$ is the unit outgoing normal to $\bar{\vect{\theta}}\e{cc}(\lambda)$. With this choice, \cref{eq:cc_perturbation_general} becomes
\begin{equation}
\label{eq:cc_perturbation_normal}
\delta\theta\e{cc}
=
\underbrace{
\vect{n}\cdot\vect{\alpha}_{od}
}_{\text{direct pert.}}
+
\underbrace{
2\abs{
    \ddf{\mu_{ods}^{-1}}{\vect{\theta}}
    }^{-1}
    \paac{
        (1-\kappa_{ods})(\kappa_{os}-\kappa_{ds}-\kappa_{od})
        + \Re\pac{\gamma_{ods}^*(\gamma_{os}-\gamma_{ds}-\gamma_{od})}
        }
}_{\text{perturbation from lens-lens coupling}}
.
\end{equation}
The first term, $\vect{n}\cdot\vect{\alpha}_{od}$, of \cref{eq:cc_perturbation_normal} is easily interpreted---it tells us how the foreground lenses would distort the critical curve if the critical curve were a source located in the $d\h{th}$ plane. The other terms are much less intuitive; they encode the effects of non-linear lens-lens coupling, which precisely make a critical curve more complicated than a mere source of light in the main-lens plane.

\subsubsection{Axially symmetric dominant lens}

From now on, we shall assume that the dominant lens is symmetric about the optical axis, $\kappa_{ods}(\vect{\theta})=\kappa_{ods}(\theta)$. This simplifying assumption is justified by the fact that we will be ultimately interested in two-point correlations of critical curve distortions. We thus expect the intrinsic departures from the axial symmetry to cancel statistically, due to cosmic homogeneity and isotropy. From that perspective, the results derived here remain general as long as the main lens is not too asymmetric. 

The critical curve of an axially symmetric lens is a circle, whose angular radius will be denoted $\bar{\theta}\e{cc}=\theta\e{E}$ (the Einstein radius); its out-going normal~$\vect{n}$ is the radial vector. Due to the symmetry of the problem, it is then natural to parameterise the critical curves with the polar angle, $\lambda=\ph$.

For an axially symmetric dominant lens, the convergence on the unperturbed critical curve is a constant, which we denote $\kappa\e{E}\define \kappa_{ods}(\theta\e{E})$.\footnote{For a point-like dominant lens, $\kappa\e{E}=0$. For a SIS, $\kappa_{ods}(\theta)=\frac{1}{2}\frac{\theta\e{E}}{\theta}$ and hence $\kappa\e{E}=\frac{1}{2}$. For a more general power-law profile of the form $\kappa_{ods}(\theta)=\frac{3-\nu}{2}\pa{\frac{\theta\e{E}}{\theta}}^{\nu-1}$, it reads $\kappa\e{E}=\frac{3-\nu}{2}$.} Besides, the shear is easily expressed as a function of the convergence field as (see e.g. \S~8.1.1 of ref.~\cite{1992grle.book.....S})
\begin{equation}
\gamma_{ods}(\vect{\theta})
= \pac{ \kappa_{ods}(\theta)
        - \frac{2}{\theta^2}
            \int_0^{\theta} \dd\theta'\;\theta'\kappa_{ods}(\theta')
        } \ex{2\ii\ph}
= - |\gamma_{ods}(\theta)| \, \ex{2\ii\ph} \ ,
\end{equation}
where in the second equality we have made the natural assumption that the projected density of the main lens decreases with the radius. Let $\gamma\e{E}\define|\gamma_{ods}(\theta\e{E})|$; since the amplification is infinite on the critical curve, we have $0=\mu_{ods}^{-1}(\theta\e{E})=(1-\kappa\e{E})^2-\gamma\e{E}^2$, and hence $\gamma\e{E}=\pm(1-\kappa\e{E})$. Here we shall only consider \emph{tangential} critical curves~\cite{1992grle.book.....S} for which $\gamma\e{E}=1-\kappa\e{E}$. In that case, it is also straightforward to show that
\begin{equation}
\abs{
    \ddf{\mu_{ods}^{-1}}{\vect{\theta}}
    }_{\theta\e{E}}
= (1-\kappa\e{E}+|\gamma\e{E}|)
    \abs{
        \ddf{}{\vect{\theta}}(1-\kappa_{ods}-|\gamma_{ods}|)
        }_{\theta\e{E}}
= \frac{4}{\theta\e{E}} \, (1-\kappa\e{E})^2 \ .
\end{equation}

\begin{figure}[h!]
\centering
\begin{minipage}{0.4\textwidth}
\begingroup%
  \makeatletter%
  \providecommand\color[2][]{%
    \errmessage{(Inkscape) Color is used for the text in Inkscape, but the package 'color.sty' is not loaded}%
    \renewcommand\color[2][]{}%
  }%
  \providecommand\transparent[1]{%
    \errmessage{(Inkscape) Transparency is used (non-zero) for the text in Inkscape, but the package 'transparent.sty' is not loaded}%
    \renewcommand\transparent[1]{}%
  }%
  \providecommand\rotatebox[2]{#2}%
  \newcommand*\fsize{\dimexpr\f@size pt\relax}%
  \newcommand*\lineheight[1]{\fontsize{\fsize}{#1\fsize}\selectfont}%
  \ifx\svgwidth\undefined%
    \setlength{\unitlength}{189.85998277bp}%
    \ifx\svgscale\undefined%
      \relax%
    \else%
      \setlength{\unitlength}{\unitlength * \real{\svgscale}}%
    \fi%
  \else%
    \setlength{\unitlength}{\svgwidth}%
  \fi%
  \global\let\svgwidth\undefined%
  \global\let\svgscale\undefined%
  \makeatother%
  \begin{picture}(1,0.39927568)%
    \lineheight{1}%
    \setlength\tabcolsep{0pt}%
    \put(0,0){\includegraphics[width=\unitlength,page=1]{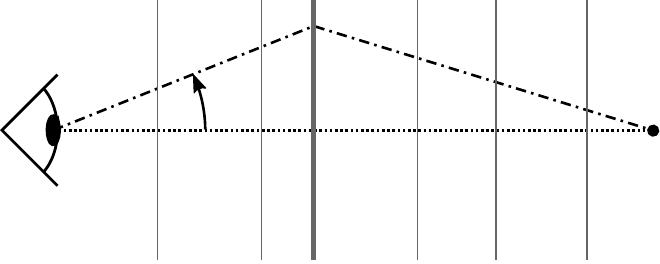}}%
    \put(0.30150819,0.13862936){\makebox(0,0)[lt]{\lineheight{1.25}\smash{\begin{tabular}[t]{l}$\theta\e{E}$\end{tabular}}}}%
    \put(0,0){\includegraphics[width=\unitlength,page=2]{critical_beam.pdf}}%
    \put(0.64327003,0.01114101){\color[rgb]{0.4,0.4,0.4}\makebox(0,0)[lt]{\lineheight{1.25}\smash{\begin{tabular}[t]{l}$l$\end{tabular}}}}%
    \put(0.55491728,0.24150582){\makebox(0,0)[lt]{\lineheight{1.25}\smash{\begin{tabular}[t]{l}$R_l$\end{tabular}}}}%
    \put(0,0){\includegraphics[width=\unitlength,page=3]{critical_beam.pdf}}%
  \end{picture}%
\endgroup%

\end{minipage}
\hfill
\begin{minipage}{0.49\textwidth}
\caption{Double-cone geometry of the critical beam. The physical radius $R_l$ at $l$ increases until $l=d$ and then decreases.}
\label{fig:radius_critical_beam}
\end{minipage}
\end{figure}

The above properties drastically simplify \cref{eq:cc_perturbation_normal} which becomes
\begin{empheq}[box=\fbox]{multline}
\label{eq:cc_perturbation_axisymmetric}
\delta\theta\e{cc}(\ph)
= \Re\pac{\ex{-\ii\ph}\cplx{\alpha}_{od}(\ph)}
    + \frac{1}{2}\frac{\theta\e{E}}{1-\kappa\e{E}} \, \Re
    \Big\{ \kappa_{os}(\ph)-\kappa_{ds}(\ph)-\kappa_{od}(\ph)\\
        - \ex{-2\ii\ph}\pac{\gamma_{os}(\ph)-\gamma_{ds}(\ph)-\gamma_{od}(\ph)}
        \Big\} ,
\end{empheq}
where we have used the complex notation for $\vect{\alpha}_{od}$. Note that $\delta\theta\e{cc}(\ph)$ now represents the radial perturbation to the critical curve, whose polar equation reads $\theta\e{cc}(\ph)=\theta\e{E}+\delta\theta\e{cc}(\ph)$. At this point, it may be useful to write down the explicit expressions for the line-of-sight terms evaluated along the unperturbed critical rays,
\begin{gather}
\label{eq:alpha_od_cc}
\cplx{\alpha}_{od}(\ph)
= \sum_{l<d} \frac{D_{ld}}{D_{od}} \, \hat{\cplx{\alpha}}_l(R_l\ex{\ii\ph}) \ ,
\\
\label{eq:kappa_gamma_cc}
\kappa_{ij}(\ph)
= \sum_{\substack{i<l<j\\l\neq d}}
    \frac{\Sigma_l(R_l\ex{\ii\ph})}{\Sigma\h{crit}_{ilj}} \ ,
\qquad
\gamma_{ij}(\ph)
= \sum_{\substack{i<l<j\\l\neq d}}
    \frac{Q_l(R_l\ex{\ii\ph})}{\Sigma\h{crit}_{ilj}} \ ,
\end{gather}
where $R_l$ denotes the physical radius of the unperturbed critical beam at $l$ (\cref{fig:radius_critical_beam}),
\begin{equation}
R_l \define
\theta\e{E}
\times
\begin{cases}
D_{ol} & l\leq d \ ,
\\[2mm]
\displaystyle
D_{ol} - \frac{D_{dl}D_{os}}{D_{ds}}
=
\frac{1+z_d}{1+z_l}\frac{D_{od}D_{ls}}{D_{ds}} & l\geq d \ .
\end{cases}
\end{equation}

\bigskip

\begin{figure}[t]
    \centering
    \includegraphics[width=\columnwidth]{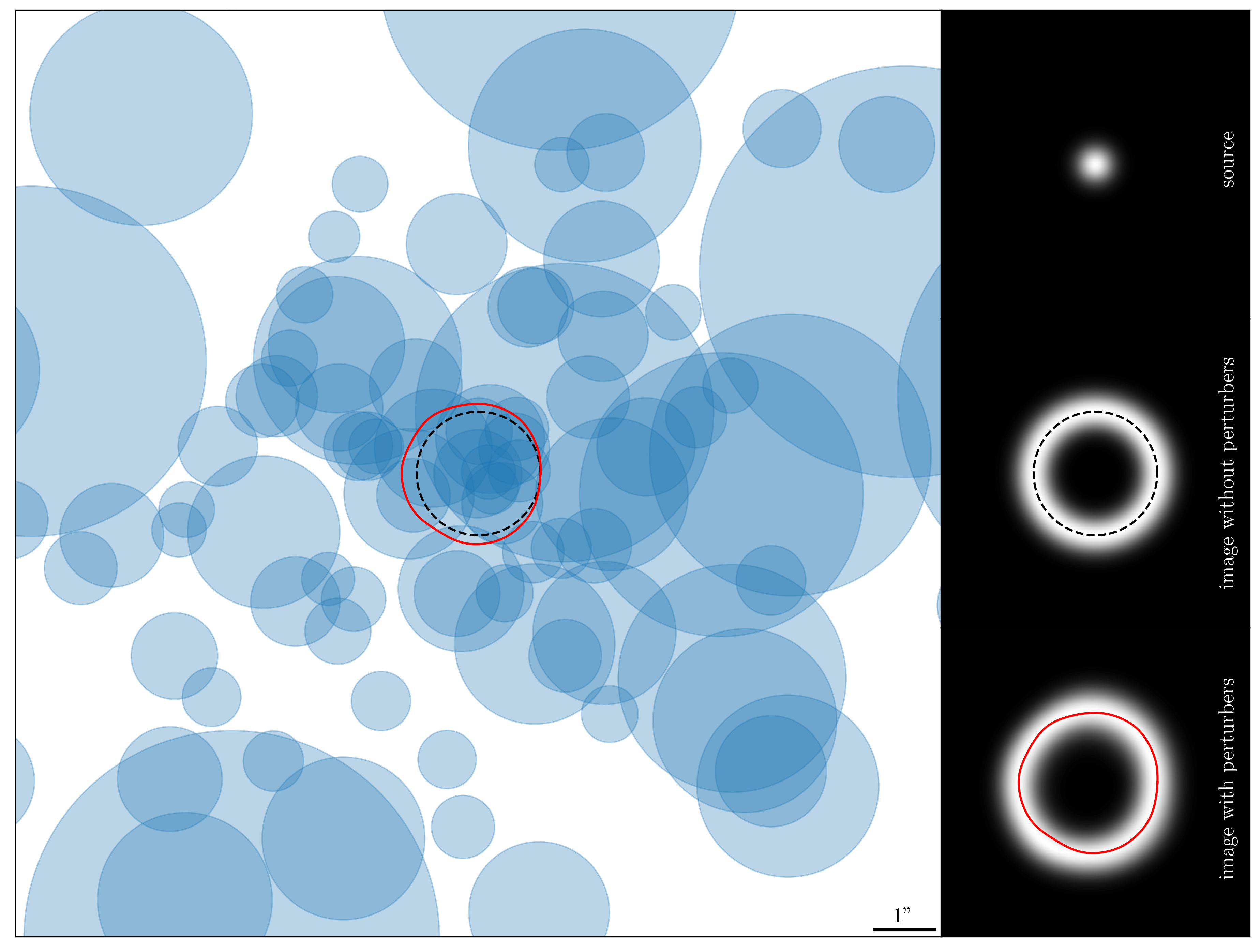}
    \caption{Distortions of images and critical curves for a SIS dominant lens ($\theta\e{E}=1", \kappa\e{E}=1/2$) with 100 line-of-sight perturbers. The perturbers are identical cored haloes with density profile $\Sigma(x)=\Sigma_0[1+(x/r\e{c})^4]^{-3/2}$, with $\Sigma_0=\Sigma\h{crit}_{ods}/10$ while $r\e{c}$ is the core radius. The total mass in the haloes is equal to the mass enclosed in the dominant-lens Einstein radius, which eventually fixes $r\e{c}$ here. The haloes are randomly placed within a tube of length $D_{os}$ and radius $5\theta\e{E}D_{os}$. Blue disks in the left panel indicate the apparent position and apparent core size of the perturbers in a $15"\times 15"$ field. The unperturbed critical curve is indicated by a black dashed line and the perturbed critical curve by a red solid line. The right panel shows the perturbed and unperturbed images of a Gaussian source with $\sigma=0.2"$ aligned with the main lens.}
    \label{fig:discrete_perturbers}
\end{figure}

The result of \cref{eq:cc_perturbation_axisymmetric} is exemplified in \cref{fig:discrete_perturbers}, where a SIS is perturbed by $100$ identical haloes randomly distributed along the line of sight. These perturbers lead to non-trivial distortions of the critical curve which could not be described with the usual convergence, shear, or even flexion. This illustrates the relevance of the general DL treatment introduced in this article when perturbers are lying near the line of sight.

\subsubsection{Critical modes: Fourier modes of the critical curve}
\label{subsubsec:critical_modes_def}

Because the perturbed critical curve is nearly circular, we may describe its distortions using a Fourier decomposition of $\delta\theta\e{cc}(\ph)$. Namely, we shall call $n\h{th}$ \emph{critical mode} the complex number
\begin{equation}
\label{eq:critical_modes}
c_n
\define
\frac{1}{2\pi\theta\e{E}} \int_0^{2\pi} \dd\ph \; \ex{\ii n\ph} \, \delta\theta\e{cc}(\ph)
=
\frac{1}{2\ii\pi\theta\e{E}} \int_{\unitcircle} \dd u \; u^{n-1} \; \delta\theta\e{cc}(u) \ ,
\end{equation}
for any integer $n$. In the second equality, we turned the angular integral into a complex integral along the unit circle~$\unitcircle$, with $u \define \ex{\ii\ph}$. This complex formulation will turn out to be convenient for practical computations. Note that, since $\delta\theta\e{cc}(\ph)\in\mathbb{R}$, we have $c_{-n}=c_n^*$, so
\begin{equation}
\delta\theta\e{cc}(\ph)
= \theta\e{E}\sum_{n\in\mathbb{Z}} c_n \ex{-\ii n\ph}
= \theta\e{E} \pac{c_0 + \sum_{n\geq 1} 2|c_n| \cos n(\ph-\ph_n)}
\end{equation}
with $c_n=|c_n|\ex{\ii n\ph_n}$. Thus, in the following we shall focus on $n\geq 0$.

\begin{figure}[t]
    \centering
    \includegraphics[width=\columnwidth]{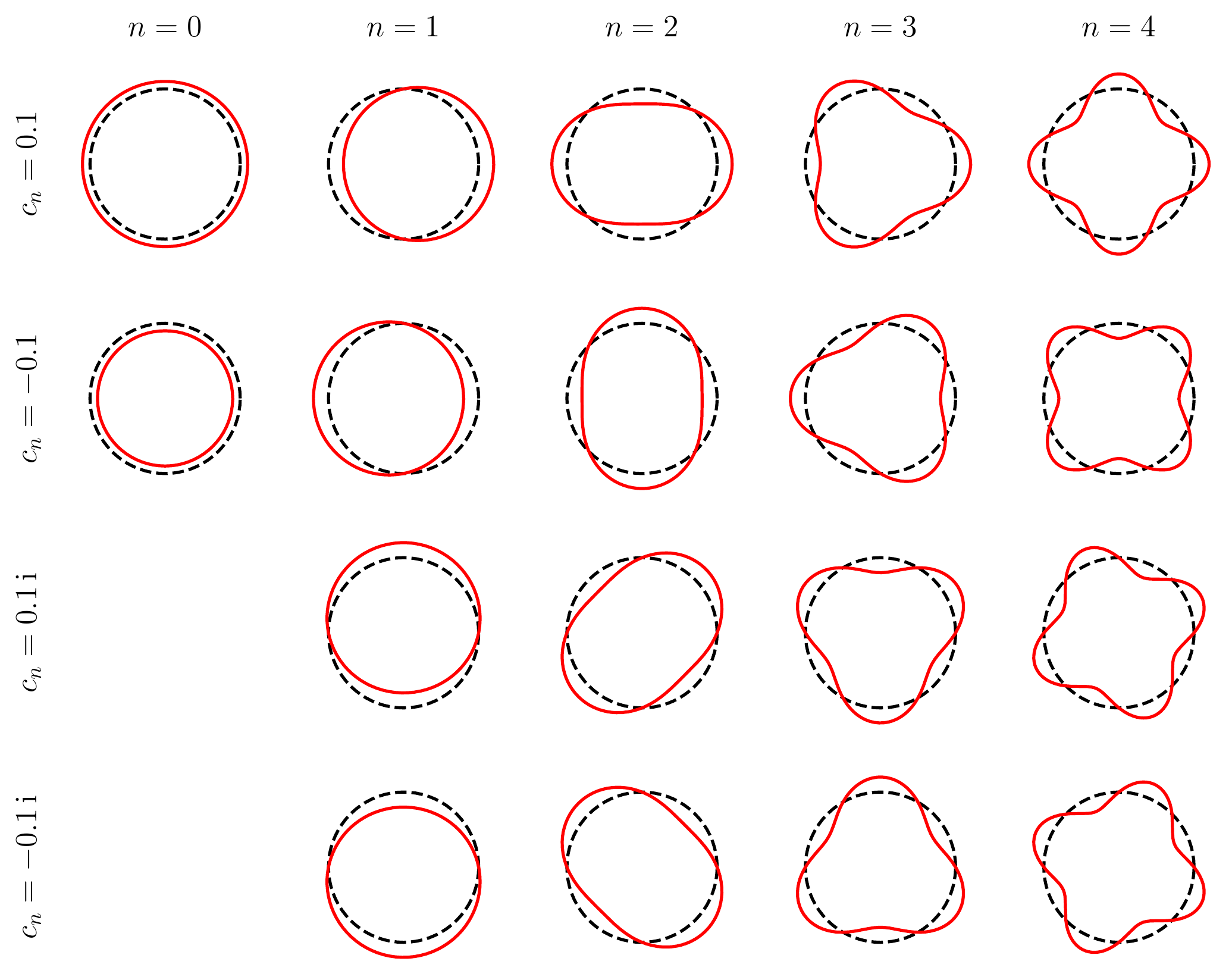}
    \caption{Illustration of the distortions associated with each of the first five critical modes~$c_{n\leq 4}$, depending on the sign of their real or imaginary parts.}
    \label{fig:illustration_modes}
\end{figure}

By definition, $c_0$ represents the increase of the mean radius of the critical curve, relative to $\theta\e{E}$; it thus encodes a notion of convergence. Similarly, $c_1$ indicates the global displacement of the critical curve, $c_2$ its ellipticity (shear measure), $c_3$ its triangularity (type-$\G$ flexion measure), etc. The effect of the first five individual modes is illustrated in \cref{fig:illustration_modes}. The critical modes will be our key observables in the remainder of this section.

Note that $c_1$ represents the global displacement of the critical curve with respect to the unlensed origin of the $d\h{th}$ plane, which is traditionally associated with the centre of the main lens. This displacement is not fully observable because the apparent position of the main lens is also affected by the foreground perturbers. We may account for this by shifting the origin of the image plane by $\vect{\alpha}_{od}(\vect{0})$. This operation is equivalent to the replacement $\cplx{\alpha}_{od}(\cplx{\theta})\rightarrow\cplx{\alpha}_{od}(\cplx{\theta})-\cplx{\alpha}_{od}(0)$ in \cref{eq:cc_perturbation_axisymmetric}, which implies
\begin{equation}
c_n \rightarrow c_n 
- \frac{1}{2\theta\e{E}} \pac{\cplx{\alpha}_{od}(0)\delta_{1n} + \cplx{\alpha}_{od}^*(0)\delta_{-1n}} \ .
\end{equation}
The resulting $c_1$ is now observable and may be thought of as a measure of the type-$\F$ flexion. We shall adopt this convention in the remainder of the article.

If the dominant lens were not axially symmetric, the critical modes would pick up an intrinsic contribution and read $c_n = c_n\h{int}+c_n\h{los}$. In principle, the distortions of a critical curve due to line-of-sight perturbers, encoded in $c_n\h{los}$, depend on the fact that the dominant lens is axially symmetric or not. However, for quasi-axially symmetric dominant lens, such a dependence would be of higher order. When considering two-point correlations of $c_n$, the intrinsic contribution $c_n\h{int}$ is expected to drop, much like the intrinsic ellipticity of galaxies statistically drops in standard cosmic shear. This justifies our simplifying assumption of an axially symmetric dominant lens.

\subsection{Critical modes in cosmology}

We now aim to express the critical modes $c_n$ of a strong lens as a function of the matter density contrast~$\delta$ along, and in the vicinity of, the line of sight. For that purpose, we proceed similarly to refs.~\cite{Fleury:2018cro, Fleury:2018odh} that were focused on the weak lensing of extended sources. Namely, we first calculate $c_n$ in the case where the perturbers are \emph{point-like} lenses (\cref{subsubsec:point-like_perturbers}). We then take the continuous limit, having the number of lenses go to infinity while their mass goes to zero, in such a way that the mass density follows $\delta$ (\cref{subsubsec:continuous_limit}).

\subsubsection{Point-like perturbers as an intermediate step}
\label{subsubsec:point-like_perturbers}

A point-like lens $l$ with mass $m$ and transverse position $\vect{y}_l$ causes at $\vect{x}$ a deflection 
\begin{equation}
\label{eq:deflection_PL}
\hat{\alpha}_l(\cplx{x})
= \frac{4G m}{\cplx{x}^*-\cplx{y}_l^*}
\end{equation}
in complex notation. Because the point-lens model is only a technical intermediate, we shall assume that all the perturbers have the same mass~$m$ for simplicity. From \cref{eq:deflection_PL}, we immediately deduce the projected density and projected field quadrupole
\begin{equation}
4\pi G \Sigma_l(\cplx{x})
= \pd{\hat{\alpha}_l}{\cplx{x}}
= 0 \ ,
\qquad
4\pi G Q_l(\cplx{x})
= \pd{\hat{\alpha}_l}{\cplx{x}^*}
= - \frac{4G m}{(\cplx{x}^*-\cplx{y}_l^*)^2} \ .
\end{equation}

\paragraph{Perturbation of the critical curve} Substituting these in \cref{eq:alpha_od_cc,eq:kappa_gamma_cc}, we obtain the following compact expression for the perturbation of the critical-curve radius,\footnote{In this expression we have already performed the shift of the main plane's origin, i.e., we have made the replacement $\cplx{\alpha}_{od}(\cplx{\theta})\rightarrow\cplx{\alpha}_{od}(\cplx{\theta})-\cplx{\alpha}_{od}(0)$ in \cref{eq:cc_perturbation_axisymmetric}.}
\begin{equation}
\delta\theta\e{cc}(u)
= \theta\e{E}
        \sum_{l<d} \bar{\kappa}_{old} \,\Re\pac{
            \frac{ u^2}{\cplx{w}_l(u-\cplx{w}_l)}
        }
    + \frac{\theta\e{E}}{2(1-\kappa\e{E})}
        \sum_{l\neq d} (\bar{\kappa}_{ols}-\bar{\kappa}_{dls}-\bar{\kappa}_{old})\,
         \Re\pac{
            \frac{\, u^2}{(u-\cplx{w}_l)^2}
        } ,
\end{equation}
where we introduced the following notation. First, $\cplx{w}_l\define \cplx{y}_l/R_l$ is the transverse position of the $l\h{th}$ lens normalised by the radius~$R_l$ of the unperturbed critical beam in the $l\h{th}$ plane (see \cref{fig:radius_critical_beam}). Thus, $|\cplx{w}_l|<1$ means that $l$ is inside the critical beam, while $|\cplx{w}_l|>1$ means that it is outside. Second, the three quantities $\bar{\kappa}_{ilj}$ are defined as
\begin{equation}
\bar{\kappa}_{ilj}
=
\begin{cases}
\displaystyle
\frac{1}{\Sigma\h{crit}_{ilj}} \, \frac{m}{\pi R_l^2} & i<l<j \ ,
\\
0 & \text{otherwise.}
\end{cases}
\end{equation}
The use of the symbol $\bar{\kappa}$ is justified by the fact that $\bar{\kappa}_{ilj}$ coincides with the $(ilj)$-convergence of a homogeneous matter plane at $l$ with density $m/(\pi R_l^2)$, which is the density of the $l\h{th}$ lens when averaged over the critical beam's cross section.

\paragraph{Critical modes due to point perturbers} The interest of expressing $\delta\theta\e{cc}$ as a function of $u=\ex{\ii\ph}$, is that the critical modes $c_n$ are easily computed as complex contour integrals using the residue theorem. Specifically, the two necessary integrals are found to read
\begin{align}
\label{eq:complex_integral_direct}
\mathcal{D}_n(\cplx{w})
&\define
\frac{1}{2\ii\pi} \int_{\unitcircle} \dd u \; u^{n-1} \,
2\Re\pac{ \frac{u^2}{\cplx{w}(u-\cplx{w})}}
\\
&=
\begin{cases}
\pac{\cplx{w}^n + (\cplx{w}^*)^{-n}}\Theta(1-|\cplx{w}|)
& n \leq 1, \\
\cplx{w}^n \,\Theta(1-|\cplx{w}|) - (\cplx{w}^*)^{-n}\,\Theta(|\cplx{w}|-1)
& n \geq 2,
\end{cases}
\\
\label{eq:complex_integral_coupling}
\mathcal{C}_n(\cplx{w})
&\define
\frac{1}{2\ii\pi} \int_{\unitcircle} \dd u \; u^{n-1} \,
2\Re\pac{\pa{\frac{u}{u-\cplx{w}}}^2}
\\
&=
\begin{cases}
2\cplx{w}^n \Theta(1-|\cplx{w}|)
& n \leq 1,
\\
(n+1)\cplx{w}^n\Theta(1-|\cplx{w}|) + (n-1)(\cplx{w}^*)^{-n}\Theta(|\cplx{w}|-1)
& n \geq 2,
\end{cases}
\end{align}
where $\Theta$ denotes the Heaviside function. These integrals already reveal two important properties of the critical modes~$c_n$. First, the cases $n\leq 1$ and $n\geq 2$ must be treated separately. Second, there is a clear dichotomy between the effect of interior perturbers, i.e., the lenses located inside the critical beam ($|\cplx{w}_l|<1$), and that of the exterior lenses ($|\cplx{w}_l|>1$). This fact is reminiscent of the results of refs.~\cite{Fleury:2017owg, Fleury:2018cro, Fleury:2018odh}. Note also that the case $|\cplx{w}_l|=1$ (perturber on the critical beam) is ill-defined. This is because in the DL approximation, we assume that the lenses $l\neq d$ can be treated as weak perturbers. Point lenses that are too close from the critical beam do not satisfy this requirement.

In terms of the complex functions $\mathcal{D}_n(\cplx{w})$ (for \textbf{d}irect foreground displacement) and $\mathcal{C}_n(\cplx{w})$ (for lens-lens \textbf{c}oupling), the critical modes simply read
\begin{equation}
\label{eq:critical_modes_PL}
c_n = \frac{1}{2}\sum_{l<d} \bar{\kappa}_{old} \, \mathcal{D}_n(\cplx{w}_l)
        + \frac{1}{4(1-\kappa\e{E})}
            \sum_{l\neq d} (\bar{\kappa}_{ols}-\bar{\kappa}_{dls}-\bar{\kappa}_{old}) \, \mathcal{C}_n(\cplx{w}_l)
            \ .
\end{equation}
Note that only interior lenses ($|\cplx{w}_l|<1$) contribute to the first two modes $n=0, 1$. For $n=1$ (global displacement of the critical curve), the result crucially depends on our choice of origin for the main lens plane. In particular, the fact that $\vect{\alpha}_{od}(\vect{0})$ was subtracted from $\vect{\alpha}_{od}(\vect{\theta})$ is the reason why the exterior perturbers do not contribute to $c_1$.

\subsubsection{Continuous limit}
\label{subsubsec:continuous_limit}

Now that we have an explicit expression~\eqref{eq:critical_modes_PL} for the critical modes~$c_n$ in the presence of point-like perturbers, we may take the continuous limit. The procedure is the following. If a perturber lies at a comoving distance~$\chi$ from the observer and at a comoving transverse position $\vect{\zeta}$ from the optical axis, then we may replace its mass\footnote{Since we are working with a homogeneous cosmological background, the mass $m$ is allowed to be negative, in order to model under-densities as well as over-densities. See ref.~\cite{Fleury:2018cro} for a detailed discussion.}~$m$ by its infinitesimal counterpart $\dd^3 m=\bar{\rho}_0\delta(\eta_0-\chi, \chi, \vect{\zeta})\,\dd\chi \dd^2\vect{\zeta}$, where $\bar{\rho}_0, \eta_0$ denote today's mean matter density and conformal time. The fact that $\delta$ is evaluated at a time $\eta_0-\chi$ ensures that we stay on the background light-cone. We shall omit this time dependence from now on, simply writing $\delta(\chi,\vect{\zeta})$ for short. The discrete sums over $l$ can then be turned into integrals over $\chi, \vect{\zeta}$. For an easier connection with the \cref{eq:critical_modes_PL}, instead of $\vect{\zeta}$ we may prefer $\vect{w}\define \vect{\zeta}/r(\chi)$ which is the comoving transverse position normalised by the comoving radius of the critical beam,
\begin{equation}
\label{eq:comoving_radius_critical_beam}
r(\chi) \define (1+z) R(\chi)
=
\theta\e{E}
\times
\begin{cases}
f_K(\chi) & \chi\leq\chi_d \ ,\\[2mm]
\displaystyle
\frac{f_K(\chi_d)f_K(\chi_s-\chi)}{f_K(\chi_s-\chi_d)} & \chi\geq \chi_d \ .
\end{cases}
\end{equation}
Summarising, with the above set of rules the continuous limit consists in the replacement
\begin{equation}
\sum_{i<l<j}\bar{\kappa}_{ilj}
\rightarrow
4\pi G \bar{\rho}_0 \int_{\chi_i}^{\chi_j} \dd\chi \;
	(1+z) \, \frac{f_K(\chi-\chi_i)f_K(\chi_j-\chi)}{f_K(\chi_j-\chi_i)} \,
	\int_{\mathbb{R}^2} \frac{\dd^2\vect{w}}{\pi} \;
		\delta\pac{\chi, r(\chi)\vect{w}} .
\end{equation}

Performing that replacement in \cref{eq:critical_modes_PL}, we finally obtain the cosmological expression of the critical modes,
\begin{empheq}[box=\fbox]{equation}
\label{eq:critical_modes_cosmic}
c_n
= 4\pi G \bar{\rho}_0
\int_0^{\chi_s} \dd\chi \, (1+z)
\int_{\mathbb{R}^2} \frac{\dd^2\vect{w}}{2\pi} \;
\pac{
    W_{\mathcal{D}}(\chi) \, \mathcal{D}_n(\cplx{w})
    + \frac{W_{\mathcal{C}}(\chi) \, \mathcal{C}_n(\cplx{w})}{2(1-\kappa\e{E})}}
\delta\pac{\chi, r(\chi) \vect{w}} ,
\end{empheq}
where the complex functions $\mathcal{C}_n$ and $\mathcal{D}_n$ are the same as in \cref{eq:complex_integral_direct,eq:complex_integral_coupling}, and we have encapsulated the various distance ratios in two weight functions,
\begin{align}
\label{eq:W_D}
W_{\mathcal{D}}(\chi)
&=
\begin{cases}
\displaystyle
\frac{f_K(\chi)f_K(\chi_d-\chi)}{f_K(\chi_d)} & 0\leq\chi\leq\chi_d \ ,\\
0 & \text{otherwise,}
\end{cases}
\\[5mm]
\label{eq:W_C}
W_{\mathcal{C}}(\chi)
&=
\begin{cases}
\displaystyle
\frac{f_K(\chi)f_K(\chi_s-\chi)}{f_K(\chi_s)}
- \frac{f_K(\chi)f_K(\chi_d-\chi)}{f_K(\chi_d)}
& 0\leq\chi\leq\chi_d \ ,\\[4mm]
\displaystyle
\frac{f_K(\chi)f_K(\chi_s-\chi)}{f_K(\chi_s)}
- \frac{f_K(\chi-\chi_d)f_K(\chi_s-\chi)}{f_K(\chi_s-\chi_d)}
& \chi_d\leq\chi\leq\chi_s \ ,\\
0 & \text{otherwise.}
\end{cases}
\end{align}

\begin{figure}[t]
\centering
\begin{minipage}{0.65\textwidth}
\includegraphics[width=\columnwidth]{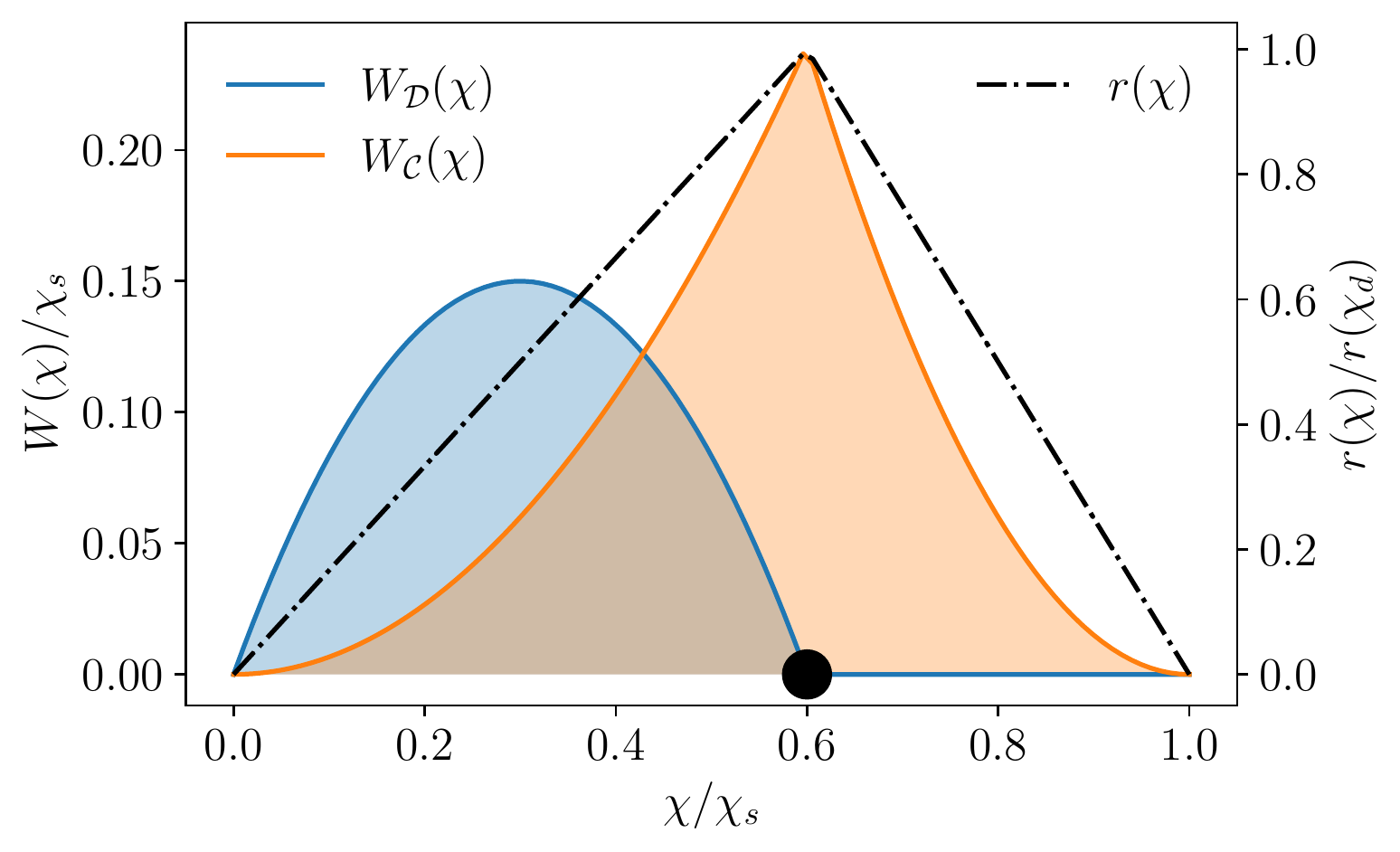}
\end{minipage}
\hfill
\begin{minipage}{0.3\textwidth}
\caption{Weight functions and comoving beam radius characterising the contribution of inhomogeneities to the critical modes~$c_n$ in \cref{eq:critical_modes_cosmic}. The comoving position of the dominant lens, $\chi_d=0.6\,\chi_s$ here, is indicated with a black disk.}
\label{fig:weight_functions}
\end{minipage}
\end{figure}

Now let us pause and analyse the structure of \cref{eq:critical_modes_cosmic}. The inhomogeneities~$\delta(\chi,\vect{\zeta})$ of the Universe contribute to $c_n$ in two ways---a direct foreground contribution via $W_{\mathcal{D}}(\chi)\mathcal{D}_n(\cplx{w})$ and a lens-lens-coupling contribution via $W_{\mathcal{C}}(\chi)\mathcal{C}_n(\cplx{w})$. While the direct foreground contribution is essentially independent of the properties of the dominant lens, the lens-lens coupling contribution is multiplied with $(1-\kappa\e{E})^{-1}$, which makes it sensitive to the mass profile of the main lens. The weight functions $W_{\mathcal{D}}(\chi), W_{\mathcal{C}}(\chi)\geq 0$ indicate how much some inhomogeneity contributes depending on its comoving distance~$\chi$ from the observer, while $\mathcal{C}_n(\cplx{w}), \mathcal{D}_n(\cplx{w})$ indicate how it contributes depending on its transverse position~$\vect{\zeta}=r(\chi)\vect{w}$ with respect to the optical axis. In particular, the effect of some inhomogeneity depends on whether it lies inside the critical beam ($|\vect{\zeta}|<r$) or outside ($|\vect{\zeta}|>r$). To make things more concrete, \cref{fig:weight_functions} illustrates $W_{\mathcal{D}}(\chi), W_{\mathcal{C}}(\chi), r(\chi)$ for $\chi_d=0.6\, \chi_s$ and $K=0$ (spatially flat Universe). We can see that while $W_{\mathcal{D}}$ is maximum mid-way between the observer and the dominant lens, $W_{\mathcal{C}}$ strongly peaks near the main lens. Of course, since the critical beam size is maximum at $\chi=\chi_d$, the relative contribution of interior lenses is also maximum there.

\subsection{Two-point correlations of the critical modes}

Since the critical modes $c_n$ take the form of a line-of-sight integral of the density contrast~$\delta$, their correlation functions across the sky must be related to the matter power spectrum, much like the standard cosmic convergence and shear. Besides, as mentioned in \cref{subsubsec:comparison_literature}, for realistic lenses the critical modes should contain both an intrinsic component and a line-of-sight component. Only the latter should be correlated on large scales.

\subsubsection{Effective critical modes in a given direction}

Consider a direction~$\vect{\vartheta}$ in the sky, and assume that there are many strong-lensing systems around that direction for which the critical curve (and their modes) could be measured. Apart from their position $\vect{\vartheta}$, the critical modes of a strong-lensing system implicitly depend on four parameters: its Einstein radius~$\theta\e{E}$; its convergence~$\kappa\e{E}$ at the Einstein radius; the distance to the lens~$\chi_d$ and distance to the source~$\chi_s$. Let us call $\vect{\Pi}\define(\theta\e{E}, \kappa\e{E}, \chi_d, \chi_s)$ the set of these four parameters, and $p(\vect{\Pi})$ their joint probability density function. We then define the \emph{effective critical mode} in the direction $\vect{\vartheta}$ as
\begin{equation}
\label{eq:critical_modes_effective}
\bar{c}_n(\vect{\vartheta})
\define
\int \dd^4\vect{\Pi} \; p(\vect{\Pi}) \, c_n(\vect{\vartheta};\vect{\Pi}) \ .
\end{equation}
Note that the direction $\vect{\vartheta}$ did not explicitly appear in the expression~\eqref{eq:critical_modes_cosmic} of $c_n$. This is because $\vect{\vartheta}$ corresponds to the optical axis which was taken as an origin of angular positions in all the previous calculations. The explicit dependence on $\vect{\vartheta}$ is thus easily restored by changing the second entry of the density contrast as $r(\chi)\vect{w}\rightarrow f_K(\chi)\vect{\vartheta}+r(\chi)\vect{w}$ in \cref{eq:critical_modes_cosmic}.

The effective modes $\bar{c}_n$ are analogous to the notions of effective convergence and shear in the standard weak-lensing formalism, once averaged over the distribution of sources~\cite{2001PhR...340..291B}. The main difference here is the additional marginalisation over $\theta\e{E}, \kappa\e{E}, \chi_d$.

\subsubsection{Two-point correlations in the flat-sky approximation}

\begin{figure}[t]
\centering
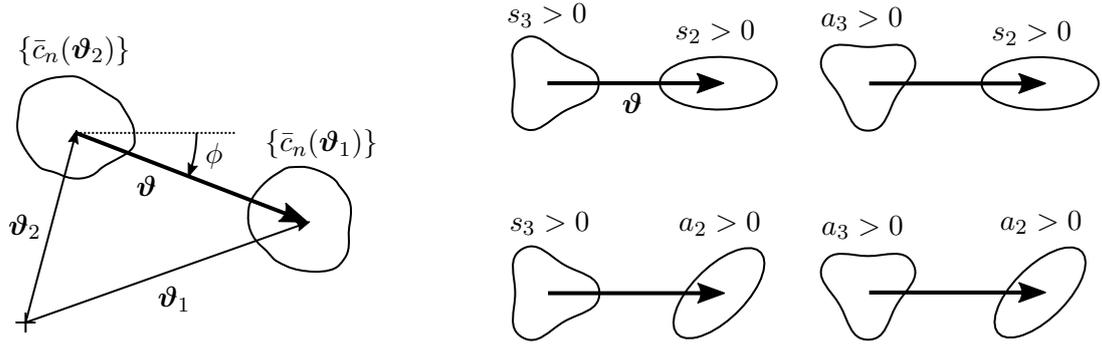
\caption{\textit{Left}: definition of the connecting vector~$\vect{\vartheta}=\vect{\vartheta}_1-\vect{\vartheta}_2$ and its polar angle~$\phi$. \textit{Right}: geometrical configurations probed by the correlation functions of critical modes with $n_1=3, n_2=2$.}
\label{fig:correlations_sketch}
\end{figure}

Let $\vect{\vartheta}_1, \vect{\vartheta}_2$ be two directions of the assumed flat sky in which effective critical modes, $\bar{c}_{n_1}(\vect{\vartheta}_1), \bar{c}_{n_2}(\vect{\vartheta}_2)$, are measured. Let us call $\vect{\vartheta}=\vect{\vartheta}_1-\vect{\vartheta}_2$ the vector separating those directions and $\phi$ the associated azimuthal angle, i.e., such that $\cplx{\vartheta}=\vartheta\ex{\ii\phi}$ (\cref{fig:correlations_sketch}). Each complex mode~$\bar{c}_n$ at the ends of $\vect{\vartheta}$ may be decomposed into a symmetric part~$s_n$ and an anti-symmetric part~$a_n$ with respect to the axis spanned by $\vect{\vartheta}$,
\begin{equation}
\bar{c}_{n}
=
\ex{\ii n \phi} \pac{ \Re\pa[1]{\ex{-\ii n \phi}\bar{c}_n} + \ii\,\Im\pa[1]{\ex{-\ii n \phi}\bar{c}_n} }
\define
\ex{\ii n \phi} \pa{ s_n + \ii a_n } \ .
\end{equation}
The notion of symmetry or anti-symmetry comes here from the geometric patterns associated with each component. For $n=1$, $s_1, a_1$ represent, respectively, the displacements parallel and transverse to $\vect{\vartheta}$; for $n=2$, $s_2$ corresponds to the ``plus'' component of shear, while $a_2$ represents the ``cross'' component, with respect to $\vect{\vartheta}$; $s_n, a_n$ generalise these notions to any $n$.

Given a couple of integers $n_1, n_2$, we may then define two correlation functions,
\begin{align}
\xi^s_{n_1 n_2}(\vartheta)
&\define \ev{s_{n_1}(\vect{\vartheta}_1)\,s_{n_2}(\vect{\vartheta}_2)} \ ,
\\
\xi^a_{n_1 n_2}(\vartheta)
&\define \ev{a_{n_1}(\vect{\vartheta}_1)\,a_{n_2}(\vect{\vartheta}_2)} \ ,
\end{align}
where $\ev{\ldots}$ denotes ensemble average. Due to statistical homogeneity and isotropy, there are no cross correlations between the symmetric and anti-symmetric components, $\ev{s_{n_1}(\vect{\vartheta}_1)a_{n_2}(\vect{\vartheta}_2)}=0$. Indeed, for a given configuration $(s_{n_1}, a_{n_2})$, there exists an equally probable Universe with $(s_{n_1}, -a_{n_2})$. This is also the reason why the argument of the correlation functions is $\vartheta=|\vect{\vartheta}|$.

Finally, just like in cosmic shear, we may actually prefer to compute the sum and difference of the symmetric and anti-symmetric correlation functions,
\begin{equation}
\label{eq:correlation_functions_pm}
\xi^\pm_{n_1 n_2}(\vartheta)
\define \xi_{n_1 n_2}^s(\vartheta) \pm \xi_{n_1 n_2}^a(\vartheta) \ .
\end{equation}
The reason for this preference is that $\xi^\pm_{n_1 n_2}(\vartheta)$ better capture the typical correlation patterns of critical modes. For instance, $\xi_{nn}^+$ is sensitive to the coherent distortions of two neighbouring systems induced by some distant matter lump.

The explicit calculation of the correlation functions~$\xi^\pm_{n_1 n_2}(\vartheta)$ is rather tedious, but it mostly follows the finite-beam calculations of ref.~\cite{Fleury:2018odh}. We shall thus give only the final result here, while details are provided in \cref{app:calculation_correlation_functions}. The plus and minus correlation functions may be expressed in terms of a common \emph{power spectrum}~$P_{n_1 n_2}(\ell)$ as
\begin{align}
\label{eq:xi_plus_result}
\xi_{n_1 n_2}^+(\vartheta)
&= \int \frac{\ell \dd\ell}{2\pi} \;
    J_{n_2-n_1}(\ell\vartheta) \, P_{n_1 n_2}(\ell)
\\
\label{eq:xi_minus_result}
\xi_{n_1 n_2}^-(\vartheta)
&= (-1)^{n_1}
	\int \frac{\ell \dd\ell}{2\pi} \;
	J_{n_1+n_2}(\ell\vartheta) \, P_{n_1 n_2}(\ell) \ ,
\end{align}
where $J_n$ denotes the $n\h{th}$ Bessel function. The power spectrum itself reads
\begin{empheq}[box=\fbox]{align}
\label{eq:power_spectrum_result}
P_{n_1 n_2}(\ell)
&= (4\pi G \bar{\rho}_0)^2 \int_0^\infty \dd\chi \; (1+z)^2 \,
		\bar{q}_{n_1}(\chi, \ell)\,\bar{q}_{n_2}(\chi, \ell) \,
		P_{\delta}\pac{\eta_0-\chi, \frac{\ell}{f_K(\chi)}}  ,
\\
\label{eq:q_n_bar}
\bar{q}_n(\chi,\ell)
&\define \int\dd^4\vect{\Pi} \; p(\vect{\Pi}) \, q_n(\chi, \ell; \vect{\Pi}) \ ,
\\
\label{eq:q_n}
q_n(\chi, \ell; \vect{\Pi})
&\define
\frac{W_{\mathcal{D}}(\chi; \vect{\Pi})}{f_K^2(\chi)}
\frac{\delta_{n1}-2J'_n[\ell\eps(\chi;\vect{\Pi}]}{\ell\eps(\chi;\vect{\Pi})}
\nonumber\\ & \quad
+ \frac{1}{1-\kappa\e{E}}\frac{W_{\mathcal{C}}(\chi; \vect{\Pi})}{f_K^2(\chi)}
        \paac{
            J_n[\ell\eps(\chi;\vect{\Pi})]
            + J_n''[\ell\eps(\chi;\vect{\Pi})]
		    } ,
\\
\label{eq:eps}
\eps(\chi; \vect{\Pi})
&\define \frac{r(\chi;\vect{\Pi})}{f_K(\chi)} \ ,
\end{empheq}
where $P_\delta(\eta, k)$ denotes the matter power spectrum. The quantity denoted $\eps(\chi;\vect{\Pi})$ is the angular size, from the observer's point of view, of the critical beam characterised by the set of parameters~$\vect{\Pi}$ at $\chi$. Due to the double-cone geometry of the critical beam (\cref{fig:radius_critical_beam}), $\chi\mapsto\eps(\chi;\vect{\Pi})$ is equal to $\theta\e{E}$ to the main lens ($\chi\leq\chi_d$), and then linearly decreases to reach $0$ at the source ($\chi_d\leq \chi\leq\chi_s$). The functions $W_{\mathcal{D}}, W_{\mathcal{C}}$ were defined in \cref{eq:W_D,eq:W_C}.

\begin{figure}[t]
\centering
\includegraphics[width=0.49\columnwidth]{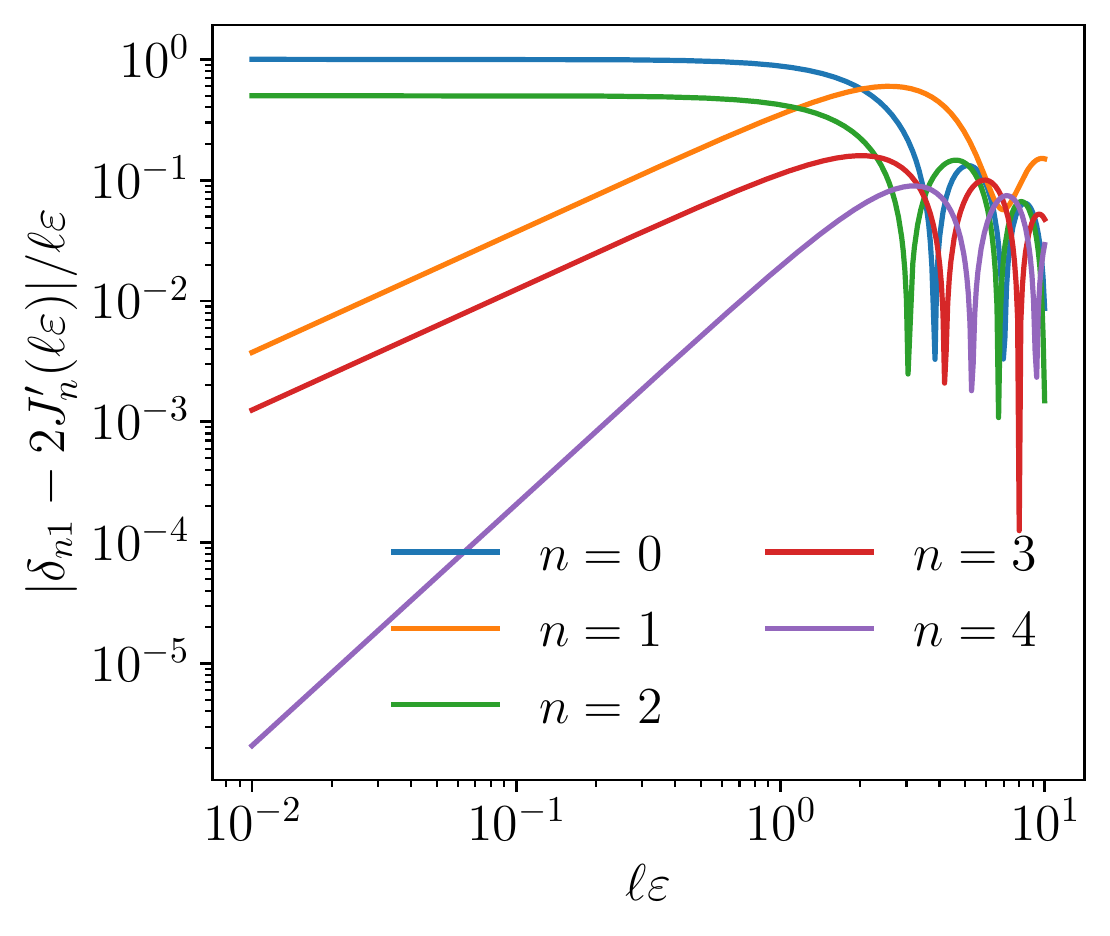}
\hfill
\includegraphics[width=0.49\columnwidth]{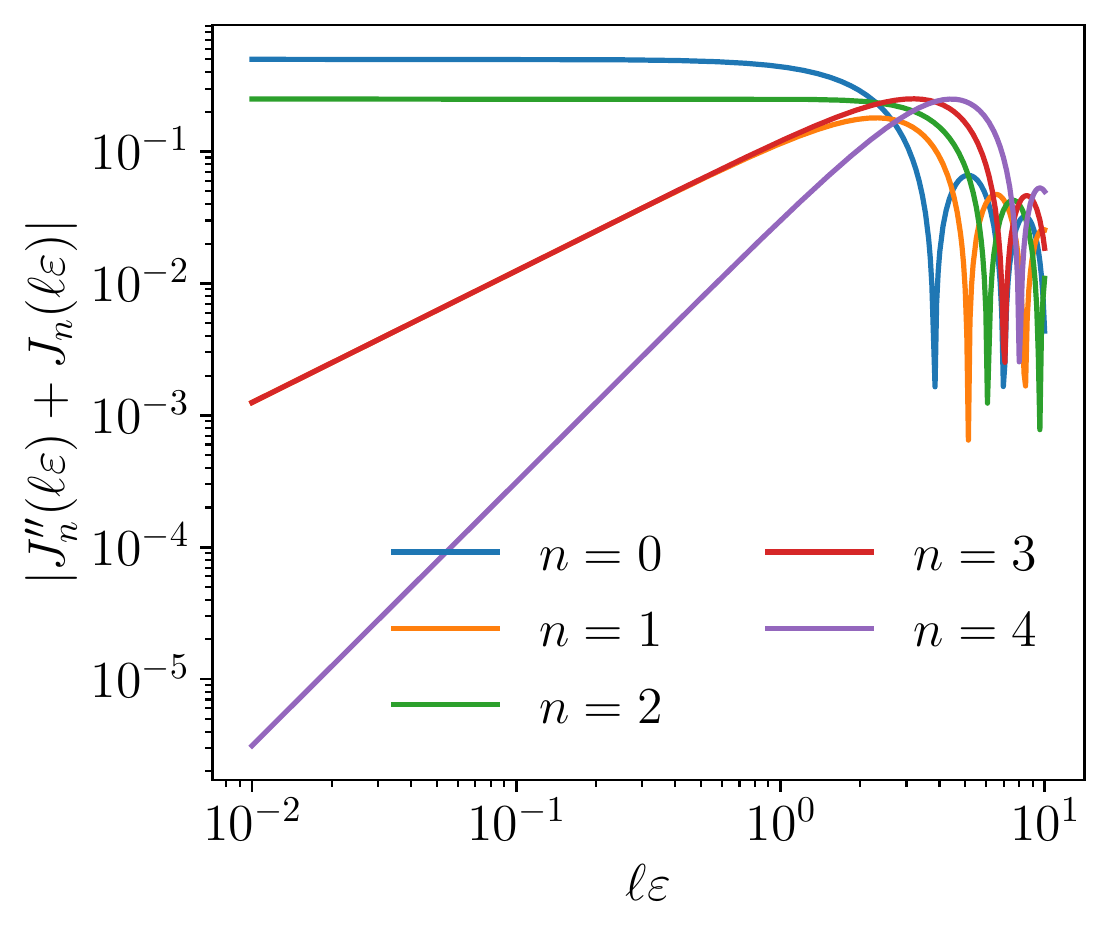}
\caption{Bessel pre-factors of $W_{\mathcal{D}}$ and $W_{\mathcal{C}}$ in the expression~\eqref{eq:q_n} of $q_n$. For practical purposes $\ell\eps\ll 1$, where these functions are manifestly well approximated by power laws.}
\label{fig:Bessel_kernels}
\end{figure}

The general structure of the power spectrum~\eqref{eq:power_spectrum_result}---a weighted line-of-sight integral of the matter power spectrum---is reminiscent of the standard cosmic-shear or cosmic-convergence spectra~\cite{Fleury:2018cro}, as well as the spectra of higher-order distortion modes~\cite{Fleury:2018odh}. The respective Bessel pre-factors of $W_{\mathcal{D}}$ and $W_{\mathcal{C}}$ are depicted in \cref{fig:Bessel_kernels} for $n\leq 4$.

The full expression of $q_n$, in particular the Bessel terms illustrated in \cref{fig:Bessel_kernels}, properly allows for correlations on scale comparable to the typical Einstein radii of the observed strong-lensing systems. However, in practice, near-future surveys will only be capable of probing correlations of Einstein rings on relatively large scales, such that $\ell\eps\ll 1$. As a rule of thumb, considering that \textit{Euclid} may provide us with $N\sim 10^5$ exploitable Einstein rings~\cite{2015ApJ...811...20C} over a celestial area of $\Omega\sim 15\,000\U{deg}^2$, implies that $\ell\e{max}\sim\pi\sqrt{N/\Omega}\sim 450$. For Einstein rings with $\theta\e{E}\sim \mathrm{arcsec}$, we conclude that $(\ell\eps)\e{\max}\sim 2\times 10^{-3}\ll 1$. In that regime, the Bessel functions behave as power laws, $J_n(x\ll 1)\approx x^n/(2^n n!)$, which yields
\begin{equation}
q_n(\chi, \ell)
\approx
\begin{cases}
- \frac{W_{\mathcal{D}}(\chi)}{f_K^2(\chi)}
- \frac{1}{2(1-\kappa\e{E})} \frac{W_{\mathcal{C}}(\chi)}{f_K^2(\chi)}
& n=0 \ ,
\\[3mm]
\pac{
\frac{3W_{\mathcal{D}}(\chi)}{f_K^2(\chi)}
+ \frac{1}{1-\kappa\e{E}} \frac{W_{\mathcal{C}}(\chi)}{f_K^2(\chi)}
} \frac{\ell\eps(\chi)}{8}
& n=1 \ ,
\\[3mm]
\pac{
-\frac{W_{\mathcal{D}}(\chi)}{f_K^2(\chi)}
+ \frac{n-1}{2(1-\kappa\e{E})} \frac{W_{\mathcal{C}}(\chi)}{f_K^2(\chi)}
} \frac{[\ell\eps(\chi)]^{n-2}}{2^{n-1}(n-1)!}
& n\geq 2 \ ,
\end{cases}
\end{equation}
where we omitted the dependencies in $\vect{\Pi}$ to alleviate the notation. Since $\ell\eps\ll 1$, $q_{n\geq 2} \propto (\ell\eps)^{n-2}$ quickly decreases as $n$ increases. This was expected---higher-order distortions to critical curves are produced by nearby perturbations that are not correlated at large distances.

We note that both $q_0$ and $q_2$ are independent of $\ell\eps$, which implies that they are not affected by such a damping. This was also expected, since the critical modes $\bar{c}_0, \bar{c}_2$ must be understood, respectively, as convergence and shear as measured from critical curves. Similarly, the modes $\bar{c}_1, \bar{c}_3$, which respectively represent the global shift of a the ring relative to the main lens, and the triangularity of the ring, are associated with the type-$\F$ and type-$\G$ flexions.

\subsubsection{Critical curves as cosmic probes; breaking the mass-sheet degeneracy}

Provided that we dispose of an efficient technique to determine critical curves from strong-lensing images, the critical modes may be used as alternative probes for weak lensing, and hence for cosmology. Two-point correlations of critical modes would be complementary with galaxy-based weak lensing for two reasons. On the one hand, critical modes would not be subject to the same observational uncertainties as galaxy shapes. On the other hand, because they rely on strongly lensed systems, they would naturally probe structures at higher redshifts compared to galaxies.

\begin{figure}[t]
    \centering
    \includegraphics[width=0.95\columnwidth]{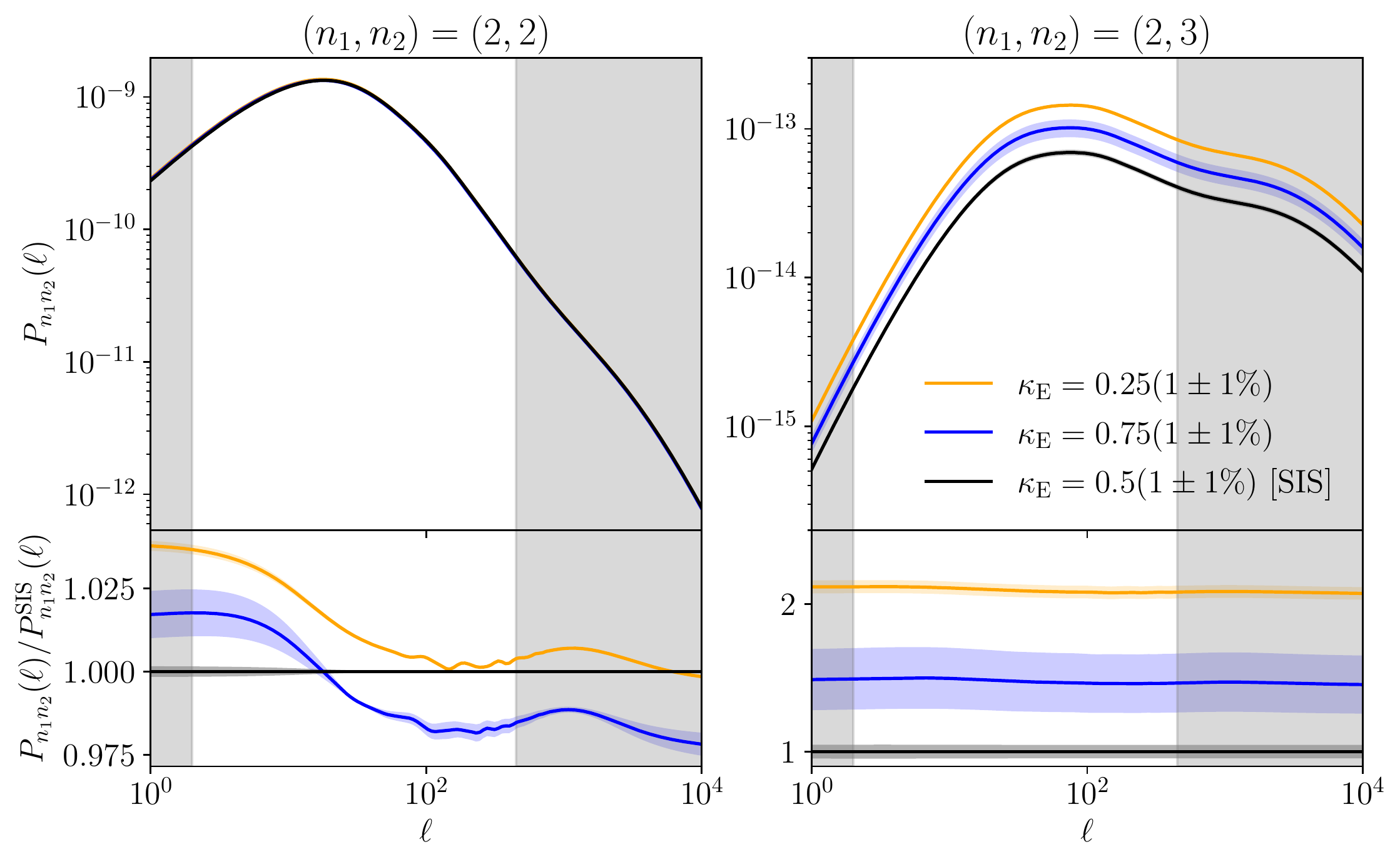}
    \caption{Power spectra $P_{n_1 n_2}(\ell)$ associated with the two-point correlation functions of the critical modes $n_1, n_2$ for a \textit{Euclid}-like survey. \textit{Left panel}: ellipticity-ellipticity spectrum $(n_1, n_2)=(2,2)$. \textit{Right panel}: ellipticity-triangularity spectrum $(n_1, n_2)=(2,3)$. The plots are made for three different assumptions for the Einstein-radius convergence of the dominant lens; from top to bottom, $\kappa\e{E}=0.25, 0.75, 0.5$. Shaded regions around each curve indicate the effect of varying $\kappa\e{E}$ by $1\%$. The bottom panels show the ratios of the spectra for a given $\kappa\e{E}$ with the SIS case, $\kappa\e{E}=0.5$. Grey-shaded regions indicate the multipoles that will \emph{not} be accessible for a \textit{Euclid}-like survey; we used $\ell\e{min}=2$ and $\ell\e{\max}=\pi\sqrt{N/\Omega}\approx 450$, where $N=10^5$ is the expected number of Einstein rings observed by \textit{Euclid} and $\Omega=15\,000\U{deg}^2$ is the celestial area that it will cover.}
    \label{fig:spectra_critical_modes}
\end{figure}

As an illustration, we show in \cref{fig:spectra_critical_modes} the expected power spectra $P_{22}(\ell)$ and $P_{23}(\ell)$ for a \textit{Euclid}-like survey.\footnote{The expressions of the power spectra derived in this paper rely on the Limber and flat-sky approximations; see \cref{app:calculation_correlation_functions}. As such, they cannot be strictly applied to estimate the correlations at large angles. However, recent studies have found that wide-angle corrections to the weak-lensing shear power spectrum should remain below cosmic variance for a future \textit{Euclid}-like survey~\cite{Kilbinger_2017}. Therefore, we do not expect the conclusions of this section to be qualitatively altered by such corrections.} Since this is a strictly theoretical estimate, the survey specifications only intervene through the expected distributions~$p(\vect{\Pi})$ for the source, lens, and Einstein-radius distributions. For those, we freely adapted the results of ref.~\cite{2015ApJ...811...20C}. Apart from $\theta\e{E}$, the only property of the main lenses which $P_{n_1 n_2}(\ell)$ depend on, is the Einstein-radius convergence~$\kappa\e{E}$. We assume for simplicity that it is identical for all the lenses. \Cref{fig:spectra_critical_modes} shows the power spectra for three different values of $\kappa\e{E}=0.25, 0.75, 0.5$. The matter power spectrum~$P_\delta$ is obtained with \textsc{camb}\footnote{\href{https://camb.info}{https://camb.info}} which integrates \textsc{halofit} for non-linear scales.

%
%

We first notice that the $P_{23}$ signal is 4 orders of magnitude weaker than the $P_{22}$ signal. This is due to the fact that the $n=3$ critical mode comes with a weight function $q_3\sim \ell\theta\e{E}\times q_2 \sim 5\times 10^{-4}\times q_2$ for $\theta\e{E}\sim \mathrm{arcsec}$ and $\ell\sim 100$. The signal is expected to be even weaker for higher-order modes, since $q_{n\geq 2}\sim (\ell\theta\e{E})^{n-2}$ in that regime. The actual detectability of $P_{23}$ with future surveys will depend on our ability to accurately measure the critical modes $c_2, c_3$; this will be addressed in future work.

The second important observation from \cref{fig:spectra_critical_modes} is that $P_{23}$ is much more sensitive to changes in $\kappa\e{E}$ than $P_{22}$. For example, decreasing $\kappa\e{E}$ from $0.5$ to $0.25$ leads to a few-percent variation in $P_{22}$, whereas it increases $P_{23}$ by more than $100\%$. This may be understood as follows. Each weight functions $q_n$ is the sum of a $W_{\mathcal{D}}$-term which does not depend on $\kappa\e{E}$, and a $W_{\mathcal{C}}$-term which is proportional to $(1-\kappa\e{E})^{-1}$. In the present set-up, it turns out that the former dominates the latter in $q_2$, while the latter dominates the former in $q_3$. The dominance of the $W_{\mathcal{C}}$-terms over the $W_{\mathcal{D}}$-terms in $q_n$ is expected to be even more pronounced as $n$ increases. In that sense, $P_{22}$ stands out as a power spectrum relatively insensitive to the Einstein-radius convergence~$\kappa\e{E}$ of the dominant lenses.

This coincidence may be an excellent opportunity to constrain the distribution of $\kappa\e{E}$ for a population of strong lenses, thereby statistically breaking the mass-sheet degeneracy~\cite{1985ApJ...289L...1F}. If a given lens model~$\vect{\alpha}(\vect{\theta})$ provides a good fit to an image, then
$
\vect{\alpha}_\lambda(\vect{\theta})
\define (1-\lambda)\,\vect{\theta} + \lambda\,\vect{\alpha}(\vect{\theta})
$
provides an equally good fit for any $\lambda\in(0,1]$. This is the mass-sheet degeneracy. However, $\vect{\alpha}_\lambda$ has a different convergence~$\kappa\e{E}^\lambda$ at its Einstein radius, namely
\begin{equation}
1-\kappa\e{E}^\lambda = \lambda\,(1-\kappa\e{E}) \ . 
\end{equation}
Therefore, any direct measurement of $\kappa\e{E}$ breaks the mass-sheet degeneracy by fixing $\lambda$. If a future survey such as \textit{Euclid} allowed us to measure at least $P_{22}$ and $P_{23}$, then $P_{22}$, which is quite insensitive to $\kappa\e{E}$, would serve as a calibration (or a cosmological probe much like standard cosmic shear), while $P_{23}$ could be used to constrain the distribution of $\kappa\e{E}$. Any measurement of the higher-order spectra $P_{24}, P_{33}, P_{34}, \ldots$ would further improve those constraints. The sensitivity of $P_{23}$ to small changes of $\kappa\e{E}$ is illustrated by the shaded regions around each curve of \cref{fig:spectra_critical_modes}. This sensitivity greatly varies with $\kappa\e{E}$; for a $1\%$ change in $\kappa\e{E}$, $P_{23}$ is found to vary by $2.5\%, 5\%, 13\%$, respectively, for $\kappa\e{E}=0.25, 0.5, 0.75$. Assuming that $\kappa\e{E}\approx 0.5$ for most lenses, this suggests that $1\%$ constraints on $\kappa\e{E}$ would require a $5\%$ measurement of $P_{23}$.

\section{Summary and outlook}
\label{sec:conclusion}

In this article, we have comprehensively revisited the problem of line-of-sight corrections in strong gravitational lensing. We have proposed a general framework to accurately model such effects, as well as several applications which open promising research directions.

The general theoretical framework was laid out in \cref{sec:dominant_lens}. It relies on the \emph{dominant-lens (DL) approximation}, which allows one to treat the case where the lensing of some light source is dominated by a single deflector (the dominant or main lens) while the other deflectors in the Universe can be treated as perturbations. On the one hand, the DL approximation is \emph{novel} in that it is not restricted to the tidal regime, where line-of-sight perturbers are collectively modelled by mere convergence and shear parameters, which has been the standard approach so far. On the other hand, the DL framework is qualitatively simpler to handle than the fully general multi-plane lensing formalism. In particular, the DL equation~\eqref{eq:lens_equation_dominant} yields a direct relation between the image position~$\vect{\theta}$ and the source position~$\vect{\beta}$. For completeness, we also derived the expression of strong-lensing time delays~\eqref{eq:time_delay_dominant} in the DL regime.

A first set of applications of the DL framework was proposed in \cref{sec:parametric}. In that section, we have adopted a parametric approach, where line-of-sight corrections are encapsulated in a few numbers. In particular, we have shown how to recover the standard tidal approximation as a special case of the DL approximation. We have insisted on the fact that secondary tidal deflectors generally yield 3 convergence parameters $(\kappa_{os}, \kappa_{od}, \kappa_{ds})$, and 3 complex shear parameters $(\gamma_{os}, \gamma_{od}, \gamma_{ds})$ in addition to the dominant lens. Motivated by recent proposals in the literature, we have investigated the possibility of measuring these line-of-sight parameters from strong-lensing images. We have argued that degeneracies had been overlooked in previous works, leading to over-optimistic results therein. Nevertheless, we have found that a special combination of the line-of-sight shears, namely $\gamma\e{LOS}=\gamma_{od}+\gamma_{os}-\gamma_{ds}$ is independent of the properties of the main lens. This conclusion confirms and sharpens the idea that future surveys may allow us to \emph{measure weak lensing with Einstein rings}. Such an approach would be complementary to the standard cosmic shear of galaxy surveys, by allowing us to probe higher redshifts while being affected by different systematic uncertainties. We believe that this novel avenue is promising and worth subsequent endeavour.

The DL approximation is more complete than the tidal regime. This is explicitly demonstrated in \cref{subsec:flexion}, where we have consistently supplemented the external convergences and shears with flexion (8 additional complex parameters). This also illustrates how the number of parameters rapidly grows and may become intractable as one refines the description of line-of-sight effects. For that reason, we have concluded that a parametric method may not be the most suitable, which is why we have switched to another approach in \cref{sec:critical_curves}.

In the second set of applications of the DL framework, presented in \cref{sec:critical_curves}, we have chosen to focus on a particular feature of strong-lensing systems, namely their \emph{critical curve}, which is the set of points where the lensing magnification is formally infinite. Although critical curves do not encode all the information about a strong-lensing system, they turned out to be a convenient tool to study line-of-sight corrections; namely, the Fourier modes of critical curves (\emph{critical modes}~$c_n$) may be seen as probes of shear ($c_2$), flexion ($c_1, c_3$), etc. We have derived the expression~\eqref{eq:power_spectrum_result} of the power spectrum~$P_{n_1 n_2}(\ell)$ of the critical modes, and estimated $P_{22}, P_{23}$ for a \textit{Euclid}-like survey. An important result is that $P_{23}$ is very sensitive to the (projected) density~$\kappa\e{E}$ of the dominant lens at the level of its Einstein radius. Thus, the two-point correlations of the distortions of critical curves are not only a potential new cosmic probe, but also a promising probe of the properties of galactic dark-matter haloes. This constitutes a second novel research avenue opened by the present work.

The parameter~$\kappa\e{E}$ was recently placed under the spotlight, because of its troublesome role in the measurement of the Hubble-Lemaître constant~$H_0$ from time-delay cosmography~\cite{1964MNRAS.128..307R}. Due to the mass-sheet degeneracy~\cite{1985ApJ...289L...1F}, the measured value of $H_0$ crucially depends on the accurate modelling of the main lens, in particular via the value of $\kappa\e{E}$. Simplistic assumptions on the main-lens mass profile~$\kappa(\vect{\theta})$ may have led the H0LiCOW collaboration~\cite{Wong:2019kwg} to biased measurements of $H_0$~\cite{Kochanek:2019ruu, Kochanek:2020crs}. Direct measurements of $\kappa\e{E}$ are possible from the stellar velocity dispersion within the main lens; the precision of this method is currently limited~\cite{2020A&A...643A.165B}, but it is expected to improve with greater statistics in the future~\cite{Birrer:2020jyr}. The results of \cref{sec:critical_curves} provide another, independent, method to constrain $\kappa\e{E}$, at least statistically, from the weak distortions of critical curves beyond their ellipticity. Shall the feasibility of this method be confirmed, it may play a significant role in the current debate over the value of $H_0$, while providing key insights into the distribution of dark matter in galactic haloes~\cite{2020ApJ...892L..27B}.

\section*{Acknowledgements}

We thank Simon Birrer for discussions and for his detailed review of the manuscript. Many thanks to Théo Duboscq for spotting several typos in the published version. PF received the support of a fellowship from ``la Caixa'' Foundation (ID 100010434). The fellowship code is LCF/BQ/PI19/11690018.
This research is not supported by the National Research Foundation (South Africa). \Cref{fig:spectra_critical_modes} used the freely available code \href{https://camb.info}{\textsc{camb}} to generate matter power spectra. The codes and notebooks used to generate the figures of this article are available via \href{https://github.com/pierrefleury/LOS}{https://github.com/pierrefleury/LOS}.

\appendix

\section{On time delays in the dominant-lens regime}
\label{app:time_delays}

\subsection{Derivation of \cref{eq:time_delay_dominant}}
\label{subsec:derivation_time_delay_dominant}

In this sub-section, we shall demonstrate that the multi-plane time-delay function
\begin{equation}
\label{eq:time_delay_multi-plane_appendix}
T(\{\vect{x}_l\})
=
\sum_{l=1}^N
\pac{
    \frac{1}{2} \, \tau_{l(l+1)} |\vect{\beta}_{o(l+1)}-\vect{\beta}_{ol}|^2
    - (1+z_l) \hat{\psi}_l(\vect{x}_l)
    } ,
\end{equation}
with $\tau_{ij}\define (1+z_i) D_{oi}D_{oj}/D_{ij}$, reduces to
\begin{multline}
\label{eq:time_delay_dominant_appendix}
T(\vect{\theta},\vect{\beta})
=
\frac{1}{2} \tau_{ds}\abs{\vect{\theta}-\vect{\alpha}_{od}-\vect{\beta}}^2
- (1+z_{d}) \hat{\psi}_{d}\pac{D_{od}(\vect{\theta}-\vect{\alpha}_{od})}
\\
- \sum_{l<d} (1+z_l)\hat{\psi}_l(D_{ol}\vect{\theta})
- \sum_{l>d} (1+z_l)\hat{\psi}_l\pac{D_{ol}(\vect{\theta}-\vect{\alpha}_{od l})}
+ \order(\eps^4) \ .
\end{multline}
in the dominant-lens regime. In our derivation, we will make extensive use of the identity
\begin{empheq}[box=\fbox]{equation}
\label{eq:identity_tau}
\forall i<j<k \qquad
\frac{1}{\tau_{ik}} = \frac{1}{\tau_{ij}} + \frac{1}{\tau_{jk}} \ .
\end{empheq}
See appendix~C or ref.~\cite{Fleury:2020cal} for a general geometric proof of \cref{eq:identity_tau}.

\paragraph{Potential terms} Within the dominant-lens approximation, the terms of the form $\hat{\psi}_l(\vect{x}_l)$ in \cref{eq:time_delay_multi-plane_appendix} are readily put in the form of \cref{eq:time_delay_dominant_appendix}; just like in \cref{subsec:dominant_lens_equation}, at second order in $\eps$ we can make the substitution:
\begin{equation}
\vect{x}_l =
\begin{cases}
D_{ol} \, \vect{\theta} & l<d \ ,\\
D_{od} \pac{\vect{\theta}-\vect{\alpha}_{od}(\vect{\theta})} & l=d \ ,\\
D_{ol} \pac{\vect{\theta}-\vect{\alpha}_{odl}(\vect{\theta})} & l>d \ ,
\end{cases}
\end{equation}
which yields the potential terms of \cref{eq:time_delay_dominant_appendix}. The difficult part lies in the geometrical terms.

\paragraph{Geometrical terms} From the multi-plane lens recursion~\eqref{eq:lens_recursion}, we have
\begin{align}
\vect{\beta}_{ol} - \vect{\beta}_{o(l+1)}
&=
\sum_{m=1}^l \pac{
                \frac{D_{m(l+1)}}{D_{o(l+1)}}
                - \frac{D_{ml}}{D_{ol}}
                } \hat{\vect{\alpha}}_m
\\
&=
\sum_{m=1}^l \pac{
                \frac{(1+z_m)D_{om}}{\tau_{m(l+1)}}
                -  \frac{(1+z_m)D_{om}}{\tau_{ml}}
                } \hat{\vect{\alpha}}_m
\\
&=
\frac{1}{\tau_{l(l+1)}}
\sum_{m=1}^l (1+z_m) D_{om} \hat{\vect{\alpha}}_m
\qquad \text{using \cref{eq:identity_tau}.}
\end{align}

We note that the above sum is free from dominant-lens terms as long as $l<d$. Thus, for any $l<d$, $|\vect{\beta}_{o(l+1)} - \vect{\beta}_{ol}|^2=\order(\eps^4)$ can be dropped, so that the sum of geometrical terms in \cref{eq:time_delay_multi-plane_appendix} may be performed over $l\geq d$. The calculation then consists in expanding the sum at first order in $\eps^2$, and then performing a few manipulations
\begin{align}
T\e{geo}
&\define
\frac{1}{2}
\sum_{l=d}^N \tau_{l(l+1)} |\vect{\beta}_{o(l+1)}-\vect{\beta}_{ol}|^2
\\
&=
\frac{1}{2}
\sum_{l=d}^N \frac{1}{\tau_{l(l+1)}}
\abs{\sum_{m=1}^l (1+z_m) D_{om} \hat{\vect{\alpha}}_m}^2
\\
&=
\frac{1}{2}
\sum_{l=d}^N \frac{1}{\tau_{l(l+1)}}
\Bigg[
    \abs{(1+z_d) D_{od} \hat{\vect{\alpha}}_d}^2
    + 2(1+z_d) D_{od} \hat{\vect{\alpha}}_d
    \cdot
    \sum_{\substack{m=1\\m\neq d}}^l (1+z_m) D_{om} \hat{\vect{\alpha}}_m
\Bigg]
\label{eq:computation_T_geo_1}
\end{align}
up to $\order(\eps^4)$ terms. Let us start with the first term on the right-hand side of \cref{eq:computation_T_geo_1}. We may notice that
$
(1+z_d) D_{od} \hat{\vect{\alpha}}_d
= \tau_{ds} \vect{\alpha}_{ods}
$;
since this term does not depend on $l$, the sum over $l$ only concerns the $1/\tau_{l(l+1)}$. Using the identity~\eqref{eq:identity_tau} we find
\begin{equation}
\sum_{l=d}^N \frac{1}{\tau_{l(l+1)}} \abs{(1+z_d) D_{od} \hat{\vect{\alpha}}_d}^2
=
\abs{\tau_{ds} \vect{\alpha}_{ods}}^2
\underbrace{
\sum_{l=d}^N \frac{1}{\tau_{l(l+1)}}
}_{1/\tau_{ds}}
= 
\tau_{ds}\abs{\vect{\alpha}_{ods}}^2 .
\end{equation}
We now move to the $\order(\eps^2)$ terms, i.e. the second term in the right-hand side of \cref{eq:computation_T_geo_1}. We may compute the associated double sum by inverting its order
\begin{align}
\sum_{l=d}^N \frac{1}{\tau_{l(l+1)}}
\sum_{\substack{m=1\\m\neq d}}^l (1+z_m) D_{om} \hat{\vect{\alpha}}_m
&=
\sum_{m<d}
\pac{
    \sum_{l\geq d} \frac{1}{\tau_{l(l+1)}}
    }
(1+z_m) D_{om} \hat{\vect{\alpha}}_m
\nonumber\\ & \quad
+\sum_{m>d}
\pac{
     \sum_{l\geq m} \frac{1}{\tau_{l(l+1)}}
    }
    (1+z_m) D_{om} \hat{\vect{\alpha}}_m
\\
&=
\frac{1}{\tau_{ds}}
\sum_{m<d} (1+z_m) D_{om} \hat{\vect{\alpha}}_m
+ \sum_{m>d} \vect{\alpha}_{oms} \ ,
\end{align}
where we exploited \cref{eq:identity_tau} once again to compute the sums over $l$.

Putting the $\order(\eps^0)$ and $\order(\eps^2)$ terms back together, and renaming $m$ into $l$, we thus have
\begin{equation}
T\e{geo}
=
\frac{1}{2} \, \tau_{ds}
    \abs{
        \vect{\alpha}_{ods}
        + \frac{1}{\tau_{ds}}
            \sum_{l<d} (1+z_l) D_{ol} \hat{\vect{\alpha}}_l
        + \sum_{l>d} \vect{\alpha}_{ols}
        }^2 .
\label{eq:computation_T_geo_2}
\end{equation}
The last step of the computation consists in substituting the lens equation~\eqref{eq:lens_equation_dominant}, which allows us to express the main displacement angle as
\begin{equation}
\vect{\alpha}_{ods}
= (\vect{\theta}-\vect{\beta})
    - \sum_{l<d} \vect{\alpha}_{ols} - \sum_{l>d} \vect{\alpha}_{ols} \ ,
\end{equation}
thereby cancelling the $l>d$ terms in \cref{eq:computation_T_geo_2}; as for the contributions of foreground lenses to $\vect{\alpha}_{ods}$, they combine with the second term of \cref{eq:computation_T_geo_2} to give
\begin{align}
\frac{1}{\tau_{ds}}
\sum_{l<d} (1+z_l) D_{ol} \hat{\vect{\alpha}}_l
- \sum_{l<d} \vect{\alpha}_{ols}
&=
\sum_{l<d} \pac{
                \pa{\frac{1}{\tau_{ls}}-\frac{1}{\tau_{ld}}} (1+z_l)D_{ol}
                -\frac{D_{ls}}{D_{os}}
                } \hat{\vect{\alpha}}_l
\\
&=
\sum_{l<d} \pac{
                \pa{
                    \frac{D_{ls}}{D_{os}}
                    -\frac{D_{ld}}{D_{od}}
                    }
                -\frac{D_{ls}}{D_{os}}
                } \hat{\vect{\alpha}}_l
\\
&=
\sum_{l<d} \vect{\alpha}_{old}
\\
&\define \vect{\alpha}_{od} \ ;
\end{align}
whence the final result
\begin{equation}
T\e{geo} = \frac{1}{2} \, \tau_{ds} |\vect{\theta}-\vect{\beta}-\vect{\alpha}_{od}|^2 \ ,
\end{equation}
which, together with the potential terms, indeed yields \cref{eq:time_delay_dominant_appendix} and thereby ends our proof.

\subsection{Fermat's potential in the dominant-lens regime}

By virtue of Fermat's principle, we expect the time-delay function $T(\vect{\theta},\vect{\beta})$ to be stationary for physical rays, i.e. rays that respect the lens equation. Precisely, we expect to find $\partial T/\partial\vect{\theta}\propto \vect{L} \define \vect{\theta} - \vect{\beta} - \vect{\alpha}(\vect{\theta})$, where $\vect{\alpha}(\vect{\theta})$ is given by \cref{eq:displacement_dominant} in the dominant-lens regime. This would indeed imply that $\partial T/\partial\vect{\theta}=\vect{0}\Leftrightarrow \vect{L}=\vect{0}$, which is the lens equation. However, this is not true for $T$ given by \eqref{eq:time_delay_dominant}.

The reason for that failure is that \cref{eq:time_delay_dominant} actually holds \emph{for physical rays only}; in other words, the lens equation has already been applied in order to get that expression of $T$, as it clearly appears in the derivation of \cref{subsec:derivation_time_delay_dominant}. An important consequence is that any function of the form
\begin{equation}
\phi(\vect{\theta},\vect{\beta})
= T(\vect{\theta},\vect{\beta}) 
    + \Delta(\vect{\theta},\vect{\beta}) \ ,
\qquad \text{such that} \quad
\Delta[\vect{\theta},\vect{\theta}-\vect{\alpha}(\vect{\theta})]=0 \ ,
\end{equation}
coincides with $T$ on physical rays. However, their gradients do not necessarily agree, even for physical rays, because while $\Delta(\vect{\theta},\vect{\beta})$ is required to vanish for physical rays nothing guarantees that its gradient $\partial\Delta/\partial\vect{\theta}$ does.

Let us now derive the expression of the Fermat potential $\phi$ whose gradient is indeed proportional to $\vect{L}$. For that purpose, let us first compute the gradient of $T=T\e{geo}+T\e{pot}$ as given in \cref{eq:time_delay_dominant}. The geometrical term yields
\begin{align}
\pd{T\e{geo}}{\vect{\theta}}
&= \pd{}{\vect{\theta}}
    \pac{ \tau_{ds}
            \abs{
                \vect{\theta}-\vect{\beta}
                -\vect{\alpha}_{od}(\vect{\theta})
                }^2
        }
\\
&= \tau_{ds} \pa{1-\mat{\Gamma}_{od}}
    \pac{ \vect{\theta}-\vect{\beta}-\vect{\alpha}_{od}(\vect{\theta}) } \ .
\end{align}
As for the potential part, let us split into $T\e{pot}=T\e{pot}^< + T\e{pot}^d + T\e{pot}^>$, where the first term contains the potential of the foreground lenses ($l<d$), the second term the potential of the dominant lens, and the third term the potential of the background lenses ($l>d$). Calculations exploiting several times the identity~\eqref{eq:identity_tau} then yield
\begin{align}
\pd{T\e{pot}^<}{\vect{\theta}}
&= \pd{}{\vect{\theta}}
    \pac{
        -\sum_{l<d} (1+z_l) \hat{\psi}_l(D_{ol}\vect{\theta})
        }
= \tau_{ds}
    \pac{
        \vect{\alpha}_{od}
        -
        \sum_{l<d} \vect{\alpha}_{ols}(\vect{\theta})
        } ,
\\
\pd{T\e{pot}^d}{\vect{\theta}}
&= \pd{}{\vect{\theta}}
    \paac{
        -(1+z_d)\hat{\psi}_d[D_{od}(\vect{\theta}-\vect{\alpha}_{od})]
        }
= -\tau_{ds} \pa{\mat{1}- \mat{\Gamma}_{od}}
                \vect{\alpha}_{ods}(\vect{\theta}-\vect{\alpha}_{od}) \ ,
\\
\pd{T\e{pot}^>}{\vect{\theta}}
&= \pd{}{\vect{\theta}}
    \paac{
        -\sum_{l<d} (1+z_l) \hat{\psi}_l[D_{ol}(\vect{\theta}-\vect{\alpha}_{odl})]
        }
\\
&= \tau_{ds}
    \pac{
        - \sum_{l<d} \vect{\alpha}_{ols}(\vect{\theta}-\vect{\alpha}_{odl})
        + \textcolor{red4}{
            (\mat{1}-\mat{\Gamma}_{ods})
            \sum_{l>d} \frac{\tau_{ls}}{\tau_{dl}}
                        \vect{\alpha}_{ols} (\vect{\theta}-\vect{\alpha}_{odl})
            }
        } .
\label{eq:calculation_Fermat_potential_1}
\end{align}
Gathering all the above derivatives would then be proportional to $\vect{L}$ if not for the \textcolor{red4}{last term} of \cref{eq:calculation_Fermat_potential_1}. We may note, however, that up to $\order(\eps^4)$ terms that interloper reads
\begin{align}
(\mat{1}-\mat{\Gamma}_{ods})
\sum_{l>d} \frac{\tau_{ls}}{\tau_{dl}}
\vect{\alpha}_{ols}
&=
\pd{\vect{L}}{\vect{\theta}} \,
    \sum_{l>d} \frac{\tau_{ls}}{\tau_{dl}}
        \vect{\alpha}_{ols}
\\
&=
\pd{}{\vect{\theta}}
\pa{
    \vect{L} \cdot
    \sum_{l>d} \frac{\tau_{ls}}{\tau_{dl}}
        \vect{\alpha}_{ols}
    }   
- \pa{
        \sum_{l>d} \frac{\tau_{ls}}{\tau_{dl}}
        \pd{\vect{\alpha}_{ols}}{\vect{\theta}}
        } \vect{L}
\end{align}
so that finally
\begin{equation}
\pd{T}{\vect{\theta}}
= \tau_{ds}
    \pa{
        \mat{1} - \mat{\Gamma}_{od}
        - \sum_{l>d} \frac{\tau_{ls}}{\tau_{dl}}
            \pd{\vect{\alpha}_{ols}}{\vect{\theta}}
        } \vect{L}
    + \pd{}{\vect{\theta}}
        \underbrace{
        \pac{
            \vect{L}\cdot\sum_{l>d} \frac{\tau_{ds}\tau_{ls}}{\tau_{dl}}
            \vect{\alpha}_{ols}(\vect{\theta}-\vect{\alpha}_{odl}) }
            }_{\define -\Delta(\vect{\theta},\vect{\beta})} ,
\end{equation}
and we conclude that the Fermat potential~$\phi$ whose gradient yields the lens equation reads
\begin{equation}
\phi(\vect{\theta},\vect{\beta})
\define T(\vect{\theta},\vect{\beta}) + \Delta(\vect{\theta},\vect{\beta})
= T(\vect{\theta},\vect{\beta}) 
+ \pac{\vect{\theta}-\vect{\beta}-\vect{\alpha}(\vect{\theta})} \cdot
    \sum_{l>d} \frac{\tau_{ds}\tau_{ls}}{\tau_{dl}}
            \vect{\alpha}_{ols}(\vect{\theta}-\vect{\alpha}_{odl}) \ ,
\end{equation}
which indeed coincides with $T$ for physical rays where $\Delta = 0$.

\subsection{Tidal regime for the secondary deflectors}
\label{subsec:derivation_time_delay_tidal}

In this section we consider the case where all the deflectors but the dominant one can be treated in the tidal regime. This means that for $l\neq d$ the projected tidal matrix is constant, the deflection angle is linear, and the projected potential is quadratic,
\begin{align}
\mat{\Sigma}_l
&= \cst
\\
\hat{\vect{\alpha}}_l(\vect{x})
&= \hat{\vect{\alpha}}_l(\vect{x}_0) + 4\pi G\mat{\Sigma}_l (\vect{x}-\vect{x}_0)
\\
\hat{\psi}_l(\vect{x})
&= \hat{\psi}_l(\vect{x}_0) 
+ \hat{\vect{\alpha}}_l(\vect{x}_0)\cdot(\vect{x}-\vect{x}_0)
+ \frac{1}{2} (\vect{x}-\vect{x}_0)\cdot 4\pi G\mat{\Sigma}_l\,(\vect{x}-\vect{x}_0) \ ,
\end{align}
for any couple $(\vect{x}, \vect{x}_0)$ in the region under consideration.

\subsubsection{Geometrical term}

Let us apply the above to the first, geometrical, term of $T(\vect{\theta}, \vect{\beta})$ in \cref{eq:time_delay_dominant_appendix}. We shall directly introduce the partially unlensed direction~$\vect{\beta}'=\vect{\beta}+\vect{\alpha}_{os}(\vect{\beta}')$, which represents the direction in which the source would be observed in the absence of the main lens only.

At order 2 in $\eps$, that is at linear order in secondary deflections like $\vect{\alpha}_{od}$, we have
\begin{align}
T\e{geo}
&\define \frac{1}{2} \tau_{ds} \abs{\vect{\theta}-\vect{\alpha}_{od}(\vect{\theta})-\vect{\beta}}^2
\\
&= \frac{1}{2} \tau_{ds}
    \abs{
        \pa{1-\mat{\Gamma}\e{od}}\pa{\vect{\theta}-\vect{\beta}'}
        -\vect{\alpha}_{od}(\vect{\beta}')
        +\vect{\alpha}_{os}(\vect{\beta}')
    }^2
\\
&= \frac{1}{2} \tau_{ds} \pa{\vect{\theta}-\vect{\beta}'}
    \cdot \pa{1-2\mat{\Gamma}_{od}} \pa{\vect{\theta}-\vect{\beta}'}
    - \tau_{ds} \pa{\vect{\theta}-\vect{\beta}'}
        \cdot\pac{\vect{\alpha}_{od}(\vect{\beta}')-\vect{\alpha}_{os}(\vect{\beta}')} .
\end{align}

\subsubsection{Potential terms}

\paragraph{Main potential term} The main potential term in \cref{eq:time_delay_dominant_appendix} may be written in terms of $\vect{x}'_d=D_{od}(\mat{1}-\mat{\Gamma}_{od})\vect{\theta}$, which accounts for the shifted origin of the main lens plane,
\begin{equation}
T\e{pot}^d
= -(1+z_d) \hat{\psi}_d\pac{D_{od} (\mat{1}-\mat{\Gamma}_{od})\vect{\theta}} \ .
\end{equation}

\paragraph{Foreground potential terms} We then consider the $l<d$ potential terms in \cref{eq:time_delay_dominant_appendix}
\begin{align}
T\e{pot}^<
&\define -\sum_{l<d} (1+z_l) \hat{\psi}_l(D_{ol} \vect{\theta})
\\
&= -\sum_{l<d} (1+z_l)
    \pac{
        \hat{\psi}_l(D_{ol} \vect{\beta}')
        + D_{ol}(\vect{\theta}-\vect{\beta}')\cdot\hat{\vect{\alpha}}_l(D_{ol}\vect{\beta}')
        + \frac{1}{2} D_{ol}^2 (\vect{\theta}-\vect{\beta}')\cdot
                                4\pi G \mat{\Sigma}_l (\vect{\theta}-\vect{\beta}')
        } .
\end{align}
A few manipulations using the identity~\eqref{eq:identity_tau} yield
\begin{align}
(1+z_l) D_{ol} \hat{\vect{\alpha}}_l(D_{ol}\vect{\beta}')
&= \tau_{ds} \pac{\vect{\alpha}_{ols}(\vect{\beta}')-\vect{\alpha}_{old}(\vect{\beta}')}
\\
(1+z_l) D_{ol}^2 \, 4\pi G \mat{\Sigma}_l
&= \tau_{ds} \pa{\mat{\Gamma}_{ols}-\mat{\Gamma}_{old}} \ ,
\end{align}
where $\mat{\Gamma}_{ilj}$ was defined in \cref{eq:partial_amplification}. Therefore, the foreground potential terms amount to
\begin{multline}
T\e{pot}^<
= -\sum_{l<d} (1+z_l) \hat{\psi}_l(D_{ol} \vect{\beta}')
\\
    - \tau_{ds} (\vect{\theta}-\vect{\beta}')
                \cdot \pac{\vect{\alpha}_{os}^<(\vect{\beta}')-\vect{\alpha}_{od}(\vect{\beta}')}
    - \frac{1}{2} \tau_{ds} (\vect{\theta}-\vect{\beta}')
                            \cdot\pa{\mat{\Gamma}_{os}^<-\mat{\Gamma}_{od}}
                            (\vect{\theta}-\vect{\beta}') \ ,
\end{multline}
where the symbols~$\vect{\alpha}_{os}^<, \mat{\Gamma}_{os}^<$ refer to the foreground-lens contributions to $\vect{\alpha}_{os}, \mat{\Gamma}_{os}$. 

\paragraph{Background potential terms} The calculation of the $l>d$ potential terms in \cref{eq:time_delay_dominant_appendix} is similar to the foreground-term case, except that the original argument of $\hat{\psi}_l$ is more complicated. We indeed start from
\begin{align}
T\e{pot}^>
&\define -\sum_{l>d} (1+z_l) \hat{\psi}_l[D_{ol}(\vect{\theta}-\vect{\alpha}_{odl})]
\\
&= -\sum_{l>d} (1+z_l)
    \Big[
        \hat{\psi}_l(D_{ol} \vect{\beta}')
        + D_{ol}(\vect{\theta}-\vect{\beta}'-\vect{\alpha}_{odl})
            \cdot\hat{\vect{\alpha}}_l(D_{ol}\vect{\beta}')
        \nonumber\\&\hspace{5cm}
        + \frac{1}{2} D_{ol}^2 (\vect{\theta}-\vect{\beta}'-\vect{\alpha}_{odl})\cdot
                                4\pi G \mat{\Sigma}_l (\vect{\theta}-\vect{\beta}'-\vect{\alpha}_{odl})
    \Big] .
\end{align}
The first step consists in re-writing $\vect{\alpha}_{odl}$ as follows,
\begin{equation}
\vect{\alpha}_{odl}
= \frac{D_{dl} D_{os}}{D_{ol} D_{ds}} \, \hat{\vect{\alpha}}_{ods}
= \frac{\tau_{ds}}{\tau_{dl}} \, (\vect{\theta}-\vect{\beta}') + \order(\eps^2) \ .
\end{equation}
Since $\vect{\alpha}_{odl}$ is systematically multiplied with $\order(\eps^2)$ terms, we can thus use the last expression in our computations.

Performing again a few manipulations using \cref{eq:identity_tau}, we find
\begin{align}
D_{ol}(\vect{\theta}-\vect{\beta}'-\vect{\alpha}_{odl})
    \cdot\hat{\vect{\alpha}}_l(D_{ol}\vect{\beta}')
&= \tau_{ds} (\vect{\theta}-\vect{\beta}') \cdot \vect{\alpha}_{ols}(\vect{\beta}') \ ,
\\
D_{ol}^2 (\vect{\theta}-\vect{\beta}'-\vect{\alpha}_{odl})\cdot
    4\pi G \mat{\Sigma}_l (\vect{\theta}-\vect{\beta}'-\vect{\alpha}_{odl})
&= \tau_{ds} (\vect{\theta}-\vect{\beta}')\cdot\pa{\mat{\Gamma}_{ols} - \mat{\Gamma}_{dls}}
    (\vect{\theta}-\vect{\beta}') \ ,
\end{align}
so that
\begin{multline}
T\e{pot}^>
= -\sum_{l>d} (1+z_l) \hat{\psi}_l(D_{ol} \vect{\beta}')
    - \tau_{ds} (\vect{\theta}-\vect{\beta}')
                \cdot \vect{\alpha}_{os}^>(\vect{\beta}')
    - \frac{1}{2} \tau_{ds} (\vect{\theta}-\vect{\beta}')
                            \cdot\pa{\mat{\Gamma}_{os}^> - \mat{\Gamma}_{ds}}
                            (\vect{\theta}-\vect{\beta}') \ ,
\end{multline}
with $\vect{\alpha}_{os}^>, \mat{\Gamma}_{os}^>$ the background-lens contributions to $\vect{\alpha}_{os}, \mat{\Gamma}_{os}$.

\subsubsection{Final result}

Gathering all the terms computed above into $T=T\e{geo}+T\e{pot}^d+T\e{pot}^<+T\e{pot}^>$, we finally obtain
\begin{multline}
T(\vect{\theta},\vect{\beta})
= \frac{1}{2} \tau_{ds}
    (\vect{\theta}-\vect{\beta}')
    \cdot\pa{\mat{1}-\mat{\Gamma}_{od}-\mat{\Gamma}_{os}+\mat{\Gamma}_{ds}}
    (\vect{\theta}-\vect{\beta}')
    - (1+z_d) \hat{\psi}_d\pac{D_{od} (\mat{1}-\mat{\Gamma}_{od})\vect{\theta}} \\
    - \sum_{l\neq d} (1+z_l) \hat{\psi}_l(D_{od}\vect{\beta}') \ ,
\end{multline}
which is indeed the sum of \cref{eq:time_delay_tidal_main,eq:time_delay_tidal_correction}, and thereby concludes our proof.

\section{Calculation of the correlation functions of critical modes}
\label{app:calculation_correlation_functions}

In this appendix, we derive the expressions~\eqref{eq:xi_plus_result} and \eqref{eq:xi_minus_result} of the correlation functions~$\xi^\pm_{n_1 n_2}$ of the effective critical modes, as well as the associated power spectrum~\eqref{eq:power_spectrum_result}. The derivations are quite similar to those presented in appendix~A of ref.~\cite{Fleury:2018odh}; they are performed in the flat-sky and Limber approximations.

\subsection{Preliminaries: a simpler expression for $\xi_{n_1 n_2}^\pm$}

For two directions~$\vect{\vartheta}_1, \vect{\vartheta}_2$ in the flat sky separated by $\vect{\vartheta}=\vect{\vartheta}_1-\vect{\vartheta}_2=\vartheta(\cos\phi, \sin\phi)$, we defined the symmetric and anti-symmetric components of each critical mode $\bar{c}_n$ at the ends of $\vect{\vartheta}$ as
\begin{equation}
\bar{c}_n = (s_n + \ii a_n) \ex{\ii n\phi} \ ,
\qquad
s_n, a_n \in \mathbb{R}\ .
\end{equation}
The plus and minus correlation functions were then defined as
\begin{equation}
\xi^\pm_{n_1 n_2}(\vartheta)
= \ev{s_{n_1}(\vect{\vartheta}_1) s_{n_2}(\vect{\vartheta}_2)
    \pm a_{n_1}(\vect{\vartheta}_1) a_{n_2}(\vect{\vartheta}_2)} .
\end{equation}
The first step of the computation consists in expressing $\xi^\pm_{n_1 n_2}$ in terms of $\bar{c}_n$. For that purpose, we use the fact that the cross correlations between $s_n$ and $a_n$ vanish for symmetry reasons, which implies that
\begin{align}
\xi_{n_1 n_2}^+(\vartheta)
&\define \ex{-\ii(n_1-n_2)\phi}
\ev{\bar{c}_{n_1}(\vect{\vartheta}_1) \, \bar{c}_{n_2}^*(\vect{\vartheta}_2)}
\\
\xi_{n_1 n_2}^-(\vartheta)
&\define \ex{-\ii(n_1+n_2)\phi}
\ev{\bar{c}_{n_1}(\vect{\vartheta}_1) \, \bar{c}_{n_2}(\vect{\vartheta}_2)} \ .
\end{align}
These expressions will be easier to handle in the following.

\subsection{Introducing the matter power spectrum in Limber's approximation}

Taking the expression~\eqref{eq:critical_modes_effective} of $\bar{c}_n$, in which we substitute \cref{eq:critical_modes_cosmic} for $c_n$, and where we introduce the Fourier transform of the density contrast~$\delta$, we have
\begin{multline}
\label{eq:c_n_Fourier}
\bar{c}_n(\vect{\vartheta})
= 4\pi G\bar{\rho}_0 \int\dd^4\vect{\Pi} \; p(\vect{\Pi}) \int_0^\infty \dd\chi \; (1+z)
    \int_{\mathbb{R}^2} \frac{\dd^2\vect{w}}{2\pi} 
    \pac{
        W_{\mathcal{D}}(\chi, \vect{\Pi}) \mathcal{D}_n(\cplx{w})
        + \frac{W_{\mathcal{C}}(\chi, \vect{\Pi})}{2(1-\kappa\e{E})} \, \mathcal{C}_n(\cplx{w})
        }
\\
\times \int \frac{\dd^3\vect{k}}{(2\pi)^3} \; \ex{\ii \vect{k}\cdot\vect{x}} \delta(\eta, \vect{k}) \ ,
\end{multline}
where $\eta=\eta_0-\chi$, and $\vect{x}$ is the spatial position at a comoving distance $\chi$ from the observer and comoving angular position $f_K(\chi)\vect{\vartheta}+r(\chi; \vect{\Pi})\vect{w}$.

When taking the ensemble average of $\bar{c}_{n_1}(\vect{\vartheta}_1)\bar{c}_{n_2}(\vect{\vartheta}_2)$, the terms coming from the second line of \cref{eq:c_n_Fourier} yield
\begin{equation}
\int \frac{\dd^3\vect{k}_1}{(2\pi)^3}
\int \frac{\dd^3\vect{k}_2}{(2\pi)^3}\;
\ex{\ii(\vect{k}_1\cdot\vect{x}_1+\vect{k}_2\cdot\vect{x}_2)}
\ev{\delta(\eta_1, \vect{k}_1) \delta(\eta_2, \vect{k}_2)}
=
\int \frac{\dd^3\vect{k}}{(2\pi)^3} \;
\ex{\ii\vect{k}\cdot(\vect{x}_1-\vect{x}_2)}
P_\delta(\eta_1, \eta_2, k) \ ,
\end{equation}
where we introduced the power spectrum~$P_\delta$ of the density contrast and used statistical homogeneity and isotropy. Using Limber's approximation then further simplifies the above as
\begin{multline}
\int \frac{\dd^3\vect{k}}{(2\pi)^3} \;
\ex{\ii\vect{k}\cdot(\vect{x}_1-\vect{x}_2)}
P_\delta(\eta_1, \eta_2, k)
\\
\approx
\frac{\delta\e{D}(\chi_1-\chi_2)}{f_K^2(\chi_1)}
\int \frac{\dd^2\vect{\ell}}{(2\pi)^2} \;
\ex{\ii\vect{\ell}\cdot\vect{\vartheta}} \,
\ex{\ii\vect{\ell}\cdot[\eps(\chi_1; \vect{\Pi}_1)\vect{w}_1-\eps(\chi_1; \vect{\Pi}_2)\vect{w}_2]
    }
P_\delta\pac{\eta_1, \frac{\ell}{f_K(\chi_1)}} \ ,
\end{multline}
with $\eps(\chi;\vect{\Pi})\define r(\chi;\vect{\Pi})/f_K(\chi)$, so that the correlation functions take the form
\begin{align}
\label{eq:xi_+_intermediate}
\xi^+_{n_1 n_2}(\vartheta)
&= (4\pi G\bar{\rho}_0)^2
    \int \frac{\dd^2\vect{\ell}}{(2\pi)^2} \; \ex{\ii[\vect{\ell}\cdot\vect{\vartheta}-(n_1-n_2)\phi]}
\nonumber\\ &\quad\times
    \int_0^\infty \dd\chi \; (1+z)^2
        g_{n_1}(\chi, \vect{\ell}) g_{n_2}^*(\chi, \vect{\ell})
        P_\delta\pac{\eta, \frac{\ell}{f_K(\chi)}} ,
\\
\label{eq:xi_-_intermediate}
\xi^-_{n_1 n_2}(\vartheta)
&= (4\pi G\bar{\rho}_0)^2
    \int \frac{\dd^2\vect{\ell}}{(2\pi)^2} \; \ex{\ii[\vect{\ell}\cdot\vect{\vartheta}-(n_1+n_2)\phi]}
\nonumber\\ &\quad\times
    \int_0^\infty \dd\chi \; (1+z)^2
        g_{n_1}(\chi, \vect{\ell}) g_{n_2}(\chi, -\vect{\ell})
        P_\delta\pac{\eta, \frac{\ell}{f_K(\chi)}} ,
\end{align}
where the function $g_{n}(\chi, \vect{\ell})$ encodes most of the difficulties of the calculation and reads
\begin{equation}
g_n(\chi, \vect{\ell})
\define
\int\dd^4\vect{\Pi} \; p(\vect{\Pi})
\int_{\mathbb{R}^2} \frac{\dd^2\vect{w}}{2\pi} \; \ex{\ii\vect{\ell}\cdot\eps(\chi; \vect{\Pi})\vect{w}}
    \pac{
        \frac{W_{\mathcal{D}}(\chi, \vect{\Pi})}{f_K^2(\chi)} \, \mathcal{D}_n(\cplx{w})
        + \frac{W_{\mathcal{C}}(\chi, \vect{\Pi})}{2(1-\kappa\e{E}) f_K^2(\chi)}
        \, \mathcal{C}_n(\cplx{w})
        } .
\end{equation}

\subsection{Calculation of the complex integrals in $g_n$}

We may now perform the integral over $\vect{w}$ in the expression of $g_n(\chi, \vect{\ell})$. Omitting the dependencies of $\eps$ for short, and using several properties of the Bessel functions, we find
\begin{align}
\int_{\mathbb{R}^2} \frac{\dd^2\vect{w}}{2\pi}
\; \ex{\ii\vect{\ell}\cdot\eps\vect{w}} \, \mathcal{D}_n(\cplx{w})
&=
\ii^n\ex{\ii n\phi_\ell} \,
\frac{\delta_{n1}+J_{n+1}(\ell\eps)-J_{n-1}(\ell\eps)}{\ell\eps} \ ,
\\
\int_{\mathbb{R}^2} \frac{\dd^2\vect{w}}{2\pi}
\; \ex{\ii\vect{\ell}\cdot\eps\vect{w}} \, \mathcal{C}_n(\cplx{w})
&=
\ii^n\ex{\ii n\phi_\ell} \,
\frac{(n+1)J_{n+1}(\ell\eps)+(n-1)J_{n-1}(\ell\eps)}{\ell\eps} \ ,
\end{align}
where $\phi_\ell$ denotes the polar angle of $\vect{\ell}$. Thus, $g_n$ takes the form
\begin{equation}
g_n(\chi, \vect{\ell})
=
\ii^n\ex{\ii n\phi_\ell}
\bar{q}_n(\chi,\ell) \ ,
\qquad
\bar{q}_n(\chi,\ell)
\define
\int\dd^4\vect{\Pi} \; p(\vect{\Pi}) \,
q_n(\chi, \ell; \vect{\Pi}) \ ,
\end{equation}
with, including all the dependencies,
\begin{multline}
q_n(\chi, \ell; \vect{\Pi})
\define
\frac{W_{\mathcal{D}}(\chi; \vect{\Pi})}{f_K^2(\chi)}
\frac{
    \delta_{n1}
    +J_{n+1}[\ell\eps(\chi;\vect{\Pi})]
    -J_{n-1}[\ell\eps(\chi;\vect{\Pi})]
    }{\ell\eps(\chi;\vect{\Pi})}
\\
+ \frac{W_{\mathcal{C}}(\chi; \vect{\Pi})}{2(1-\kappa\e{E}) f_K^2(\chi)}
    \frac{
        (n+1) J_{n+1}[\ell\eps(\chi;\vect{\Pi})]
        + (n-1) J_{n-1}[\ell\eps(\chi;\vect{\Pi})]
        }{\ell\eps(\chi;\vect{\Pi})} \ .
\end{multline}
Due to the many identities satisfied by Bessel functions, the above may be re-written in various ways. Among them, there are ($x\define\ell\eps$)
\begin{align}
\frac{J_{n+1}(x)-J_{n-1}(x)}{x}
&= -\frac{2J'_n(x)}{x} \ ,
\\
\frac{(n+1)J_{n+1}(x)+(n-1)J_{n-1}(x)}{x}
&= \frac{J_{n-2}(x) + 2J_n(x) + J_{n+2}(x)}{4}
\\
&= \frac{2n^2 J_n(x)}{x^2} - \frac{2J'_n(x)}{x}
\\
&= J_n''(x) + J_n(x) \ .
\end{align}

For $n\geq 2$, the first term of $q_n$, proportional to $W_{\mathcal{D}}$, is identical to the finite-beam kernel derived in ref.~\cite{Fleury:2018odh} except for a factor 2. This factor comes from differences between the definition of critical modes~$c_n$ in the present article and the reduced image moments~$\mu_n$ defined in ref.~\cite{Fleury:2018odh}. Namely, those moments were defined such that $\mu_n$ was the $n\h{th}$ Fourier mode of the complex contour~$\cplx{\theta}(\ph)=\theta(\ph)\ex{\ii\ph}$ of an image. This eventually implies that the correspondence between those $\mu_n$ and the direct foreground contribution to $c_n$ is $\mu_n\leftrightarrow 2c_n\h{direct}$.

\subsection{Integrals over $\phi_\ell$}

The last step of the calculation consists in performing the integration over the polar angle~$\phi_\ell$ of $\vect{\ell}$ in \cref{eq:xi_+_intermediate,eq:xi_-_intermediate}. Accounting for the $\ii^n\ex{\ii n\phi_\ell}$ terms coming from $g_n(\chi, \vect{\ell})$, the relevant integral for $\xi^+_{n_1 n_2}$ is found to read
\begin{equation}
\ii^{n_1-n_2}
\int \frac{\dd^2\vect{\ell}}{(2\pi)^2} \; \ex{\ii[\vect{\ell}\cdot\vect{\vartheta}+(n_1-n_2)(\phi_\ell-\phi)]}
= \int_0^\infty \frac{\ell\dd\ell}{2\pi} \; J_{n_2-n_1}(\ell\vartheta) \ ,
\end{equation}
while for $\xi^-_{n_1 n_2}$ we have
\begin{equation}
(-1)^{n_2}\ii^{n_1+n_2}
\int \frac{\dd^2\vect{\ell}}{(2\pi)^2} \; \ex{\ii[\vect{\ell}\cdot\vect{\vartheta}+(n_1+n_2)(\phi_\ell-\phi)]}
= (-1)^{n_1}\int_0^\infty \frac{\ell\dd\ell}{2\pi} \; J_{n_2+n_1}(\ell\vartheta) \ .
\end{equation}
Gathering all the pieces together, we get the final result,
\begin{align}
\xi^+_{n_1 n_2}(\vartheta)
&= \int_0^\infty \frac{\ell\dd\ell}{2\pi} \; J_{n_2-n_1}(\ell\vartheta) \, P_{n_1 n_2}(\ell) \ ,
\\
\xi^-_{n_1 n_2}(\vartheta)
&= (-1)^{n_1}\int_0^\infty \frac{\ell\dd\ell}{2\pi} \; J_{n_1+n_2}(\ell\vartheta) \, P_{n_1 n_2}(\ell) \ ,
\\
P_{n_1 n_2}(\ell)
&\define
(4\pi G\bar{\rho}_0)^2
\int_0^\infty \dd\chi \; (1+z)^2
\bar{q}_{n_1}(\chi,\ell)
\bar{q}_{n_2}(\chi,\ell) \,
P_\delta\pac{\eta_0-\chi, \frac{\ell}{f_K(\chi)}} \ .
\end{align}

\bibliographystyle{JHEP.bst}
\bibliography{bibliography.bib}

\end{document}